\documentclass[12pt,a4paper]{article}            
 \usepackage[skins,theorems]{tcolorbox}
\tcbset{highlight math style={enhanced,
  colframe=red,colback=white,arc=0pt,boxrule=1pt}}
  \usepackage[bookmarksopen, bookmarksnumbered, bookmarksopenlevel=2]{hyperref}
  \usepackage{tikz}
  \usepackage{tikz-3dplot}
 \usetikzlibrary{calc}
 \usetikzlibrary{decorations} %
 \usepackage[UKenglish]{babel}
 \usepackage[toc,page]{appendix}
 \usepackage{amsmath}
 \usepackage{amssymb}
 \usepackage{graphicx}
 \usepackage{hhline}
 \usepackage[bf]{caption}
\usepackage{cite}
\usepackage[vcentermath]{youngtab}
\usepackage{geometry}
\usepackage{slashed}
\usepackage{color}
\usepackage{stackrel}
\usepackage{tikz-cd} 
\usepackage{cancel} 
\usepackage{booktabs}

\usepackage{mathtools}

\usepackage{tikz}
\usetikzlibrary{shadows} 
\usepackage[framemethod=tikz]{mdframed}

\global\mdfdefinestyle{myboxstyle}{%
  shadow=true,
  linecolor=black,
  shadowcolor=black,
  shadowsize=6pt,
  nobreak=false,
  innertopmargin=10pt,
  innerbottommargin=10pt,
  leftmargin=5pt,
  rightmargin=5pt,
  needspace=1cm,
  skipabove=10pt,
  skipbelow=15pt,
  middlelinewidth=1pt,
  afterlastframe={\vspace{5pt}},
  aftersingleframe={\vspace{5pt}},
  tikzsetting={%
draw=black,
very thick} }

\newmdenv[style=myboxstyle]{whitebox} \newmdenv[style=myboxstyle,backgroundcolor=black!20]{graybox}

\newmdenv[style=myboxstyle,nobreak=true]{blockwhitebox}
\newmdenv[style=myboxstyle,backgroundcolor=black!20,nobreak=true]{blockgraybox}
\newmdenv[nobreak=true,hidealllines=true]{blockbox}

\usepackage{empheq}
\usepackage{arydshln}

\newcommand{\bqa}{\begin{eqnarray}}
\newcommand{\eqa}{\end{eqnarray}}

\newcommand{\nn}{\nonumber}

\def\qr{r}

\def\utau{\Omega}

\def\qt{q_T}
\def\qu{q_U}
\def\qv{q_V}
\def\T{T}
\def\U{U}
\def\V{V}
\def\imt{{\rm Im}}
\def\ret{{\rm Re}\,}

\hypersetup{
    pdftitle={},
    pdfauthor={},
    pdfsubject={}
}
\numberwithin{equation}{section}
\numberwithin{table}{section}

\makeatletter

\newcommand{\be}{\begin{equation}}
\newcommand{\ee}{\end{equation}}
\newcommand{\beq}{\begin{equation}}
\newcommand{\eeq}{\end{equation}}
\newcommand{\ba}{\begin{aligned}}
\newcommand{\ea}{\end{aligned}}

\newcommand{\bea}{\begin{eqnarray}}
\newcommand{\eea}{\end{eqnarray}}

\newcommand{\cF}{\mathcal{F}}

\newcommand{\cS}{\mathcal{S}}

\newcommand\bi{\begin{itemize}}
\newcommand\ei{\end{itemize}}

\def\unit{{1\kern-.65ex {\rm l}}}
\def\1{{1\kern-.65ex {\rm l}}}

\def\IZ{\mathbb{Z}}
\def\IP{\mathbb{P}}

\def\utau{\Omega}
\def\qt{q_T}
\def\qr{q_U}

\def\utau{\Omega}

\def\qt{q_T}
\def\qu{q_U}
\def\qv{q_V}
\def\T{T}
\def\U{U}
\def\V{V}
\def\imt{{\rm Im}}
\def\ret{{\rm Re}\,}

\newcount\hour \newcount\minute
\hour=\time \divide \hour by 60
\minute=\time
\def\now{%
\ifnum \hour<13
  \ifnum \hour=0 \advance \hour by 12 \number\hour:\else \number\hour:\fi%
     \ifnum \minute<10 0\fi%
     \number\minute%
\ A.M.%
\else \advance \hour by -12 \number\hour:%
  \ifnum \minute<10 0\fi%
  \number\minute%
  \ P.M.%
\fi%
}

\def\fnote#1#2{\begingroup\def\thefootnote{#1}\footnote{#2}
     \addtocounter{footnote}{-1}\endgroup}

\makeatother

\begin{document}

\begin{flushright}
{\tt\normalsize {CERN-TH-2021-214}}\\
{\tt\normalsize CTPU-PTC-21-41}\\
{\tt\normalsize  ZMP-HH/21-24}
\end{flushright}

\vskip 40 pt
\begin{center}
{\large \bf 
Physics of Infinite Complex Structure Limits in eight Dimensions
}

\vskip 11 mm

Seung-Joo Lee${}^{1}$, Wolfgang Lerche${}^{2}$, 
and Timo Weigand${}^{3}$

\vskip 11 mm
\small ${}^{1}${\it Center for Theoretical Physics of the Universe, \\ Institute for Basic Science, Daejeon 34126, South Korea} \\[3 mm]
\small ${}^{2}${\it CERN, Theory Department, \\ 1 Esplande des Particules, Geneva 23, CH-1211, Switzerland} \\[3 mm]
\small ${}^{3}${\it II. Institut f\"ur Theoretische Physik, Universit\"at Hamburg, \\  Luruper Chaussee 149, 22607 Hamburg, Germany } \\[3 mm]
\phantom{\small ${}^{3}$}{\it Zentrum f\"ur Mathematische Physik, Universit\"at Hamburg, \\ Bundesstrasse 55, 20146 Hamburg, Germany  }   \\[3 mm]

\fnote{}{Email: seungjoolee at ibs.re.kr, wolfgang.lerche at cern.ch, 
timo.weigand at desy.de}

\end{center}

\vskip 7mm

\begin{abstract}

We investigate infinite distance limits in the complex structure moduli space of F-theory compactified on K3 to eight dimensions. While this is among the simplest possible arenas to
test ideas about the Swampland Distance Conjecture, it is nevertheless non-trivial enough to improve our
understanding of the physics for these limiting geometries, including phenomena of emergence. It also provides a perspective on
infinite distance limits from the viewpoint of open strings.
The paper has two quite independent themes. In the main part we 
show that all degenerations of elliptic K3 surfaces at infinite distance as analysed in the companion paper \cite{Kulikov} can be interpreted as (partial) decompactification or emergent string limits in F-theory, in agreement with the Emergent String Conjecture. We present a unified geometric picture of the possible towers of states that can become light  
and illustrate our general claims via the connection between Kulikov models of degenerating K3 surfaces and the dual heterotic string.
As an application we classify the possible maximal non-abelian Lie algebras and their Kac-Moody and loop extensions that can arise in the infinite distance limits. 
In the second part we discuss the infinite distance behaviour of certain exact quartic gauge couplings. We encounter a tension with the hypothesis that effective couplings should be fully generated by integrating out massive states. We show 
that by appropriately renormalizing the string coupling, at least partial emergence can be achieved.

\end{abstract}

\vfill

\thispagestyle{empty}
\setcounter{page}{0}
\newpage

\tableofcontents

\setcounter{page}{1}
\newpage

\section{Introduction}

Understanding the structure of the landscape of consistent quantum gravity theories, as opposed to the swampland of effective theories without a UV completion involving gravity,
is one of the most ambitious goals in modern theoretical physics. Numerous criteria defining the boundary between both types of theories have been proposed within the Swampland Program initiated in \cite{Vafa:2005ui}.
Among these, the Swampland Distance Conjecture \cite{Ooguri:2006in} arguably plays a central role: It underlies the de Sitter Conjecture in asymptotic regions of moduli space \cite{Ooguri:2018wrx}, 
serves as inspiration for the Anti-de Sitter Conjecture \cite{Lust:2019zwm}, the Spin-2 Conjecture \cite{Klaewer:2018yxi} and the Gravitino Mass Conjecture \cite{Cribiori:2021gbf,Castellano:2021yye}, predicts a holographically dual CFT Distance Conjecture \cite{Baume:2020dqd,Perlmutter:2020buo} and, quite generally, illustrates the tension between parametrically large field traversions
in cosmology and validity of the effective field theory, to name but a few aspects.\footnote{For further details and a collection of related works we refer to the reviews \cite{Brennan:2017rbf,Palti:2019pca,vanBeest:2021lhn,Grana:2021zvf}.} The conjecture asserts that at infinite distance in moduli space, a tower of states becomes light exponentially fast in any quantum gravity theory.
From a conceptual point of view, the fascination behind this claim lies in the fact that every quantum gravity theory must enter a new phase at infinite distance in its moduli space where the original effective description breaks down.

Given its special status within the Swampland Program, the Distance Conjecture clearly begs for a more fundamental explanation. In \cite{Lee:2019wij} it was proposed that the relevant towers of states, which become light at the parametrically fastest rate on the infinite boundaries of moduli space, are either a (dual) Kaluza-Klein tower or the tower of excitations of an emergent asymptotically weakly-coupled fundamental string.
If generally true this would drastically simplify our global picture of the quantum gravity landscape, to the extent that it would reduce the boundaries of moduli space in the asymptotic regime to well controlled theories.\footnote{A related, but independent idea stressing the role of strings for the asymptotics of moduli space is the Distant Axionic String Conjecture of \cite{Lanza:2020qmt,Lanza:2021udy}.}
This would demystify the breakdown of the effective theory at infinite distance and also offer a new perspective on the Emergence Proposal of \cite{Harlow:2015lma,Heidenreich:2018kpg,Grimm:2018ohb}.\footnote{An explanation of the Swampland Distance Conjecture based on entropy arguments was recently proposed in \cite{Hamada:2021yxy}.}

This Emergent String Conjecture has been successfully tested for a number of different corners of string and M-theory in various dimensions and in situations with as little as four supercharges. Apart from higher dimensional compactifications \cite{Lee:2019xtm}, this includes $N=1$ supersymmetric compactifications of F-theory to six \cite{Lee:2018urn} and four \cite{Lee:2019tst,Klaewer:2020lfg} dimensions and of M-theory to five \cite{Lee:2019wij} and four \cite{Xu:2020nlh} dimensions, as well as 4d $N=2$ supersymmetric  compactifications of Type IIA \cite{Lee:2019wij} and Type IIB \cite{Baume:2019sry}.
In all these cases, the moduli space whose infinite distance limits were investigated corresponds to the K\"ahler moduli space of the underlying Calabi-Yau variety. Intricate features of K\"ahler geometry guarantee that whenever the parametrically lightest tower of states does not correspond to a Kaluza-Klein tower, a unique (oftentimes solitonic) string emerges. The geometry works in such a way that its excitations lie parametrically at the same scale as a Kaluza-Klein tower and there exists a duality frame in which the emergent string is weakly coupled. Depending on the nature of the moduli space under investigation \cite{Baume:2019sry,Klaewer:2020lfg}, this rests on a remarkable conspiracy of classical and quantum effects. 

By contrast, an explicit confirmation or falsification of the Emergent String Conjecture in the complex structure moduli space of string compactifications has so far proven to be very difficult. 
In \cite{Grimm:2018ohb,Grimm:2018cpv,Joshi:2019nzi,Gendler:2020dfp,Palti:2021ubp} the Swampland Distance Conjecture as such has been quantitatively confirmed for Type IIB compactifications near the asymptotic boundary of complex structure moduli space, by arguing for the appearance of a tower of BPS states from D3-branes wrapping asymptotically vanishing 3-cycles.
The mathematical structure of the boundaries of complex structure moduli spaces has been scrutinised from the point of view of asymptotic Hodge theory in a series of further works \cite{Grimm:2020cda,Grimm:2021ikg,Bastian:2021eom}. This program in particular has 
lead to important insights into the possibilities of moduli stabilisation in flux compactifications \cite{Grimm:2019ixq,Bastian:2021hpc,Grimm:2021ckh}, a topic of central relevance in string theory. 
Irrespective of this progress, it is fair to say that a clear physics interpretation of the infinite towers of states, or even an identification of the parametrically leading towers, is yet to be obtained. By mirror symmetry, Type IIB string theory on Calabi-Yau threefolds is equivalent to corresponding Type IIA compactifications, for which the asymptotically massless states at infinite distance in K\"ahler moduli space have a clear interpretation as Kaluza-Klein states or emergent string excitations \cite{Lee:2019wij}.
It is therefore desirable to obtain a comparable understanding of the asymptotic physics directly from the point of view of the complex structure moduli space.

\begin{figure}[t!]
\centering
 \includegraphics[height=7cm]{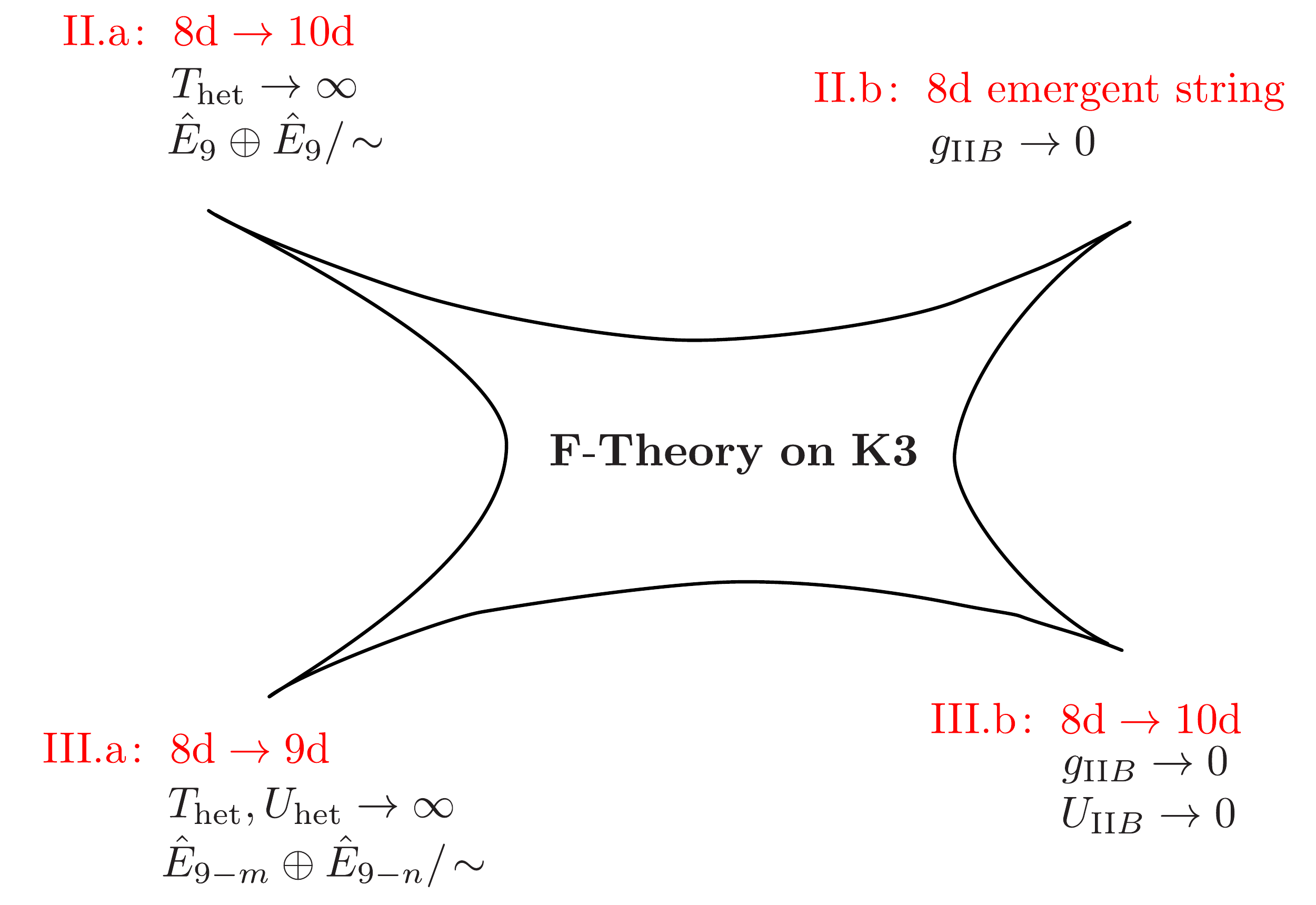}
\caption{Infinite distance complex structure degenerations for F-theory on elliptic K3 surface, and the physics in the various asymptotic regimes.\label{fig:Fmodspace}}
\end{figure}

In this paper we embark on this program in the arguably simplest non-trivial context, namely of F-theory compactified to eight dimensions on elliptically fibered K3 surfaces. While at first sight eight-dimensional theories might look straightforward to analyse, especially when viewed as compactifications of the heterotic string on $T^2_{\rm het}$, the structure of infinite distance limits turns out to be surprisingly intricate from the geometric point of view. 
The focus of our work is therefore on 
{\it understanding} the geometry of complex structure degenerations and its match to physics at infinite distance. Our analysis builds on the refinement of the so-called Type III Kulikov models for elliptic K3 surfaces that was elaborated on in the companion paper \cite{Kulikov}, as well as in the independent series of works of \cite{Brunyantethesis,alexeev2021compactifications}. 
We will obtain perfect agreement between
the refined Kulikov models and the possible decompactification or emergent string limits as suggested by the Emergent String Conjecture. For an illustration, see Figure \ref{fig:Fmodspace}.
We will give a more detailed summary of the main points of this analysis, which are the subject of Sections \ref{Sec_LoopAlgebras} - \ref{sec_Maximal9d}, in the second part of this introduction.

Moreover, in the largely independent Section \ref{emergence} of this paper, we comment on some other aspects of emergence, profiting from the canonical, highly constrained structure of the theory in eight dimensions. Specifically, we address the interesting proposal  \cite{Harlow:2015lma,Heidenreich:2017sim,Heidenreich:2018kpg,Grimm:2018ohb,Corvilain:2018lgw,Palti:2019pca} that gauge couplings
(and by courageous extrapolation, all couplings) emerge in the IR just from integrating out loops of massive states. In particular,  their divergence in the decompactification limit arises from
integrating out an asymptotically massless tower of KK states or string excitations.
 In geometric duality frames, such as F-theory, singularities of gauge couplings generally arise from 
 degenerating geometries, for example, from D-branes tied to
collapsing or coincident cycles; the prime example being the conifold singularity of Calabi-Yau manifolds.  One customarily says that the geometry automatically integrates out the asymptotically massless states.

However, these singularities arise at tree-level of the geometric compactifications, and the question arises whether there
exist other duality frames where the singularity, or rather the RG flow for that matter, literally arises from integrating out of loops of massive states. For the conifold singularity in Type II strings compactified on Calabi-Yau threefolds, it is known that the
relevant dual perturbative frame is the heterotic string, for which the singularity in the coupling arises at one-loop order due to a massless state circulating in the loop \cite{Strominger:1995cz}.

While this picture is very appealing, it is not a priori guaranteed that for any kind of large distance singularity in the moduli space there exists a duality frame for which the singularity arises from perturbative loops of asymptotically massless states. In fact, already in the early literature \cite{Ooguri:2006in} it was pointed out that this picture may fail for the special case of Kaluza-Klein gauge symmetries, 
for which the large distance singularity arises from
 to the diverging volume of the compactification space.
For such a case one expects at most partial emergence, in the sense that only part of the divergence may arise at the quantum level \cite{Heidenreich:2017sim,Heidenreich:2018kpg,Palti:2019pca}.

 We will address this issue for eight-dimensional theories, which are very restricted in that the perturbative dual frame is unambiguously the heterotic string compactified on $T_{\rm het}^2$, while non-perturbative effects are absent.
 An extra benefit is that it is the quartic, and not the quadratic gauge couplings that are dimensionless, and this disentangles the issues of gauge fields becoming dynamical at low energies, metric on moduli space, and the running of one-loop diagrams. That is, the quadratic gauge couplings have positive mass dimension and so vanish in the UV; in this sense, gauge dynamics emerges in the IR.
But these couplings are not renormalized at one loop order, so one cannot really say that it is the integrating-out of massive states that makes these fields dynamical.

As mentioned, the relevant one-loop exact, BPS saturated gauge couplings in eight dimensions are quartic, and these share a lot (but not all) of the properties of the well-investigated quadratic couplings of $N=2$ supersymmetric strings in four dimensions.
We will focus only on such couplings of a certain kind, namely on those of the $U(1)$ gauge fields that arise via KK reduction from ten dimensions. Our point will be to show that certain 1-loop amplitudes, in fact, do not exhibit any singularity in the decompactification limit, but rather all the divergence arises at tree level.  This seems to be in tension even with just partial emergence.\footnote{
As we will mention later, this phenomenon appears also in four dimensions and is therefore not an artifact of eight dimensions.} 

Subsequently we refine the discussion and argue how this situation can be ameliorated, by sweeping part of the problem into a renormalization of the string coupling. In this way, at least partial emergence can be recovered.

The lesson is that the notion of emergence as the phenomenon of literally integrating out states at the quantum level seems to be overly restrictive, at least in the current framework.\footnote{A way out might be that the quantum emergence of KK gauge fields is inherited from an emergence 
of gravity itself in the higher dimensional theory \cite{Palti:2019pca}.  We thank Irene Valenzuela for a discussion of this possibility.}     Because of dualities, there is no absolute distinction between what may be called tree-level or one-loop effect anyway. 
Indeed the key features of the Swampland Distance Conjecture, such as the appearance of massless towers of states at infinite distances in moduli space, can be well captured in a purely geometric formulation at tree level, irrespective of whether or not there exists a duality frame for which the singularity arises at the quantum level.

\subsection*{Physical interpretation of Kulikov models}

After this general overview, we now summarize in more detail the results of Sections \ref{Sec_LoopAlgebras} - \ref{sec_Maximal9d}, which concern the infinite distance limits in the complex structure moduli space of F-Theory on K3, and in particular their physical interpretation. 

According to the classic theory of semi-stable degenerations \cite{FriedmanMorrison}, a K3 surface generally splits into a union of several  surface components at infinite distance in complex structure moduli space.
The possible degenerations go by the name of Kulikov Type II and Type III models \cite{Kulikov1,Kulikov2,PerssonPink,Persson}.\footnote{Type I models appear at finite distance  in moduli space and are not of interest to us here.}  The theory of asymptotic Hodge structures guarantees the existence of one (Type III) or two (Type II) 
transcendental elliptic curves whose calibrated volumes vanish at infinite distance. From the point of view of M-theory on the degenerating space, we therefore expect one or two towers of asymptotically massless states
from M2-branes wrapped an arbitrary number of times along these vanishing cycles. This parallels the situation for Type IIB compactifications on Calabi-Yau threefolds, where towers of states appear from wrapped D3-branes along
asymptotically vanishing 3-cycles, as in \cite{Grimm:2018ohb,Grimm:2018cpv,Joshi:2019nzi,Gendler:2020dfp,Palti:2021ubp}. 
However, it turns out that these states form in general only part of the of the asymptotically light towers. 
This makes it crucial to understand the nature of the asymptotically vanishing cycles and to interpret the physics of the associated states.

It is at this stage that a refined geometric picture of the degeneration comes into play. 
As is well-known to string theorists already from the works \cite{Morrison:1996na,Morrison:1996pp,Aspinwall:1997ye}, the elliptic Type II Kulikov degenerations fall into two qualitatively different sectors, called Type II.a and II.b in \cite{Clingher:2003ui}.

Type II.a models describe the famous stable degeneration limits where the elliptic K3 breaks up into the union of two rational elliptic surfaces intersecting over an elliptic curve. Such limits are  dual to the $T^2$ compactification of the heterotic string in the limit of large volume, $T_{\rm het} \to  i \infty$. 
Indeed, we will argue that the two towers of wrapped M2-branes have a clear interpretation in F-theory in terms of string junctions: these form the two imaginary roots that enhance the non-abelian gauge algebra to the double-loop algebra 
\be
G_{\infty} =  (\hat E_9 \oplus \hat E_9)/\sim   \qquad \quad \text{(Type II.a)}   \,,
\ee
 where the quotient indicates that the two imaginary roots of the first double-loop algebra $\hat E_9$ and those of the second $\hat E_9$ are identified. 
These string junctions are the F-theoretic incarnation of two towers of Kaluza-Klein states in the dual heterotic frame, as expected in the stable degeneration limit. This establishes the Type II.a limits as complete decompactification limits from 8d to 10d.

By contrast, the Type II.b Kulikov models represent limits in which the 10d axio-dilaton diverges as $\tau_{\rm IIB} \to  i \infty$, corresponding to weakly coupled Type IIB orientifold limits. 
While it is true that the two towers of states from wrapped M2-branes along the two vanishing cycles become asymptotically light, they do not furnish the only parametrically leading light states. 
Rather, an M2-brane wrapped on a degenerating 1-cycle in the fiber gives rise to an asymptotically tensionless string, which is identified with the perturbative Type IIB string. Its (denser) tower of excitations sits at the same scale as the M2-brane states, and the limit is to be interpreted as an emergent string limit \cite{Lee:2019wij} in 8d, rather than a decompactification limit.

While the Type II limits have been known from the classic works \cite{Morrison:1996pp,Aspinwall:1997ye,Clingher:2003ui}, the remaining Type III limits at infinite distance have only recently been understood systematically for elliptic K3 surfaces \cite{Kulikov,Brunyantethesis,alexeev2021compactifications}, by analysing their associated Weierstrass models. In particular, the refined classification in terms of Type III.a and III.b limits 
has been established in our companion paper \cite{Kulikov}  in this way.
In both cases, the Weierstrass model of the degenerate K3 surface breaks up into a chain of possibly degenerate elliptic fibrations, where the generic elliptic fibers of the $i^{\rm th}$ components are of Type I$_{n_i}$ in the language of Kodaira's classification of elliptic fibrations.\footnote{Recall that a smooth elliptic fiber is of Type I$_0$, while a fiber of Type I$_k$ with $k>1$ is a nodal curve and the result of shrinking the $(k,0)$ cycle.}
It is guaranteed that $n_i >0$ for the middle components and hence their fibrations degenerate, while three possibilities arise for the end components:
\bi
\item Both end components are rational elliptic surfaces (Type III.a of first kind);
\item one end coomponent is a rational elliptic surface and the other one is a surface with generic I$_{n>0}$ fibers (Type III.a of second kind);
\item both end components are surfaces with generic I$_{n>0}$ fibers (Type III.b).
\ei

Type III.a models are similar to the Type II.a models described above, in the following sense: 
At the intersection of the rational elliptic end component(s) with the adjacent surface component with generic fibers of Type I$_{n>0}$, one vanishing transcendental torus is localised.
M2-branes wrapped thereon can be interpreted as string junctions responsible for the enhancement of an $E_{9-n}$ algebra to the affine algebra $\hat E_{9-n}$. For Type III.a limits with two rational elliptic end components, the symmetry algebra hence contains an affine algebra  
\be
G_{\infty} \supset (\hat E_{9-n} \oplus \hat E_{9-m})/\sim \,, \quad  1\leq n,m \leq 9 \,, \qquad \quad \text{(Type III.a of first kind)}  \,,  
\ee
while in models with only a single rational elliptic component only one affine algebra factor can arise. 
The imaginary root responsible for the affinisation of the symmetry algebra gives rise to a single tower of Kaluza-Klein modes, and the limit is therefore a partial decompactification limit from 8d to 9d.

Type III.b limits, on the other hand, are weak coupling limits in which $\tau_{\rm IIB} \to  i \infty$, while at the same time the complex structure of the Type IIB torus diverges. Such limits must therefore be full decompactifications from 8d to 10d.
This might seem, at first sight, to be in tension with the fact that one finds only a single, rather than two independent light towers from wrapped M2-branes. This tower can be interpreted as the winding tower along the small 1-cycle on $T^2_{\rm IIB}$. 
The missing second tower must correspond to a proper supergravity Kaluza-Klein tower  that cannot be detected as easily from geometry, even though the physical interpretation of the limit is unambiguous.

As explained before, the limits of Type II.a and III.a correspond to decompactification limits in the heterotic duality frame, where the infinite distance limit is to be understood as the limit
\bi
\item $T_{\rm het} \to i \infty$ with $U_{\rm het}$ finite (Type II.a);
\item $T_{\rm het} \to i \infty$ with $T_{\rm het}/U_{\rm het}$ finite (Type III.a). 
\ei
In the original F-theory duality frame, by contrast, the infinite directions in moduli space can be interpreted as non-compact directions in the moduli space of 7-branes.
While naively one might think that the open string or brane moduli space, on a compactification space of finite volume and at finite values of the complex structure, should be compact, this is in general not true if one includes
mutually non-local 7-branes. The non-compact directions appear when certain such $(p,q)$ branes approach each other. More precisely, as explained in \cite{DeWolfe:1998eu}, the affinisation of an $E_n$ algebra to $\hat E_n$ appears, in the brane picture, when a single 7-brane of certain $(p,q)$-type approaches a stack of branes with some gauge algebra $E_n$. In such a situation the elliptic fiber degenerates to a non-minimal type in the sense of Kodaira's classification, i.e., the vanishing orders of the Weierstrass sections $f$ and $g$ both exceed the critical values of $4$ and $6$. The Type II and Type III Kulikov models are obtained by removing these singularities by a chain of blowups in the base.
This establishes the following generic correspondence:  \\
\begin{whitebox}
\begin{minipage}{0.5cm}
$~$
\end{minipage}
\begin{minipage}{4cm}
Non-minimal \\ Kodaira fibers \\ in codimension one
\end{minipage}
\begin{minipage}{1cm}
$\Longleftrightarrow$
\end{minipage}
\begin{minipage}{4cm}
Affine or loop extensions of Lie algebras
\end{minipage}
\begin{minipage}{1cm}
$\Longleftrightarrow$
\end{minipage}
\begin{minipage}{4cm}
Decompactification limits in a dual frame at infinite distance
\end{minipage}
\end{whitebox}
 \vspace{4mm}
 
 In Section \ref{Sec_LoopAlgebras}
 we begin by reviewing the affinisation of Lie algebras in F-theory in the language of $(p,q)$ 7-branes, as established in \cite{DeWolfe:1998eu}.
 In Section \ref{sec_Kulikov} we elaborate on the physical interpretation of the possible infinite distance limits of elliptic K3 surfaces, as probed by F-theory.
 An important test of the resulting picture is provided in Section \ref{sec_het}, where we systematically construct infinite distance limits for a Weierstrass model with minimal non-abelian gauge algebra $E_7 \oplus E_8$; the geometric results there can be compared quantitatively with the dual heterotic description via the mirror map worked out in \cite{Malmendier:2014uka}.
 In Section \ref{sec_Maximal9d} we classify the possible maximal enhancements of non-abelian gauge algebras in 9d, as determined by the Type III.a Kulikov models, and find complete agreement with the previous results of \cite{Cachazo:2000ey,Font:2020rsk,Bedroya:2021fbu}.

\section{Review: Affine Algebras and infinite Distance Limits}   \label{Sec_LoopAlgebras}

Infinite distance limits in the complex structure moduli space of F-theory on elliptic K3 surfaces can be approached from several different angles. 
The most systematic one is via the theory of semi-stable degenerations of K3 surfaces, and we will follow this route beginning with Section \ref{sec_Kulikov}.
To facilitate the physical interpretation of the geometric results, it is beneficial to first sharpen our intuition on the possible phenomena that we expect
to encounter at infinite distance in complex structure moduli space.

An obvious non-compact direction is the 10d string coupling, $g_s$, i.e. the imaginary part of the axio-dilaton, $\tau = C_0 + \frac{i}{g_s}$. It is geometrised in F-theory as the complex structure modulus of the elliptic fiber.
In absence of other competing effects, infinite distance limits along the direction $\tau \to i \infty$ in the moduli space will describe weak coupling limits. These can be superimposed with infinite distance limits in the complex structure of the torus $T^2_{\rm IIB}$ of the associated Type IIB orientifold.

A less evident type of non-compact directions occurs in genuinely non-perturbative configurations with mutually non-local 7-branes, 
and this corresponds to the formation of certain affine, or more generally loop algebras. Before developing a clear geometric description for such limits in the subsequent sections, we will
first take a complementary viewpoint as provided via the formalism of string junctions, which we now briefly review following the discussion in \cite{DeWolfe:1998eu}.

\begin{figure}[t!]
\centering
 \includegraphics[height=3.5cm]{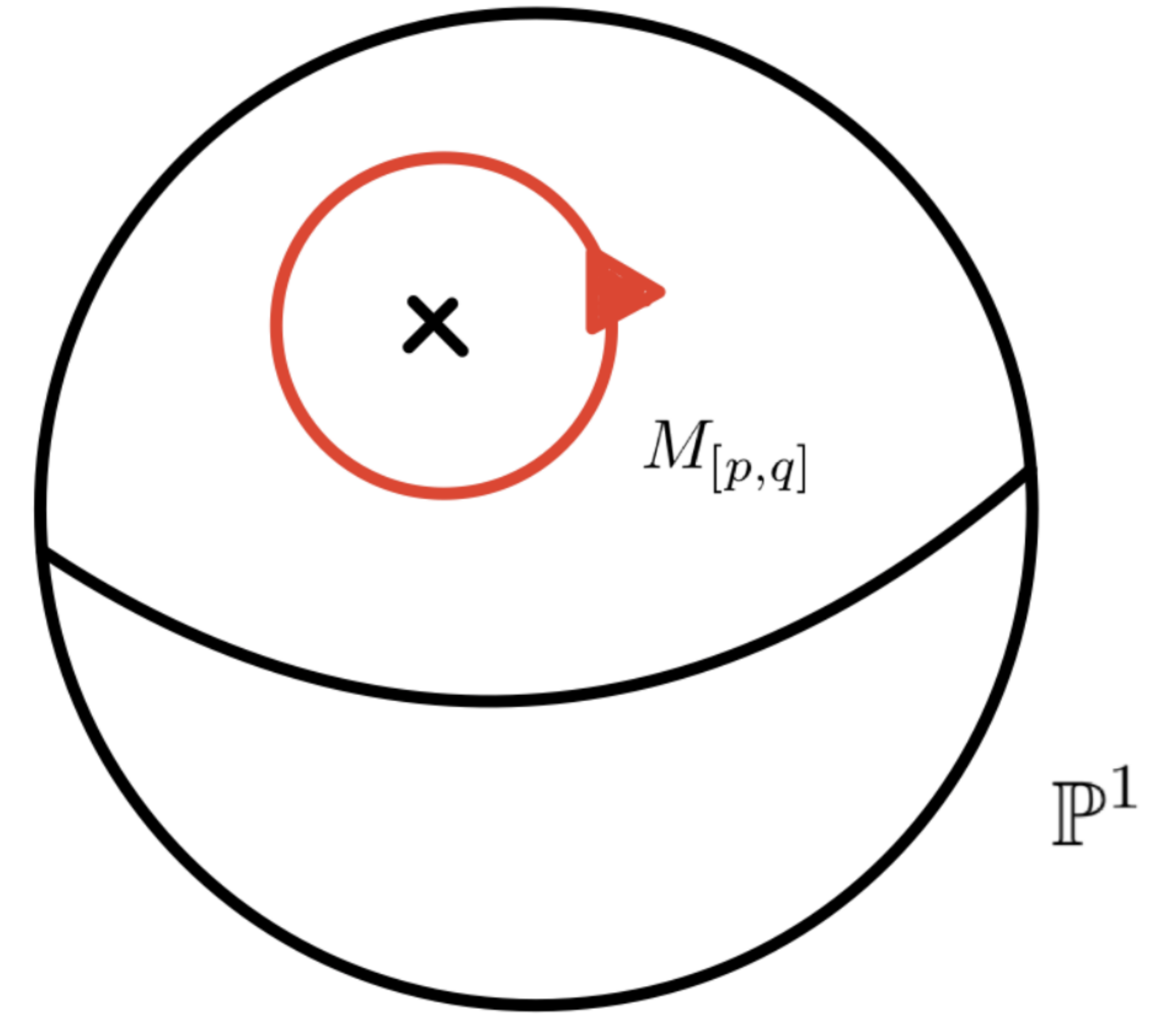}
\caption{Monodromy around a $[p,q]$ 7-brane in F-theory on K3.\label{Mpq}}
\end{figure}

When we compactify F-theory to eight dimensions, the physical compactification space,
as seen from the Type IIB string perspective,
 is a rational curve which forms the base $B$
of an elliptic K3 surface $X$. The 24 singular fibers of the K3 surface correspond to the locations of 7-branes of general $[p,q]$-type.
Upon encircling such a  location counter-clockwise, as depicted in Figure \ref{Mpq}, a general $(r,s)$ string undergoes an $SL(2,\mathbb Z)$ monodromy of the form
\bea
 \left(\begin{matrix} r  \\ s  \end{matrix}\right) \rightarrow  M_{[p,q]} \left(\begin{matrix} r  \\ s  \end{matrix}\right)   \,, \qquad   M_{[p,q]} = \left( \begin{matrix}  1 + pq & - p^2 \\q^2 & 1-pq \end{matrix}\right) \,. 
\eea

Suitable combinations of such $[p,q]$ type 7-branes are well known to realize Lie algebras of ADE type \cite{Gaberdiel:1997ud}.
Adopting the conventions of \cite{DeWolfe:1998zf}, we consider the following set of 7-branes $X_{[p,q]}$ to serve as building blocks:
\bea
A = X_{[1,0]} \,, \qquad  B = X_{[1,-1]} \,, \qquad C = X_{[1,1]} \,. 
\eea
Then gauge theories of ADE type are constructed as configurations of branes including the ones in Table \ref{table_Kodaira}.\footnote{Recall that as for characterizing the monodromies, one considers
a brane combination $X_{[p_1,q_1]} \ldots X_{[p_n,q_n]}$ as a chain and takes the branch cut induced by the $SL(2,\mathbb Z)$ backreaction of the collection of $[p,q]$ 7-branes as a straight line flowing down from $X_{[p,q]}$.
The monodromy associated with a brane configuration written as such a chain $X_{[p_1,q_1]} \ldots X_{[p_n,q_n]}$ is then computed as the matrix product
$M_{[p_n,q_n]} \cdot \ldots \cdot M_{[p_1,q_1]}$.}

The singularity in moduli space that leads to an enhancement to some 
finite Lie algebra of ADE type corresponds to a finite distance motion in the 7-brane moduli space. This can be formulated as a finite distance deformation
in the complex structure moduli space of the associated elliptic K3 surface $X$. The corresponding singularities in the elliptic fiber
were famously classified by Kodaira and N{\'e}ron and can be realized by tuning the vanishing orders of the characteristic functions $f$
and $g$ of the Weierstrass model for $X$, as recalled in Table \ref{tab_Kodairavanishing}.
 This raises the natural question whether also infinite distance deformations in this brane moduli space are possible.
As we will argue, such deformations correspond to the formation of infinite dimensional loop algebras, rather than ordinary ADE Lie algebras.

In \cite{DeWolfe:1998zf} it was explained how an exceptional Lie algebra\footnote{Here $E_1 = A_1$, $E_2 = A_1 \oplus u(1)$, $E_3 =  A_2 \oplus A_1$, $E_4 = A_4$, $E_5 = D_5$.}  $E_N$ can be enhanced  to a loop algebra, more precisely an affine algebra of Kac-Moody type, by addition of a single extra brane. Specifically, 
for $N = 1 \ldots, 8$, the finite exceptional Lie algebra $E_N$  can be enhanced into the affine algebra $\hat E_N$ by adding to the brane stack 
$A^{N-1} BCC$ an extra 7-brane $X_{[3,1]}$:
\bea \label{hatENenh1}
 && E_N   \stackrel{+X_{[3,1]}}{\longrightarrow} \hat E_N  \,, \qquad N = 1, \ldots, 8 \,,\\
&&\hat E_N = E_N X_{[3,1]} = A^{N-1} B C C X_{[3,1]} = A^{N-1} B C B C \,.
 \eea
A second series of affine enhancements can be constructed as
\bea \label{hatTildeENenh1}
&& \hat{\tilde E}_N    \stackrel{+X_{[4,1]}}{\longrightarrow} \hat{\tilde E}_N  \,, \qquad N = 0, \ldots, 8 \,,    \\
&&\hat{\tilde E}_N =  (A^N X_{[2,-1]} C) X_{[4,1]} \,,
\eea
which for $2\leq N \leq 8$ turns out to be equivalent to the series (\ref{hatENenh1}).
However for $N=1$ and $N=0$ this yields independent enhancements, with $\tilde E_1 = u(1)$ and $\tilde E_0 = \emptyset$. 

\begin{table}
  \centering
  \begin{tabular}{c|c|c}
$G$      &   branes     & Monodromy $M_G$ \\  \hline
$A_{N}$ & $A^{N+1}$ & $\left( \begin{matrix}  1 & - N-1 \\0 & 1 \end{matrix}\right)$ \\
$D_N$       & $A^N BC$     & $ \left( \begin{matrix}  -1 & N-4 \\0 & -1 \end{matrix}\right) $ \\
 $E_N$ &     $A^{N-1} BCC$     &  $ \left( \begin{matrix}  -2 & 2N - 9 \\-1 & N-5 \end{matrix}\right) $
\end{tabular}
   \caption{Brane content and monodromies for ADE groups on elliptic K3 surfaces.  \label{table_Kodaira}}
  \end{table}

\begin{figure}[t!]
\centering
 \includegraphics[height=3.5cm]{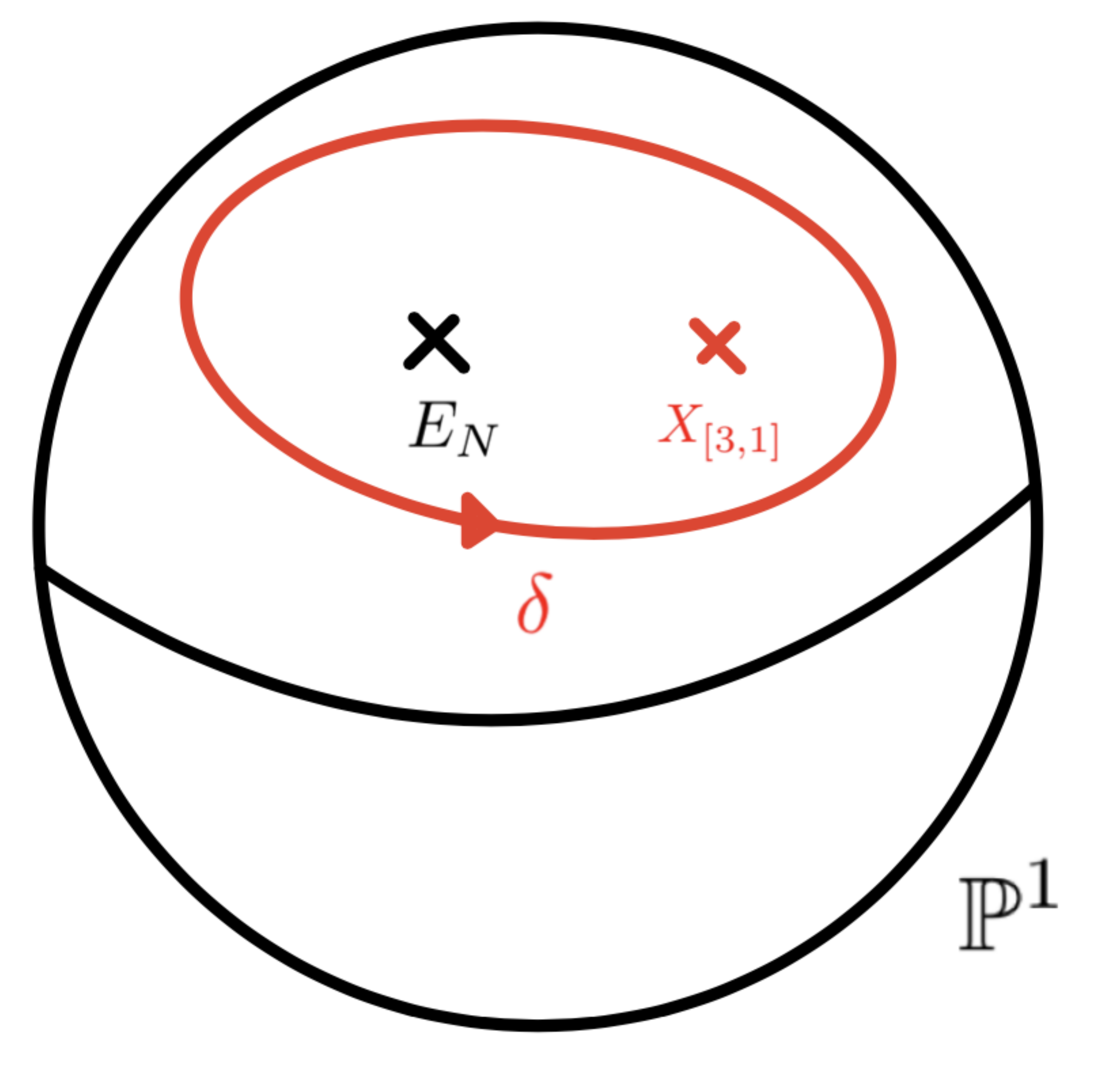} \qquad 
 \includegraphics[height=3.5cm]{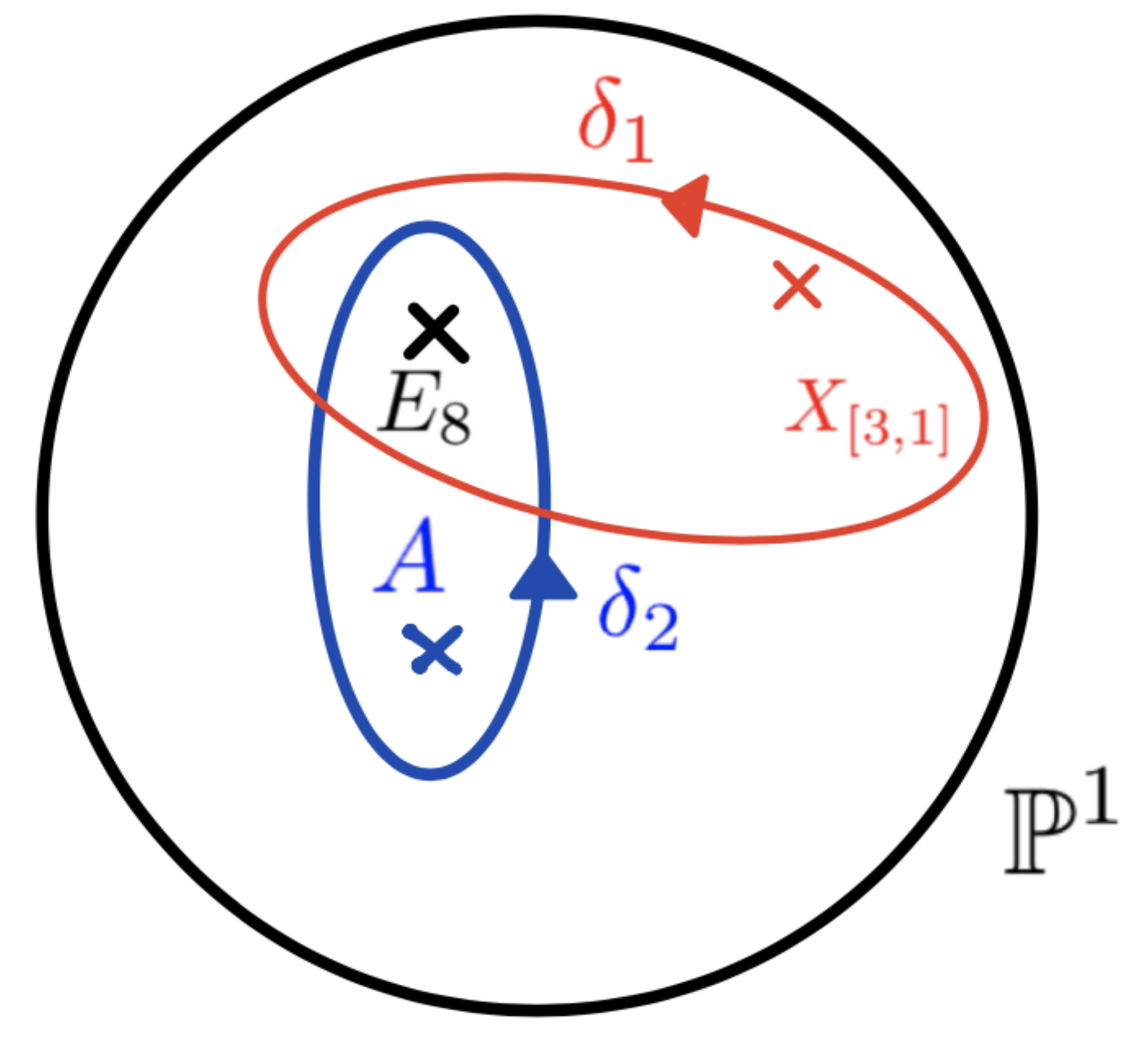}
\caption{Monodromy and imaginary junctions for affine Lie algebras $\hat E_N$ (left) and for the loop algebra $\hat E_9$ (right). \label{Stringjunctions_fig}}
\end{figure}

The resulting brane configuration induces an $SL(2,\mathbb Z)$ monodromy of the form\be
\qquad M_{\hat E_N} =  M_{\hat{\tilde E}_N} = \left( \begin{matrix} 1  &9-N \\ 0 & 1  \end{matrix}   \right) \,.
\ee
In particular, it leaves invariant the charges of a $(1,0)$ string encircling the configuration because\footnote{For simplicity of notation, here and in the sequel we will not explicity refer to $\hat{\tilde E}_N$ as opposed to $\hat{E}_N$ unless the difference is important.}

\be
M_{\hat E_N}   \delta_1  = \delta_1    \,, \qquad \delta_1  = \left( \begin{matrix} -1  \\ 0  \end{matrix}   \right)   \,.
\ee
Such a $(1,0)$ string winding around a non-trivial path encircling the branes gives rise to a particle in the eight-dimensional field theory, which becomes massless in the  
limit for which all branes in the  $\hat E_N$ configuration coincide, i.e. precisely when $E_N$ enhances to $\hat E_N$. See Figure \ref{Stringjunctions_fig}.
The string junctions associated with this state satisfy the relations \cite{DeWolfe:1998zf}
\bea
\delta_1 \cdot \delta_1 = 0   \,, \qquad \delta_1 \cdot \alpha^{E_N}_j = 0\
\eea
within the string junction lattice, where $\alpha^{E_N}_j$ is a string junction associated to one of the simple roots of the Lie algebra $E_N$.
Group theoretically both conditions identify $\delta_1$ as the imaginary root within the root lattice of the affine Lie algebra $\hat E_N$  \cite{DeWolfe:1998zf}. 
The possible rank-one affine enhancements which can be obtained in this manner are \cite{DeWolfe:1998zf}
\bea   \label{hatEandnonhat}
&& \hat E_{8}, \quad \hat E_7, \quad  \hat E_6, \qquad \hat E_5 = \hat D_5, \\ 
&& \hat E_4 = \hat A_4,  \quad \hat E_3 = \widehat{A_2 \oplus A_1},  \quad \hat E_2 = \widehat{A_1 \oplus u(1)},  \quad \hat E_1 = \hat A_1, \quad \hat{\tilde E}_1 = \widehat{u(1)} ,   \quad \hat{\tilde E}_0 = \hat{\emptyset}\,.  \nonumber
\eea
In this list  we include for  $\hat E_3$ and $\hat E_2$  the imaginary roots for both 
involved Lie algebras, but identify them such that the total rank increases just by one.

Since the condition for a string junction $J$ to give rise to a BPS state is that $J \cdot J \geq -2$, any multiple 
\be
n \, \delta_1 \,, \qquad n \in \mathbb Z
\ee
 gives rise to a BPS state.
As a result, we obtain an infinite tower of massless BPS states from the collision of an $E_N$ brane stack and an $X_{[3,1]}$ brane, or equivalently in the limit of enhancing $E_N$ to its affine algebra $\hat E_N$.

Based on a monodromy analysis,  ref. \cite{DeWolfe:1998zf} furthermore identifies a number of rank-two enhancements. The only one which turns out to make an appearance in our context is the enhancement from $E_8$ to the double-loop algebra $\hat E_9$, which is realised as the brane configuration
\bea
\hat E_9 = A \hat E_8 = A E_8 X_{[3,1]} =  A^8 B C C X_{[3,1]}  = A^8 B C B C \,.
\eea
Its monodromy matrix
\be
M_{\hat E_9} =  \left( \begin{matrix} 1  &0 \\ 0 & 1  \end{matrix}  \right)
\ee
leaves invariant a 2-dimensional lattice of string junctions spanned by 
\bea
\delta_1 = \left( \begin{matrix} -1  \\ 0  \end{matrix}   \right) \,, \qquad \delta_2 = \left( \begin{matrix} 0  \\ -1  \end{matrix}   \right) \,.
\eea
These satisfy
\bea
\delta_i \cdot \delta_j = 0 \,, \qquad \delta_1 \cdot \alpha^{E_8}_j = 0 \,
\eea
and therefore act as the two imaginary roots in the enhancement of $E_8$ to the double loop algebra $\hat E_9$ (which in fact is not an affine algebra).
In the limit of enhancement, one finds two towers of massless BPS states from the junctions
\bea
n_1 \delta_1    \,, \qquad n_2 \delta_2 \,, \qquad n_i \in \mathbb Z \,.  
\eea

The appearance of one or two infinite towers of massless states suggests that the affine or double loop enhancements occur at infinite distance in the moduli space.
We will show that this intuition is indeed correct, and that the deformations leading to loop algebras are in fact the only possible infinite distance degenerations in the complex structure moduli space of F-theory on K3 except for the weak coupling limits (possibly followed by an infinite distance degeneration in the complex structure of the Type IIB torus $T^2_{\rm IIB}$).
In the Weierstrass models, loop enhancements will be identified as deformations giving rise to certain non-minimal Kodaira singularities over codimension-one loci on the base $B$.
To remove the non-minimal singularities, a sequence of blowups leads to a degeneration of the K3 into multiple components, which we identify with certain Kulikov models \cite{Kulikov}.
From a physics perspective, the tower of states associated with the imaginary roots $\delta$ play the role of a Kaluza-Klein tower in the dual heterotic string and signals decompactification from 8d, either to 9d for affine enhancements to $\hat E_N$ with $N \leq 8$, or to 10d for the loop enhancement to $\hat E_9$. 
In general, these states may form only a subset of the parametrically leading towers of states which become light, and one of our tasks will be to identify the full set of leading towers in order to establish the correct physical interpretation of the infinite distance limits.

\begin{table}
\centering
\begin{tabular}{c|c|c|c|c|c}
Branes &Algebra &Kodaira Type &${\rm ord}(f)$ & ${\rm ord}(g)$  & ${\rm ord}(\Delta)$ \\\hline
$A^{n+1}$ & $A_n$  & I$_{n+1}$   & 0 & 0 & $n+1$ \\
$A^n B C$ & $D_n$ &  I$^\ast_{n-4}$   &2 & 3 & $n+2$ \\
$A^5 B C^2$  & $E_6$   & IV$^\ast$ & $\geq 3 $& 4 & 8 \\
$A^6 B C^2$  &   $E_7$& III$^\ast$  & 3 &$ \geq 5$ & $9$ \\
$A^7 B C^2$  &  $E_8$ & II$^\ast$ &   $\geq$ 4 & $ 5$ & $10$ \\\hline
\end{tabular}
\caption{Vanishing orders for a Weierstrass model $y^2 = x^3 + f x z^4 + g z^6$ with discriminant $\Delta = 4 f^3 + 27 g^2$ realising the ADE brane configurations of Table \ref{table_Kodaira} as singularities in the fiber according to the Kodaira classification.
\label{tab_Kodairavanishing}}
\end{table}

\section{Kulikov Models as Emergent String Limits or Decompactifications} \label{sec_Kulikov}

We now present the physical interpretation of the infinite distance limits in the  complex structure moduli space of elliptic K3 surfaces.
We begin with a quick review of the geometric results from our companion paper \cite{Kulikov} which refines the classification of infinite distance limits via Kulikov models 
in a way suitable for our purposes. The four canonical types of such geometric infinite distance, Type II.a and II.b \cite{Clingher:2003ui}  and Type III.a and III.b  \cite{Kulikov} (see also \cite{alexeev2021compactifications}) are then subsequently analysed from the point of view of F-theory.

\subsection{Refined Kulikov models for elliptic K3 surfaces} \label{sec_Kulikovrecap}

Infinite distance limits in the complex structure moduli space of an elliptic K3 surface
can be studied within the framework of semi-stable degenerations \cite{FriedmanMorrison}. 
For simplicity we will restrict ourselves to one-parameter degenerations as this turns out to be sufficient for understanding the asymptotic physics.

 A (one-parameter) degeneration ${\cal X}$ of K3 surfaces is a family 
 of K3 surfaces $X_u$, where $u \in D = \{u \in \mathbb C: |u| < 1 \}$.
The central fiber of the degeneration, $X_{u=0}$, is the degenerate surface which we aim to study.
The degeneration is semi-stable if $X_0$ is a union of surfaces each appearing with multiplicity one such that the total family ${\cal X}$ is smooth as a threefold
and the singularities of $X_0$, if there are any, are locally of  normal crossing form.
By the important  results of \cite{MumfordToroidal,Kulikov1,Kulikov2,PerssonPink}, every degeneration of K3 surfaces can be brought into the form of a Kulikov model: This means that
 the degeneration is semi-stable and in addition the family ${\cal X}$ is Calabi-Yau.
 For general K3 surfaces, the Kulikov models enjoy a classification into models of Type I (at finite distance), Type II and Type III (both at infinite distance).
 
 It is known from the general theory of degenerations of K3 surfaces \cite{FriedmanMorrison} that the central fiber $X_0$ of a Type II Kulikov model exhibits
 two transcendental elliptic curves, $\gamma_1$ and $\gamma_2$, whose calibrated volumes vanish,
 \bea  \label{gammacycles}
{\rm vol}(\gamma_i) = \int_{\gamma_i} \Omega_0 = 0\,.
 \eea
Here $\Omega_0$ denotes the (2,0) form on the degenerate surface $X_0$.
 For models of Kulikov Type III  there exists only a single such transcendental elliptic curve of asymptotically vanishing volume.
 If we compactify M-theory on such K3 surfaces, we therefore obtain one or two towers of asymptotically massless BPS particles
  from M2-branes that wrap the vanishing tori arbitrarily many times.
  However, as we will see, in general these particles form only a subset of the asymptotically massless towers, and sometimes they do not form the 
  parametrically leading towers. A full understanding of the asymptotic physics therefore requires a refined
geometric picture of the degeneration and an analysis of the nature of states which become massless in compactifications of M- or F-theory.

 Fortunately, if we restrict ourselves to 
degenerations of elliptically fibered K3 surfaces, such a geometric picture has been obtained for Type II degenerations in  \cite{Clingher:2003ui}  and recently for
Type III degenerations in our companion paper \cite{Kulikov}, as well as in~\cite{alexeev2021compactifications}.
By blowing down all exceptional curves in the fiber, one obtains a family $\cal Y$ of Weierstrass models ${Y}_u$.
In particular, to the degenerate K3 surface $X_0$ one associates a Weierstrass model $Y_0$.

Kulikov models of Type I correspond to the degenerations of the elliptic fiber which lie at finite distance in the complex structure moduli.
The degeneration $Y_0$ is  a Weierstrass model over a single rational curve $B_0$ and has singularities in the elliptic fiber of the familiar Kodaira-N\'eron types
as listed in Table \ref{tab_Kodairavanishing}.

For the infinite distance degenerations of elliptic Kulikov Types II and III, on the other hand, the base 
$B_0 = \cup_{i=0}^P B^i$ is a {\it chain} of rational curves $B^i$. The surface $Y_0$ correspondingly decomposes into a chain of surfaces
\be \label{Y0chain}
Y_0 = \cup_{i=0}^P   Y^i  \,,
\ee
where each surface component $Y^i$ is a possibly degenerate Weierstrass model over $B^i$.
By a combination of birational transformations and base changes, elliptic Type II and Type III  Kulikov models can be brought into certain canonical forms, which are distinguished as follows:
 
 \begin{enumerate}
 \item Elliptic Type II models \cite{Clingher:2003ui}: The degeneration $\cal X$ can be brought into the canonical form where $X_0 = X^1 \cup X^2$ such that $E:=X^1\cap X^2$ is an elliptic curve.
 \begin{itemize}
 \item Type II.a (see also \cite{Morrison:1996na,Morrison:1996pp}): $X^1$ and $X^2$ are two rational elliptic surfaces, i.e. elliptic fibrations with 12 singular fibers. The intersection curve $E$ is the smooth elliptic fiber
 over the intersection point $B^1 \cap B^2$ in the base.   
 \item Type II.b (see also \cite{Aspinwall:1997ye}): The base $B_0$ is a non-degenerate rational curve, and $X^1$ and $X^2$ are both rationally fibered over $B_0$. The intersection $X^1\cap X^2$ is a ramified double cover of $B_0$, which indeed describes an elliptic curve  $E$.
 \end{itemize}

 \item Elliptic Type III models \cite{Kulikov, alexeev2021compactifications}:  The Weierstrass degeneration $\cal Y$ associated with $\cal X$ can be brought into a canonical form  such that each intermediate surface $Y^i$, $i=1, \ldots, {P-1}$ appearing in (\ref{Y0chain})
 is a degenerate Weierstrass model over a rational base component $B^i$ with generic fibers of Kodaira Type I$_{n_i}$, $n_i >0$. Over special points of $B^i$, the singularity can enhance to an $A$-type singularity (in the sense of the Kodaira classification of Table \ref{tab_Kodairavanishing}).\footnote{Note that our characterisation of the singularity enhancements is in the context of the degenerate K3 surface, while the stated feature of generic I$_{n_i}$ fibers is observed in the context of the 3-fold $\mathcal Y$. Hereafter we will not explicitly distinguish the contexts unless potential confusions arise.} Without loss of generality one can assume that at least one intermediate surfaces are present (i.e. $P>1$) and that no special fibers are located in the intersection loci of the components.
 \begin{itemize}
 \item Type III.a:  One or both of the end components $Y^0$ and $Y^P$ are rational elliptic surfaces. If the end component is not rational elliptic, it is a degenerate Weierstrass model with I$_{n>0}$ fibers over generic points and precisely two singularities of $D$-type (along with possibly extra $A$-type singularities).
 
  \item Type III.b: Each of the two end components is a degenerate Weierstrass model with I$_{n>0}$ fibers over generic points and precisely two singularities of $D$-type (along, possibly, with extra $A$-type singularities). \end{itemize}
 \end{enumerate}

The family $\cal Y$ of Weierstrass model can be parametrized as 
\bea \label{Weierstrassfam-def}
 Y_u: \qquad y^2 = x^3  + f_u(s,t) x z^4 + g_u(s,t) z^6 \,,
\eea
with discriminant 
\be
\Delta_u = 4 f_u^3 + 27 g_u^2   \,.
\ee
 For $u \neq 0$, $f_u(s,t)$ and  $g_u(s,t)$ are homogeneous polynomials of degree $8$ and $12$, respectively, in homogenous coordinates $[s :t]$ of the base $B_u  = \mathbb P^1_{[s:t]}$. The degeneration occurs in the limit $u \to 0$. Note that, for notational simplicity, we will hereafter omit the subscript $u$ in the Weierstrass sections $f$ and $g$, as well as in the discriminant $\Delta$, of which $u$-dependence will be assumed in the context.  
 Then, Kulikov models of Type II.b have the property that the Weierstrass sections in the limit $u \to 0$ take the special form
 \be
f|_{u=0} = -3 h^2 \,, \qquad  g|_{u=0} = 2 h^3 \,,
 \ee
 while otherwise all vanishing orders respect the minimality bound according to Kodaira's classification.\footnote{The latter condition is satisfied if and only if the degree-$4$ polynomial $h$ has four distinct roots.} 
 As detailed in \cite{Kulikov}, models of Type II.a, III.a and III.b, on the other hand, are in one-to-one correspondence with Weierstrass models which, for $u=0$, acquire one or several non-minimal fibers. If we choose one of the non-minimal fibers to lie at $s=0$, this implies for the vanishing orders  that\footnote{The classification of vanishing orders is understood modulo base change. This means that one has to consider the maximal vanishing orders which can be obtained by transforming $u \to u^k$ for some integer $k\geq 1$.}

 \bea \label{vanorders-gen}
 {\rm ord}_{\cal Y}(f,g,\Delta)|_{u=s=0} = (4 + m, 6 + n, 12 + k) \,,    \qquad m \geq 0 \,, n \geq 0 \,, k \geq 0 \,, 
 \eea
with the following correspondence:
 \begin{itemize}
 \item If $k = 0$, blowups and base changes can bring the model into Kulikov form of Type II.a.
 \item If  $m =0$, $n =0$ and  $k > 0$, the model can be brought into Kulikov form of Type III.a or III.b.
 \item If $m >0$ and $n >0$ (and hence $k>0$), the model can be of Kulikov form of Type I (finite distance) or Type II/III.
 \end{itemize}

Models of Type III.a with only a single rational elliptic end component require, in addition to the above non-minimal singularity, precisely one of the two types of tunings that we will describe below, while Models of Type III.b require the both. The relevant tunings are as follows: 
\begin{enumerate}
\item  The Weierstrass sections restricted to the curve $u=0$ take the form 
\bea   \label{IIIafurthertuning}
f|_{u=0} = -3 s^4 \hat h^2   \,, \qquad g|_{u=0} = 2 s^6 \hat h^3   \,,
\eea
where $\hat h$ is a polynomial of degree $2$ in $[s:t]$. 
\item
If we denote by $f_-$ and $g_-$ the parts of $f$ and $g$ consisting only of the terms of degrees $d_f \leq 4$ and, respectively, $d_g \leq 6$ in $s$, whose vanishing orders in $u$ are precisely 
$(4-d_f)P$ and $(6-d_g)P$, then these parts take the form~\cite{Kulikov}
\be   \label{IIIbtuning}
f_- = - 3 t^4 (L_0 u^{2P} t^2 + L_1 u^P s t + L_2 s^2)^2 \,,   \quad 
g_- = 2 t^6 (L_0 u^{2P} t^2 + L_1 u^P s t + L_2 s^2)^3    \,.
\ee
Here $L_i \in \mathbb C$ and the parameter $P$ coincides with the required number of blowups (i.e. the number of components in (\ref{Y0chain})). For the precise statement see \cite{Kulikov}.
\end{enumerate}

With this preparation, we are now in a position to 
analyse the towers of states that become asymptotically massless in both types of limits.
This can be approached either  in the language of M-theory as before, or via duality directly in F-theory, by analysing the spectrum of branes that wrap the degenerate K3 surface.

\subsection{Type II.b Kulikov models as emergent string limits}

Let us begin with the Kulikov Type II.b limits, which were first described in this language in the F-theory literature in 
\cite{Aspinwall:1997ye}. The physics of such limits is 
 a perturbative weak coupling limit of Sen type \cite{Sen:1996vd,Clingher:2012rg}.

By definition \cite{Clingher:2003ui}, a Type II.b Kulikov model is constructed as a blowup of a singular elliptic fibration over the base $\mathbb P^1_{[s:t]}$ for which the generic elliptic fiber has degenerated to a nodal curve. This degeneration is the result of the collapse of the $(1,0)$ cycle in the elliptic fiber, which we denote by $S_A$, while the dual $(0,1)$ cycle $S_B$ of the elliptic fiber survives in the nodal fiber.
Blowing up the node replaces the nodal fiber by a Kodaira Type I$_2$ fiber consisting of a pair of rational curves intersecting, generically, in two points $p_1$ and $p_2$.
 See Figure \ref{fig:TypeIIb} for an illustration.
Over four points, $p_1$ and $p_2$ come together to form a double point. 
The two intersection points $p_1$ and $p_2$ define a bi-section, i.e. a double cover of the base branched over the four points where $p_1$ and $p_2$ coincide. The 2-cycle defined by the bi-section is  identified with the elliptic curve $E = X^1 \cap X^2$.

To understand the origin of the transcendental 2-cycles $\gamma_i$, $i=1,2$, as advertised in (\ref{gammacycles}),
note that the group $H_1(E,\mathbb Z)$ of the intersection curve $E = X^1 \cap X^2$ is spanned by two 1-cycles, which we denote by $\sigma_1$ and $\sigma_2$.
By suitably combining the collapsed one-cycle $S_A$ with each $\sigma_i$ one constructs two homologically non-trivial transcendental 2-cycles with the topology of a torus, and these two 2-cycles are the objects $\gamma_1$ and $\gamma_2$
whose calibrated volume (\ref{gammacycles}) vanishes in the degenerate situation. 

{

\begin{figure}[t!]
\centering
\includegraphics[width=8cm]{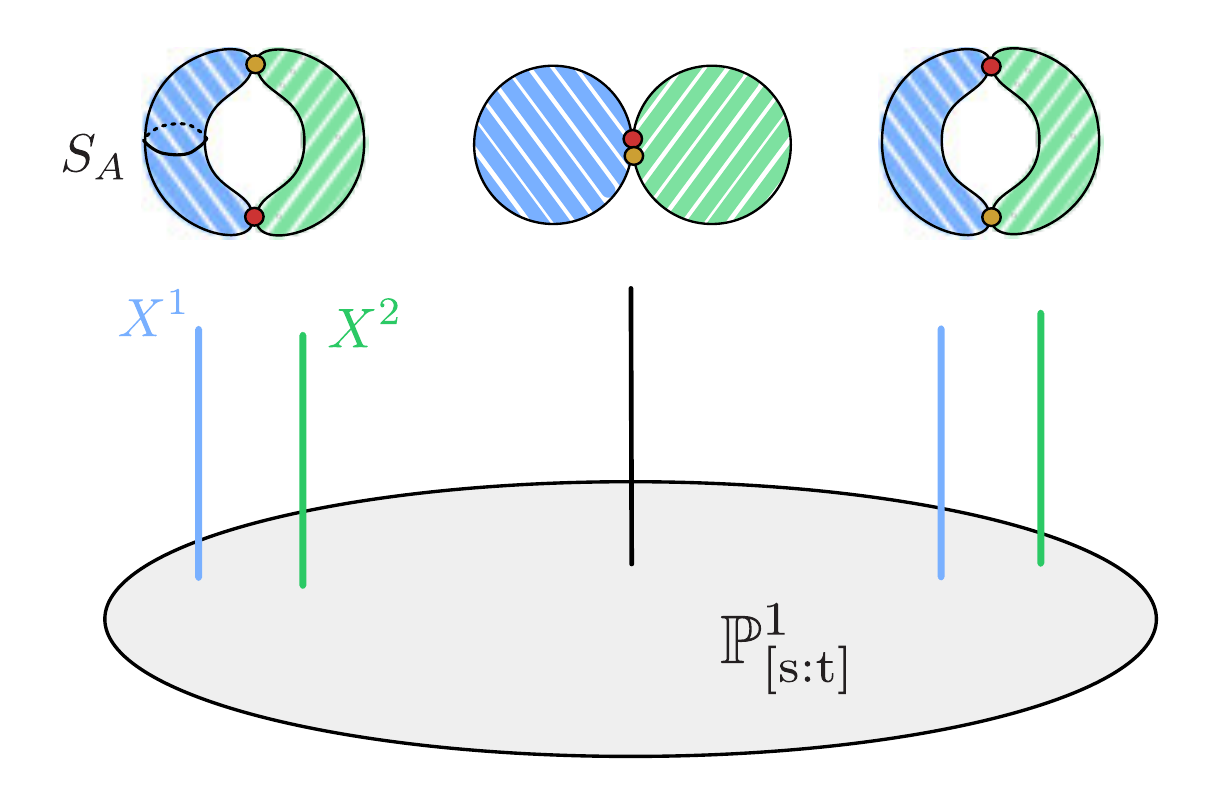}
\caption{
Kulikov Type II.b degeneration. Note that the nodes exchange under
mondromy around the singular configuration shown in the center.}
\label{fig:TypeIIb}
\end{figure}

If we compactify M-theory on $X_0=X^1 \cup X^2$, we obtain the following asymptotically massless objects in seven dimensions:
First, an M2-brane wrapping the vanishing cycle $S_A$ gives rise to an asymptotically tensionless, fundamental Type II string. This string is non-BPS because $S_A$ is actually homologically trivial.
Second, wrapping an M2-brane arbitrarily many times along either of the two calibrated 2-cycles, $\gamma_1$ and  $\gamma_2$, yields a tower of BPS particles.
Note that the existence of these BPS particles is guaranteed because the 2-cycles $\gamma_i$ have the topology of a torus and can therefore be wrapped arbitrarily often.
The BPS particles are to be identified with the winding states from wrapping the Type II string along the two 1-cycles $\sigma_1$ and $\sigma_2$ of $E$.
The latter sit parametrically at the same scale as the string excitation tower.
The winding tower is of course dual to a corresponding Kaluza-Klein tower of supergravity modes of the same parametric mass scale.
All in all, the asymptotic spectrum of the Type II.b Kulikov degeneration carries the signature of an Emergent String Limit as defined in \cite{Lee:2019wij}.

Note that if it were not for the asymptotically tensionless string obtained by wrapping an M2-brane along $S_A$, we would have incorrectly characterised the infinite distance limit
as a KK decompactification.  However the KK modes form only part of the spectrum, and due to the higher density of states it is actually the tower of string excitations which is defining the asymptotic physics as an equi-dimensional weak coupling limit. 
 
 From the perspective of F-theory in 8d, this behaviour is of course no surprise.
To realise an I$_2$ fiber over generic points of the base,  the vanishing orders of $f$, $g$ and $\Delta$ of the Weierstrass model (\ref{Weierstrassfam-def}) must  become
 \bea
{\rm ord}_{\cal Y}(f,g,\Delta))|_{u=0} = (0,0,2)  \quad \text{at generic points of } \, \,  \mathbb P^1_{[s:t]}   \,.
 \eea
The behaviour of the 10d Type IIB axio-dilaton, $\tau = C_0 + \frac{i}{g_s}$, at a generic point of the base can then be read off from the $j$-function:
  \bea
 j(\tau) \sim \frac{ f^3}{\Delta} \sim \frac{1}{u^2}   \quad    \Longrightarrow    \quad \tau \to i \infty  \qquad {\rm for} \, \, \, u\to 0 \,.
 \eea
 A convenient way to realise the vanishing orders is to consider the degeneration \cite{Sen:1996vd}
\bea
f _u= - 3 h^2 +  u \, \eta  \,,  \qquad g_u= -2 h^3 + u \, h  \,  \eta - \frac{u^2}{12} \chi    \,,  \qquad \Delta = - 9 \, u^2 \, h^2 \, (\eta^2 - h \chi) + {\cal O}(u^3)  \,,
\eea
where 
  $h$, $\eta$ and $\chi$ are  polynomials of degree $4$, $8$ and $12$.
  In the Type IIB orientifold interpretation,
the limit $u\to 0$ gives rise to an orientifold compactification on the torus $E$ and the four zeroes of $h=0$ correspond to the location of the orientifold planes.
If we disregard the overall factor of $u^2$ in $\Delta$ as 
\beq
\Delta=u^2 \Delta' \,,
\eeq
the vanishing orders at the orientifold become 
\beq
{\rm ord}(f_0,g_0,\Delta'_0)|_{h=0} = (2,3,2)\,,
\eeq
where the subscripts indicate that $u=0$ has been taken to begin with. 
 Away  from these localised regions, the theory reduces to a weakly coupled, perturbative Type IIB string, which is non-BPS in 8d since the Kalb-Ramond field $B_2$ is projected out by the orientifold projection.
 The string tower appears parametrically at the same scale as the two KK towers associated with each one-cycle of  $E$, and both scales vanish in units of the 8d Planck scale.

Finally, as remarked in \cite{Kulikov}, a more general parametrisation of weak coupling limits is to simply take
\be  \label{weakoupling-gen}
f_u = - 3 h^2 + u^a \eta \,, \qquad g_u = - 2 h^3 + u^b \rho  \,,
\ee
for integral $a \geq 1$, $b\geq 1$ and generic sections $\eta$ and $\rho$ of suitable degree. 
In this case, the general theorems of \cite{Clingher:2003ui} guarantee the existence of a birational transformation which, when followed by a suitable base change, brings the fibration
into Kulikov form  of Type II.b.

\subsection{Type II.a Kulikov models as decompactification limits} \label{sec_II.a}

The spectrum of asymptotically massless states in limits of Kulikov Type II.a is very different.
The two del Pezzo (dP$_9$) surfaces  $X^1$ and $X^2$ into which the K3 surface degenerates are each elliptically fibered over a rational curve, $B^1$ and $B^2$, respectively. 
In other words, the base $\mathbb P^1_{[s:t]}$ of the K3 surface $X_u$ splits for $u=0$ into the union of $B^1$ and $B^2$ intersecting
at a single point ${\cal P} = B^1 \cap B^2$,
and the elliptic double curve $E = X^1 \cap X^2$ represents the common elliptic fiber over ${\cal P}$. 
This geometry was described in the F-theory literature early on in  \cite{Morrison:1996na,Morrison:1996pp,Aspinwall:1997ye}.
The two 2-cycles $\gamma_i$ can therefore be constructed by fibering the two one-cycles $\sigma_i$ in $H_1(E, \mathbb Z)$ over a 1-cycle $\Sigma$ that encircles the intersection point ${\cal P}$
on either $B^1$ or $B^2$. The geometry is sketched in Figure \ref{fig:TypeIIa}.

\begin{figure}[t!]
\centering
\includegraphics[width=8cm]{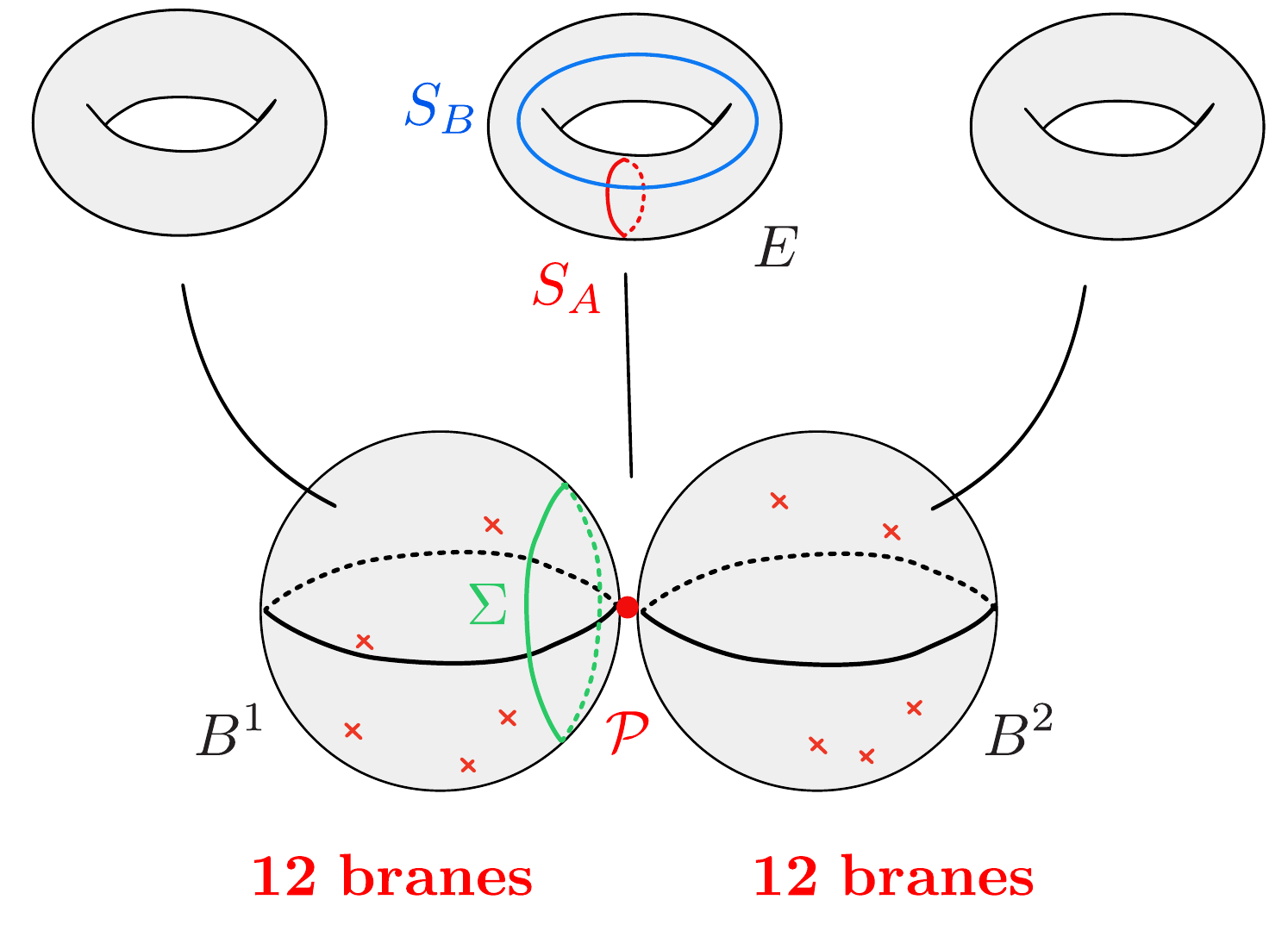}
\caption{
Kulikov Type II.a degeneration. The base degenerates into two $\mathbb P^1$s that
form the bases of two elliptic del Pezzo surfaces dP$_9$. }
\label{fig:TypeIIa}
\end{figure}

Viewed as a geometric 1-cycle on $B^1$ or $B^2$, $\Sigma$ can be slipped off the base sphere by deforming it to the anti-podal pole and is therefore trivial, but fibering both $\sigma_i$ over $\Sigma$ near ${\cal P}$ indeed gives rise to two non-trivial 2-cycles $\gamma_i$. The asymptotic vanishing of their calibrated volume (\ref{gammacycles}) for degeneration parameter $u=0$ is an intuitive consequence of the fact that $\Sigma$ 
can be contracted towards the intersection point ${\cal P}$.

The seven-dimensional compactification of M-theory on K3 is therefore characterised by the appearance of two towers of asymptotically massless particles which arise  from M2-branes wrapped any number of times on the two 2-cycles
$\gamma_1$ and $\gamma_2$. Unlike for limits of Type II.b, these towers of BPS particles are not accompanied
by a tower of excitations from an asymptotically tensionless string.
The reason is that the M2 brane cannot be wrapped along the 1-cycle $\Sigma$ on either of the base components, because viewed in isolation $\Sigma$ can simply be slipped off $B^1$ or $B^2$.

From the point of the view of the dual F-theory compactification on K3, these two towers of BPS particles correspond to two towers of asymptotically massless particles in 8d:
These arise from string junctions of the form $n_1 \, \delta_1$ and $n_2 \, \delta_2$, where $n_i \in \mathbb Z$ represents the winding numbers. If we denote by $\sigma_1$ and $\sigma_2$ the $(1,0)$
and $(0,1)$ cycles on $E$, then an M2-brane on $\gamma_1$ uplifts to a $(1,0)$ string and the cycle on $\gamma_2$
uplifts to a $(0,1)$ string, each encircling the intersection point  ${\cal P}$ along $\Sigma$.
The important difference to the M-theory viewpoint is that in F-theory, it is not possible to freely slip the $(1,0)$ and $(0,1)$ strings on $\Sigma$ to the other pole of the curves $B^1$ or $B^2$, because the strings would have to cross the 12 7-branes located on either of the two curves.

Nonetheless we can interpret the $(1,0)$ and $(0,1)$ strings on $\Sigma$ as string junctions encircling the entire configuration of 12 branes on $B^1$ or $B^2$. 
This makes contact with the discussion in Section \ref{Sec_LoopAlgebras}.
The combined monodromy associated with the 12 singular fibers on a dP$_9$ is unity, and the associated brane configuration would enhance to
the double loop algebra $\hat E_9$ if all 12 branes coincide on the dP$_9$.
The wrapped strings therefore
 correspond precisely to the two junctions $\delta_1$ and $\delta_2$ of Section \ref{Sec_LoopAlgebras}.
In fact, as we will discuss in more detail in Section \ref{sec_E7E8-ConcreteReal}, the Type II.a Kulikov model is the result of blowing up a singular Weierstrass model over $\mathbb P^1_{[s:t]}$, for
which a complex structure degeneration enforces the collision of the 12 constituent branes of an $\hat E_9$ configuration
 at up to two points. At these points, the singularity in the elliptic fiber is non-minimal
in the sense of the Kodaira classification, i.e. the Weierstrass  
vanishing orders become
\be
{\rm ord}_{\cal Y}(f,g,\Delta)|_{u=0} = (\geq 4,6,12)    \quad {\rm or} \quad (4, \geq 6,12)  \,.
\ee
Prior to blowing up the non-minimal singularity in the base,
one obtains two towers of asymptotically massless states
of the form $n_1 \delta_1$ and $n_2 \delta_2$ localised at the singularity.
The blowup gives a regularisation of the non-minimal singularity by separating the 12 branes in such a way that the resulting fiber types are of minimal form.
This comes at the cost of degenerating the base into several components. In this regularised geometry, the massless towers of states are still visible as the same types of string junctions, now encircling the intersection point ${\cal P}$, as described above, or equivalently in M-theory, from the towers
of M2-branes along $\gamma_1$ and $\gamma_2$.
Note that the situation is symmetric between both components of the Type II.a Kulikov model and in particular the string junctions on $B^1$ and $B^2$ are homologous in the junction lattice \cite{DeWolfe:1998pr}.
The total symmetry algebra of the system in the infinite distance limit is therefore
\bea \label{hatE9hatE9II}
G_\infty = (\hat E_9 \oplus \hat E_9)/\sim  \,,
\eea
where the quotient identifies the two imaginary roots $\delta_i$ from each of the first and from the second factor.

These two towers of states are interpreted as two Kaluza-Klein towers in the dual heterotic string, signalling the decompactification of the two circles of the heterotic torus $T^2_{\rm het}$.
We will elaborate on this interpretation in detail in Section \ref{sec_E7E8-ConcreteReal}.
This identifies the Type II.a limits as decompactification limits from 8d to 10d. This is of course nothing but the well-known statement that stable degenerations (of Type II.a)
realise large volume limits of the dual heterotic string  \cite{Morrison:1996na,Morrison:1996pp,Aspinwall:1997ye}. Clearly, the gauge algebra in the decompactified theory is simply 
\be
G_{\rm 10d} = E_8 \oplus E_8 \,,
\ee
the maximal {\it finite} Lie subalgebra of (\ref{hatE9hatE9II}).

\subsection{Type III.a Kulikov models as partial decompactification limits}    \label{subsec_8dto9d}

We now 
characterise the physics of F-theory on a K3 surface in an infinite distance limit of Kulikov Type III.a.
In such a limit, the theory undergoes a partial decompactification to nine dimensions. This can be traced back to the appearance of one or two factors of an $\hat E_{9-n}$ affine algebra for some values $1 \leq n \leq 9$, as we now explain from F-theory.

\begin{figure}[t!]
\centering
\includegraphics[width=8cm]{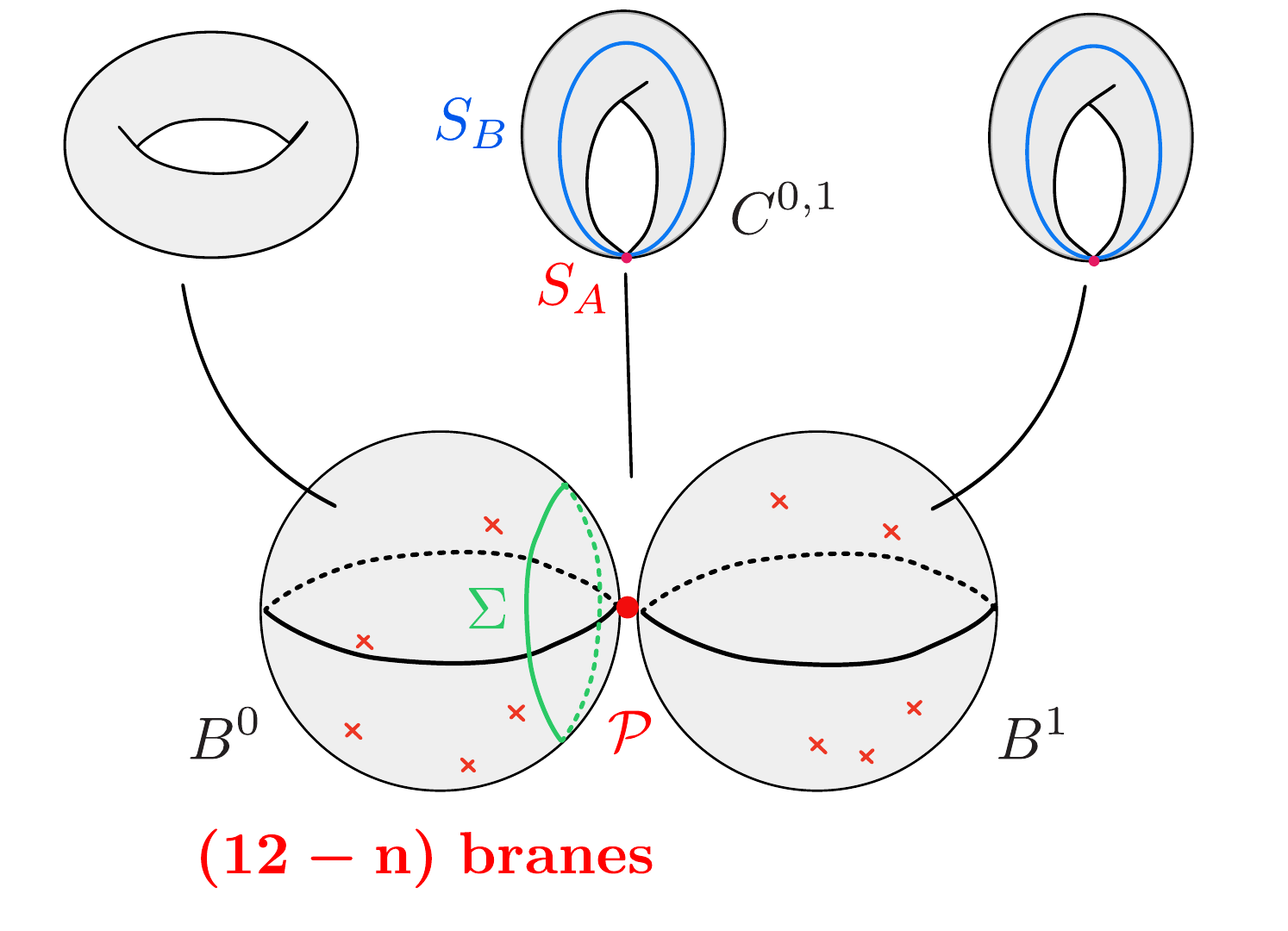}
\caption{
Rational elliptic end component of a Kulikov Type III.a degeneration intersecting a neighbouring positive $I_n$ component from the right.}
\label{fig:TypeIIIa}
\end{figure}

In models of Type III.a, at least one of the two end components of the degenerate Weierstrass model $Y_0$ given in (\ref{Y0chain}) is a rational elliptic surface, which we will often refer to as a dP$_9$ surface.
As reviewed in Section \ref{sec_Kulikovrecap} such configurations are obtained by suitable blowups of non-minimal singularities of the type (\ref{vanorders-gen}) with $m= n = 0$ and $k >0$.\footnote{Recall that the configuration could be of other Kulikov types if $k>0$ was achieved by $m>0$ and $n >0$.  For details see \cite{Kulikov}.}
Generically, this gives rise to a chain of surfaces with rational elliptic surfaces on both ends, $Y^0$ and $Y^P$. An additional tuning of the form (\ref{IIIafurthertuning}) or~\eqref{IIIbtuning} results in a configuration where
only one of the end components is rational elliptic. 

Let us first assume that there is only one rational elliptic end component, which we identify w.l.o.g. with the surface $Y^0$. Its neighbouring surface component $Y^1$ is a degenerate
fibration over $B^1$ with fibers of Type I$_{n_1 >0}$ in codimension zero. In the sequel we will set $n_1 =: n$. The two curves $B^0$ and $B^1$ intersect at a point $\cal P$. The fiber over $\cal P$ is the intersection curve 
 $C^{0,1} =Y^0 \cap Y^1$.
This fiber $C^{0,1}$ is a nodal curve, obtained by collapsing the $(1,0)$ cycle $S_A$ of the elliptic fiber. See Figure \ref{fig:TypeIIIa} for an illustration.
If $n >1$, then after blowing up the nodal point, the curve $C^{0,1}$ is replaced within $X_0$ by a cycle of $n$ intersecting rational curves intersecting like the nodes of the affine Dynkin diagram of $SU(n-1)$

What is important for us is that the $SL(2,\mathbb Z)$ monodromy picked up upon encircling the intersection point $\cal P$ along a 1-cycle $\Sigma$ on $B^0$ is represented by the matrix
\bea
M_{{\rm I}_{n}} = \left( \begin{matrix}  1 & - n-1 \\0 & 1 \end{matrix}\right) \,.
\eea 
This implies that the dP$_9$ surface $Y^0$ must generically have $(12-n)$ singular fibers away from $\cal P$. The monodromy upon encircling the location of the associated $(12-n)$
7-branes is  
\bea
 M^{-1}_{{\rm I}_{n}}  = M_{\hat E_{9-{n}}} \,,
\eea
which identifies the 7-branes on $Y^0$ away from the intersection point $\cal P$ as the constituents of an $\hat E_{9-n}$ configuration. The enhancement to $\hat E_{9-n}$ occurs in the original Weierstrass model prior to blowup as a consequence of the
non-minimal singularity (\ref{vanorders-gen}), and the blowup partially separates the branes.
We therefore claim that there is the following dictionary between  affine $\hat E_{9-n}$ algebras in F-theory and the geometry of the blowup of a Weierstrass model with non-minimal singularities: \vspace{3mm}

\begin{whitebox}
\begin{minipage}{4cm}
$\hat E_{9-n}$ enhancements \vspace{1mm} \\
 in F-theory  
\end{minipage}
\begin{minipage}{2cm}
\qquad $\Longleftrightarrow$
\end{minipage}
\qquad \begin{minipage}{10cm}
Rational elliptic end components of a Type III.a  \vspace{1mm}\\ Kulikov  Weierstrass model with \vspace{1mm} \\
$({\rm deg}(f),{\rm deg}(g),{\rm deg}(\Delta)) = (4,6,12 -n)$ \\
{ away from the other surface components}
\end{minipage}
\vspace{3mm}
\end{whitebox}
where ${\rm deg}$ denotes the sum of vanishing orders.  
By analogy with our discussion of the Type II.a models  in Section \ref{sec_II.a}, it is now clear that we can construct {\it one} independent elliptic transcendental 2-cycle $\gamma$ by transporting the (1,0) cycle $S_A$ in the elliptic fiber along the pinching 1-cycle $\Sigma$ around the intersection point $\cal P$.
This is possible because the $(1,0)$ cycle in the  elliptic fiber is left invariant by the monodromy $M_{{\rm I}_n}$ upon encircling $\cal P$.
The calibrated volume of $\gamma$ in the degenerate K3 surface $X_0$ vanishes. 
Unlike for a Type II.a degeneration, the $(0,1)$ cycle in the elliptic fiber is not left invariant by this monodromy. This explains why there is only a single, rather than two, 
homologically independent transcendental 2-torus of vanishing volume.

M2-branes wrapping $\gamma$ give rise to a tower of BPS states in the seven-dimensional M-theory compactification,
which uplifts to a tower of BPS states in the 8d F-theory.
In the F-theory language this tower is identified with the tower of string junctions $k \, \delta_1$, $k \in \mathbb Z$, where $\delta_1$ is a $(1,0)$ string that encircles the point $\cal P$.
Equivalently, it can be viewed as a string junction encircling the $(12-n)$ 7-branes located on the base $B^0$ away from the intersection point $\cal P$.

Compared to the Type II.a limit,  we therefore encounter
 only a single tower of asymptotically massless BPS particles, corresponding to the single loop enhancement to build an
  $\hat E_{9-n}$
algebra for $n \geq 1$. Such loop algebras are affine Lie algebras of Kac-Moody type. Furthermore,
this tower of BPS states can be interpreted as Kaluza-Klein tower signalling a decompactification to 9d in the language of
the dual heterotic string. We will confirm this interpretation by a careful match with 
the dual heterotic side in Section \ref{subsec_TypeIIIhet}.

If both end components $Y^0$ and $Y^P$ are dP$_9$ surfaces, the total non-abelian part of the symmetry algebra of the configuration is
\bea \label{symmetryIIII}
G_\infty = H \oplus (\hat E_{n_0} \oplus \hat E_{n_P})/\sim  \,,
\eea
where the quotient indicates that the two string junctions associated with the imaginary roots of the two affine algebras are not independent but rather are identified.
If only one end component is a dP$_9$ surface, the second affine factor does not appear.
The factor $H$ refers to the non-abelian symmetry algebra localised on all components with generic fibers of Type I$_{k>0}$.
The non-abelian part of the gauge algebra in the 9d decompactification limit is the maximal finite Lie subalgebra of (\ref{symmetryIIII}), which is given by
\bea
G_{\rm 9d} = H \oplus  E_{n_0} \oplus E_{n_P}  \,.
\eea
We will give further evidence for this picture in Section \ref{subsec_TypeIIIhet}.

There is a second difference as compared to the Kulikov Type II.a limit: In a Type III.a model, the intermediate surface components $Y^i$, $i=1,\ldots, P-1$ (and possibly one of the two end components as well), are I$_{n_i}$ surfaces with $n_i>0$. Since the enhancements at the intersection of the components are all of Kodaira Type I$_k$, the generic I$_{n_i}$ fibers over the different components are mutually local with respect to one another.
Therefore, the 10d axio-dilaton coupling vanishes along these components.
 Moreover there is a
$(1,0)$ string which is obtained from an M2-brane that wraps the vanishing $(1,0)$ cycle in the fiber of the components $Y^i$, with $n_i>0$. It becomes asymptotically tensionless in units of the 8d Planck scale, but only locally in the region of the degenerate compactification space.

Nonetheless, the limit is not an 8d weak coupling limit.
First, as just mentioned, the $(1,0)$ string becomes tensionless only away from the rational elliptic end component(s), where the generic fiber is non-degenerate and hence the $(1,0)$ cycle is not
contracted. This already shows that there does not exist a {\it globally} weakly coupled frame in which the $(1,0)$ string plays the role of the fundamental string.
Second, the mass scale of the
tower of states arising from the string junctions $k \, \delta_1$
vanishes faster than the vanishing rate of the excitations of the (local) $(1,0)$ string. This is because the tower of string junctions is composed of asymptotically tensionless $(1,0)$ strings winding, in addition, around a vanishing cycle encircling $\cal P$. This  subtle point justifies the interpretation as a decompactification, rather than as equidimensional weak coupling limit. 

We note that the local weak coupling nature along the surface components away from the rational elliptic end components
manifests itself also in the structure of Kodaira fibers over special points of the base. As reviewed in Section \ref{sec_Kulikovrecap}, it was found in \cite{Kulikov} that the only possible fiber enhancements on the surface components with 
generic I$_{n>0}$ fibers are to singularities of $A$-type or $D$-type. These are the singularities which are compatible, at least locally, with weak coupling.
On an end component with I$_{n>0}$ fibers, the $D_k$ enhancements in Table \ref{tab_Kodairavanishing} can furthermore occur also for values $0 \leq k < 4$:
Such enhancements are known to be absent, by Kodaira's classification, on non-degenerate elliptic surfaces, corresponding to a non-perturbative obstruction
to forming bound states of branes of the form $A^k B C$ for $0 \leq k < 4$. In locally weakly coupled regions, on the other hand, no such obstruction should occur, and this is in perfect agreement with the 
existence of the corresponding $D_k$-type fibers on the weakly coupled surface components. Examples of this phenomenon can be found in  \cite{Kulikov}.

Finally, in the next subsection we will provide an interpretation of the extra constraint (\ref{IIIafurthertuning}) which distinguishes the Type III.a limits with one or two rational elliptic end surfaces.

\subsection{Type III.b Kulikov models as weak coupling plus decompactification limits}

As we now show, the physics of a Kulikov Type III.b  limit is that of a weak coupling limit dual to a perturbative Type IIB orientifold,  in combination with a large complex structure limit for the torus
$T^2_{\rm IIB}$ on which the Type IIB theory is compactified. This leads to an asymptotic decompactification from 8d to 10d.

First, the codimension-zero I$_{n_i>0}$ fibers in all surface components $Y^i$ indicate the existence of a weakly coupled 
duality frame that is globally defined on the base $B_0$
in the infinite distance limit. The  I$_{n_i}$ singularities in the fibers $Y^i$ are induced by the shrinking of one-cycles, 
and since the I$_{n_i}$ singularities of all components are mutually local with respect to each other (otherwise the components $Y^i$ would intersect each other
in fibers different from I$_k$ type), they are in fact copies of the same one-cycle, $S_A$, that shrinks in all components. 
An M2-brane wrapped along $S_A$ gives rise to a critical string in the uncompactified dimensions which is asymptotically tensionless as measured in 7d Planck units. This string uplifts to the weakly coupled Type IIB string in F-theory, whose tower of excitations is asymptotically tensionless compared to the 8d Planck scale. 
In addition, one finds a tower of massless states from the winding modes of this string along the 1-cycle $\Sigma$ on each base component $B^i$ that shrinks at the intersection of two components.
The fibration of $S_A$ over $\Sigma$ defines one transcendental elliptic curve~$\gamma$ of asymptotically vanishing volume. The same tower of asymptotically massless states can hence be equivalently understood in M-theory as 
the tower of M2-branes wrapping this elliptic curve~$\gamma$ multiple times. It is clear that the mass scale associated with this latter tower of particles asymptotes to zero  faster than the excitations of the weakly coupled fundamental string from the M2-brane wrapped on $S_A$ alone. This is because
\bea
\frac{M^2_{\rm str}}{M^2_{11}} \sim {\cal V}_{11}(S_A)  \to 0   \,, \qquad \quad   \frac{M_{{\rm w}}}{M_{11}}  \sim {\cal V}_{11}(S_A) \times {\cal V}_{11}(\Sigma) \sim  {\cal V}_{11}(\gamma)   \to 0 \,,
\eea
such that 
\be
\frac{M_{{\rm w}}}{M_{\rm str} } \sim  M_{\rm str}  {\rm Vol}(\Sigma) \to 0 \,,
\ee
where ${\cal V}_{11}(\ldots)$ measures volumes in units of the 11d Planck scale, $M_{11}$.

In fact, we will identify the tower associated with $\gamma$ as the winding modes of the weakly coupled fundamental Type IIB string wrapped on an asymptotically vanishing one-cycle on $T^2_{\rm IIB}$.
The vanishing of the one-cycle indicates that the complex structure of $T^2_{\rm IIB}$ undergoes an infinite distance limit, in addition to the weak coupling limit $\tau \to i \infty$.
Since the K\"ahler volume of the compactification space remains constant in units of the Type IIB string scale,  such a large complex structure limit indicates a complete decompactification to 10d. In addition to the tower of winding states described above, there must thus arise a tower of supergravity Kaluza-Klein modes associated with the dual 1-cycle of $T^2_{\rm IIB}$ whose volume becomes inversely proportional to the one of the shrinking cycle. This tower of states is not visible as wrapped M2-branes in M-theory and needs to be inferred in a more indirect manner. 

We can understand these statements more clearly in the language of the degenerating Weierstrass model as follows.
Since all components of $Y_0$ have generic fibers of Kodaira Type I$_{n_i>0}$, we can blow down all but one components to arrive at a (highly singular) family of Weierstrass models ${\widehat{\cal Y}}$, which feature a degenerate central element $\widehat Y_0$ whose  
codimension-zero fibers are guaranteed to be of type I$_{n>0}$. Therefore its Weierstrass model can be written in the general form (\ref{weakoupling-gen}) for certain $\eta$ and $\rho$.
At the same time, the fact that the original degeneration leads to a multi-component central fiber $Y_0$ implies the existence of non-minimal Kodaira fibers in $\widehat Y_0$. They must then occur at the zeroes of $h$. This follows from the fact that at each of the four zeroes of $h$, the degree $8$ function $f$ and the degree $12$ function $g$ vanish to order $2$ and $3$, respectively, which leaves no room for additional zeroes of $f$ and $g$ at different points. A non-minimal Kodaira singularity at one of the zeroes of $h$ in turn is only possible if at least two zeroes of $h$ coincide on ${\widehat{Y}_0}$. Since from a Type IIB perspective the zeroes of $h$ 
are the locations of the O7-planes, this means that at least one pair of O7-planes coalesces in the limit.
When this happens, one must now distinguish two qualitatively different situations:
\begin{itemize}
\item Generically, i.e., without further tuning of the parameters beyond the described collision of O7-planes via that of zeros of $h$, we will be taken away from a uniform weak coupling limit by introducing strongly coupled localised codimension-one objects on the base.
This manifests itself in such a way that after performing the necessary blowups, which bring us back to the Weierstrass family ${\cal Y }$, at least one of the end components of $Y_0$ is not of I$_n$ type with positive $n$, but rather is a rational elliptic surface. That is, the degeneration is of Type III.a. The rational elliptic component is the one which contains the location of the colliding O7-planes after the blowup and the axio-dilaton is generically of ${\cal O}(1)$.
The situation is thus best described as a 9d decompactification limit of the dual heterotic string, as explained in Section \ref{subsec_8dto9d}.
\item
In contrast, by a special tuning of the parameters of the Weierstrass model underlying the family ${\widehat{\cal Y}}$, we can remain at weak coupling despite the collision of the O7-planes, namely if the weak coupling limit is taken at a faster rate than the limit leading to the collision.
This way, the blowup indeed results in a Type III.b degeneration, in which all surface components of $Y_0$ have generic I$_{n>0}$ fibers. 
The required tuning is precisely the additional condition (\ref{IIIbtuning}).
\end{itemize}

The interpretation of such infinite-distance complex structure limits in terms of weakly coupled Type IIB strings is suggested by noting that the distance of the O7-planes
measures the length of the 1-cycles of the Type IIB compactification space $T^2_{\rm IIB}$. More precisely, the non-trivial 1-cycles on $T^2_{\rm IIB}$ correspond to the one-cycles encircling pairs of zeroes of the section $h$ on the base $B_0$ of $\widehat{Y_0}$ introduced above. Colliding two of these zeroes is therefore equivalent to a complex structure degeneration of $T^2_{\rm IIB}$ for which a one-cycle shrinks.
To the extent that the K\"ahler volume in Type IIB string units is unaffected by the complex structure degenerations, this leads to a decompactification to 10d, rather than to 9d.

\section{Infinite Distance Limits in the $E_7 \times E_8$-Weierstrass Model}   \label{sec_het}

In this section, we systematically analyse a representative class of infinite distance limits in the complex structure moduli space of an elliptic K3 surface
from the point of view F-theory/ heterotic duality.
We will obtain a clear physical interpretation of the infinite distance limits in the moduli space of F-theory via
the explicitly known mirror map to the heterotic moduli in terms of Siegel modular forms.
This illustrates and  lends additional support to our claims of the previous sections concerning the asymptotic physics of the infinite distance limits.

For concreteness, we consider a one-parameter family (\ref{Weierstrassfam-def}) of K3 surfaces described by a Weierstrass model over base $\mathbb P^1_{[s:t]}$
which at a generic point $u\ \neq 0$ in moduli space gives rise to a non-abelian gauge algebra $E_7 \times E_8$. 
This model is sufficiently simple to allow for a systematic treatment, and at the same time exhibits all qualitative properties of the infinite distance limits
of Kulikov Type II.a and III.a as described in the previous section.\footnote{The $E_7 \oplus E_8$ algebra is inconsistent with limits of Type II.b or III.b.}
We will
construct these infinite distance limits by enhancing the exceptional gauge algebras to loop algebras $\hat E_{9-n}$.
This will be achieved by constructing non-minimal singularities in the Weierstrass model $\cal Y$ corresponding to the vanishing orders~\eqref{vanorders-gen} with $m=0$ or $n=0$, i.e.,
 \bea  \label{Kodairavan-k}
 {\rm ord}_{\cal Y}(f,g,\Delta)|_{u=s=0} = (4+m,6,12+k) \quad {\rm or}  \quad    (4, 6+n,12+k) 
 \,,   
 \eea
at a point on the base $\mathbb P^1_{[s:t]}$, which, here, is taken to be $s=0$.
We will see explicitly how degenerations with $k=0$ are of Type II.a and
give rise to complete decompactifications of the 8d theory to 10d, as can be read off from the appearance of two loop algebras of Type $\hat E_9$. 
For $k>0$, on the other hand, the degenerations will be of Type III.a and we will realise affine Lie algebras of type $(\hat E_{m} \times \hat E_8)/ \sim$ with $m=7,8$. The effective theories will in turn be interpreted as partial decompactifications to 9d by carefully translating the model into the dual heterotic frame via the formalism developed in  \cite{Malmendier:2014uka} and studied further in \cite{Gu:2014ova,Garcia-Etxebarria:2016ibz}.
Moreover, depending on the parametrisation of the degeneration, we will find an intricate structure of various branches leading 
to different gauge algebras in the effective 9d theory.

After introducing the $E_7 \times E_8$ Weierstrass model in Section \ref{Sec_E7E8Weier-gen}, we will explain how to
systematically construct non-minimal singularities of the form (\ref{Kodairavan-k})
 in Section \ref{Sec_E7E8BeyKod}
and outline the structure of the resulting degenerate Kulikov Type II or Type III models. 
In Sections \ref{sec_E7E8-ConcreteReal} and~\ref{subsec_TypeIIIhet} we will illustrate the general picture for a particular  parametrisation of the degeneration. This will lead to different branches of the blowup theory,
and we will analyse the gauge symmetry in the effective 9d theory both within F-theory and from the perspective of the dual heterotic theory.

\subsection{The $E_7 \times E_8$-Weierstrass model and its heterotic dual}  \label{Sec_E7E8Weier-gen}

Our starting point is the family of Weierstrass models $y^2 = x^3  + f x z^4 + g z^6$ over $\mathbb P^1_{[s:t]}$ defined by the degree 8 and 12 functions\footnote{The parameters $a$, $b$, $c$, $d$, $e$ depend on $u$, which we suppress, and for notational simplicity we also drop the subscript of $f_u$ and $g_u$, as already mentioned.}
\bea \label{Weierstrass-E7E8-v1}
f=  t^3 s^4(    a \,  t + c \,  s )  \,,    \qquad g=  t^5 s^5 (d \, s^2 + b \, s \,   t + e \, t^2) \,.
\eea
For generic values of the complex parameters $a$, $b$, $c$, $d$, $e$, the vanishing orders of $f$, $g$ and their discriminant $\Delta = 4 f^3 + 27 g^2$ are given by
\be
{\rm ord}(f,g,\Delta)|_{t=0} =(3,5,9)   \,, \qquad  {\rm ord}(f,g,\Delta)|_{s=0} =(4,5,10)   \,.
\ee
This identifies the non-abelian part of the gauge algebra in the 8d compactification of F-theory on this family of K3 surfaces as $E_7 \oplus E_8$, for generic values of the parameters.
For special choices of parameters, the non-abelian gauge algebra can enhance further.

The complex parameters $a$, $b$, $c$, $d$, $e$ are subject to the 2-parameter family of rescalings
\bea \label{rescalingWeier}
(x,y,z,t,s, a,b,c,d,e) \sim (\lambda^{2 \tau} x,\lambda^{3 \tau} y, z, \lambda^{\sigma} t, s, \lambda^{4 \tau-4 \sigma} a, \lambda^{6 \tau-6 \sigma} b,\lambda^{4 \tau-3 \sigma} c,\lambda^{6 \tau-5 \sigma} d,\lambda^{6 \tau-7 \sigma} e) \nonumber\\
\eea
for arbitrary $\tau, \sigma \in \mathbb C^\ast$.
All-in-all we are therefore left with a moduli space of complex dimension three.

This moduli space can be most conveniently characterized in terms of  the dual heterotic string compactification on $T^2_{\rm het}$.\footnote{We will drop the subscripts from now on when referring to the quantities in the heterotic duality frame.}
The duality map, which is essentially the mirror map, has been worked out in detail in \cite{Malmendier:2014uka} and was studied further in \cite{Gu:2014ova,Garcia-Etxebarria:2016ibz}. In the patch $e=1$ of moduli space, the remaining four parameters of the K3 Weierstrass model can be identified as
\bea \label{abcdid}
a = - \frac{\psi_4({\utau})}{48} \,, \quad b = - \frac{\psi_6({\utau})}{864} \,, \quad c = - 4 \chi_{10}({\utau})  \,, \quad d =  \chi_{12}({\utau}) \,, \quad e=1 \,.
\eea
Here $\psi_{k}$ and $\chi_{m}$ are Siegel modular forms\footnote{Definitions and some of the most relevant properties are reviewed in Appendix~\ref{app_seigel}.} of genus $g=2$ of the indicated weight. They depend via 
\be
\utau:=\left(\begin{matrix}
T & V\\
  V & U
\end{matrix}\right)
\ee
on the  K\"ahler modulus $T$, complex structure modulus $U$ and Wilson line modulus $\V$ of the dual heterotic string compactified on $T^2$.
For later purposes, we also introduce the variables
\be
\qt\equiv e^{2\pi iT} \,, \qquad \qr\equiv e^{2\pi iU}  \,, \qquad \qv \equiv e^{2\pi i \V}. 
\ee
Another quantity that will be relevant for us is the discriminant factor
\be   \label{Qvschi35}  
Q(a,b,c,d) \sim    \frac{\chi^2_{35}}{\chi_{10}}   \,,
\ee
where the cusp form $\chi_{35}$ is the single odd generator of the ring $M_{*}(\Gamma_2)$ of genus $g=2$ Siegel modular forms (\ref{Mring}).
The polynomial $Q$ is defined as follows \cite{Malmendier:2014uka}:
For generic values of the parameters $a$, $b$, $c$, $d$, the discriminant of the Weierstrass model takes the form
\bea
\Delta = t^9 s^{10}  \Delta'\,,
\eea
where $\Delta'$ is a polynomial of degree five in $[s:t]$. The discriminant of this polynomial factors further into two polynomials
\bea
{\rm Discr}(\Delta') =   P(a,b,c,d)  \times Q(a,b,c,d) \,.
\eea
If $P(a,b,c,d)=0$,  
the model acquires an additional Type II Kodaira fiber over some point on the base, while otherwise whenever $Q(a,b,c,d)=0$ has a single zero, there occurs an I$_2$ singularity, which is characteristic of an $SU(2)$
enhancement.\footnote{Such codimension one singularities are called Humbert surfaces, and if multiple roots of $Q(a,b,c,d)$ coincide, higher codimension intersections of Humbert surfaces occur that are called Shimura curves or complex multiplication points. The nomenclature is not important for our purposes.}
Explicitly, one finds that   \cite{Malmendier:2014uka}   
\be
\begin{split}   \label{Qexpress}   
Q(a,b,c,d) &= -16\,a^6 b\,c^3 - 216\,a^3\,b^3\,c^3 - 729\,b^5\,c^3 + 16\,a^5\,c^4 + 
 2700\,a^2\,b^2\,c^4 - 5625\,a\,b\,c^5   \cr
 & + 3125\,c^6 + 16\,a^7\,c^2\,d + 
216\,a^4\,b^2\,c^2\,d + 729\,a\,b^4\,c^2\,d 
- 3420\,a^3\,b\,c^3\,d + 
 6075\,b^3\,c^3\,d  \cr
 & + 4125\,a^2\,c^4\,d + 888\,a^4\,c^2\,d^2 - 5670\,a\,b^2\,c^2\,d^2 - 
 13500\,b\,c^3\,d^2 + 16\,a^6\,d^3 + 216\,a^3\,b^2\,d^3  \cr
 & +729\,b^4\,d^3 - 
 2592\,a^2\,b\,c\,d^3 + 16200\,a\,c^2\,d^3 + 864\,a^3\,d^4 - 5832\,b^2\,d^4 + 11664\,d^5 \,.
\end{split}
\ee

\subsection{Systematics of affine enhancements in the $E_7 \times E_8$-Weierstrass model} \label{Sec_E7E8BeyKod}

Our goal is to construct the possible infinite distance limits in the three-dimensional complex structure moduli space.
With the help of two scaling relations it can always be arranged that the parameters $(a,b,c,d,e)$ take finite values.
In this scheme, the infinite distance limits that are not of weak coupling type\footnote{Such limits occur by degenerating the elliptic fiber over generic points of the base, as systematised in \cite{Kulikov}.} can be obtained 
 by arranging for non-minimal Kodaira singularities, i.e.~vanishing orders for $(f,g,\Delta)$ of order $(4,6,12)$ or beyond,  at points in codimension one on the base.
With all parameters kept finite, there are a priori three different ways of achieving this in the family (\ref{Weierstrass-E7E8-v1}) of Weierstrass models:
\begin{enumerate}
\item
Non-minimality only at $t=0$ requires taking $c \to 0$ and $d \to 0$, while keeping $e \neq 0$.
\item
Non-minimality only at $s=0$  requires taking $e \to 0$ while keeping $(c,d) \neq (0,0)$.
\item
Non-minimality both at $s=0$ and $t=0$ requires  $c \to 0$, $d \to 0$ and $e \to 0$.
\end{enumerate}

However, it is easy to see that due to the rescaling symmetry (\ref{rescalingWeier}) these three types of limits can be {\it partially} mapped into one another, at least in certain regimes of moduli space.
Note first that since we are interested in identifying relations between asymptotically massless towers, it is convenient to express the infinite distance limits as one-parameter limits.
Consider then a general limit in which
\bea
c \sim u^\gamma   \,, \quad d \sim u^\delta   \,, \quad e \sim u^\epsilon   \,\quad {\rm with}   \, \quad u \to 0   \,,  \qquad  \gamma, \delta, \epsilon \geq 0 \,.
\eea
The two-parameter family (\ref{rescalingWeier}) of rescalings contains the one-parameter family (for $\lambda^{\tau} = \lambda^{\sigma} =: \lambda$)
\bea
(a,b,c,d,e) \sim   (a,b,\lambda c,\lambda d, \lambda^{-1} e)  \,.
\eea
Then taking $\lambda = u^\epsilon$, as long as 
\bea   \label{restriction-mod}
\gamma > \epsilon \,, \qquad \delta > \epsilon \,,
\eea
we can go to a patch in which $c \to 0$ and $d \to 0$ while $e$ stays finite and of order one in the limit $u \to 0$.
Importantly, we still have the freedom to perform a rescaling which identifies
\bea \label{residualtrafo}
(a,b,c,d) \sim (\lambda^4 a, \lambda^6 b, \lambda^{10} c, \lambda^{12} d) 
\eea
without leaving the patch $e=1$, in agreement with the modular behaviour of the Siegel modular forms (\ref{abcdid}).

For concreteness, we will be considering regimes in moduli space which fall into this class.
In the regime of small $c$ and $d$, the combinations of Siegel modular forms that appear in (\ref{abcdid}) and (\ref{Qexpress})  admit expansions in $\qt$, $\qr$ and $\qv$ of the form (c.f. Appendix~\ref{app_seigel})
\be  \label{expansions-gen}
\begin{split}
c &\sim \chi_{10} \sim   \qt \qr  (-2 + \qv + \frac{1}{\qv} )  + \ldots  \,,  \cr
d &\sim \chi_{12} \sim   \qt \qr  (10 + \qv + \frac{1}{\qv} ) + \ldots   \,,  \cr
4 a^3 + 27 b^2&\sim (\psi_4^3-\psi_6^2) \sim   \qt +\qr + \ldots \,,  \cr
Q &\sim \chi^2_{35}/\chi_{10} \sim \qt^3 \qr^3 (\qt -\qr)^2   \qv (\qv + \frac{1}{\qv})^2  + \ldots    \,.
\end{split}
\ee
The infinite series of subleading terms in $\chi_{10}$, $\chi_{12}$ as well as the terms in $\psi_4$ and $\psi_6$ underlying the quoted expression for $\psi_4^3-\psi_6^2$ can be ignored as long as
\bea \label{conditionforchi1012}
\frac{\qt}{\qv} \ll 1   \,,  \qquad \frac{\qr}{\qv} \ll 1   \,.
\eea
This condition is consistent with the condition for the modulus of Wilson line to lie within the fundamental domain given by
\bea
{\rm Im}(U){\rm Im}(T) \geq ({\rm Im}(\V))^2 \,.
\eea
Similarly, the subleading terms for $ \chi^2_{35}/\chi_{10}$ can be ignored as long as
\bea \label{conditionforchi35}
\frac{\qt}{\qv^2} \ll 1   \,,  \qquad \frac{\qr}{\qv^2} \ll 1 \,.
\eea
In the regime of small $\qv$, the condition (\ref{conditionforchi35}) is stronger than (\ref{conditionforchi1012}) and can therefore be violated within the fundamental domain. The expansion (\ref{expansions-gen}) of $Q$ can only be applied if (\ref{conditionforchi35}) holds.

After these preliminaries we now consider infinite distance limits in the patch where $e=1$ by taking $c \to 0$ and $d \to 0$.
As it turns out, the physically different infinite distance limits are distinguished by the relative vanishing order of $c$ and $d$ with respect to
the vanishing of the combination $4 a^3 + 27 b^2$.
This motivates considering the vanishing orders
\bea  \label{knmpar-def}
4 a^3 + 27 b^2 \sim u^k \,, \qquad   c \sim u^n  \,, \quad  d \sim u^m \,  \qquad n \geq 1, m \geq 1 \,, k \geq 0 \,,
\eea
where the value of $k$ coincides with that in (\ref{Kodairavan-k}).
Models with such vanishing orders require
\be
 l = {\rm min}(n,m)
 \ee
  blowups to resolve the non-minimal singularity at $t=0$.

As we will see, the physics at infinite distance can be classified as follows:
\begin{enumerate}
\item Limits with $k=0$ give rise to a Kulikov Type II degeneration of the elliptic K3 surface.
This follows from the fact that in such cases after the blowup both end components contain 
12 branes each, as expected for a $\hat E_9$ configuration.
Correspondingly, the expansion of modular forms implies that 
in this case, $\qt \to 0, \qr = {\cal O}(1)$ or vice versa, which amounts to decompactification to 10 dimensions.
\item
Limits with $k \geq 1$ are of Kulikov Type III, corresponding to a  $\hat E_{8-N}$ configuration for $N=0$ or $N=1$ on the two end components.
The expansion of the modular forms is consistent with the fact
that $\qt \to 0$ and $\qr \to 0$ at a finite ratio, signalling a partial decompactification to 9 dimensions.

For otherwise generic coefficients, we will find Type III limits with 
\bea
{\rm Im}(T) =(l-1) {\rm Im}(U) \,.  
\eea 
Here we made use of the duality symmetry to make sure that, without loss of generality, ${\rm Im}(T) \geq  {\rm Im}(U)$.
For each choice of $l$, different branches are obtained by special tunings of the remaining coefficients, with the following properties:
If $n > m$, the non-abelian part of the gauge algebra in 9 dimensions always includes an $E_8 \oplus E_8$ factor, which can be enhanced to $E_8 \oplus E_8 \oplus SU(2)$.
If $n \leq m$, the non-abelian part of the gauge group in 9 dimensions can be $E_8 \oplus E_8$ or $E_8 \oplus E_8 \oplus SU(2)$, or $E_7 \oplus E_8$ 
with possible enhancement to $E_7 \oplus E_8 \oplus SU(2)$.\footnote{As will be explained at the end of Section \ref{subsec_TypeIIIhet}, the algebra of maximal enhancement $G_{\rm 9d} = E_7 \oplus E_8 \oplus SU(3)$ occurs outside the patch of moduli space characterised by (\ref{restriction-mod}).}

\end{enumerate}

\subsection{Type II degenerations}   \label{sec_E7E8-ConcreteReal}

To illustrate the general picture, we consider
 a class of limits with $n=m=4$ for the parameters defined in (\ref{knmpar-def}).
It turns out  convenient to parametrise the coefficients appearing in (\ref{Weierstrass-E7E8-v1}) as
\bea \label{hatE7tuning5}
\begin{split}
&& a  = a_0 \,, \qquad   b &= \frac{2 i a_0^{3/2}}{3 \sqrt{3}} (1  +y +  b_1 u) + b_2 u^2 +  {\cal O}(u^3)  \\ 
&& c  =  -  a_0( i \sqrt{3} +x) u^4  + a_0  p u^5 + {\cal O}(u^6) \,, \qquad  d &= a_0^{3/2} u^4 + d_1 u^5 +  {\cal O}(u^6)  \,,    
\end{split}
\eea
where  the infinite distance limit is enforced by taking $u \to 0$.

As a warmup, we take all parameters $b_1$, $a_0$, $p$, $x$ and $y$ to be generic and of order one.
At $u=0$, the discriminant factorises as
\bea \label{DeltaprimedII}
\Delta|_{u=0} = t^{12} s^{10} \Delta' \,.
\eea
In particular, the Weierstrass vanishing orders $({\rm ord}(f),{\rm ord}(g),{\rm ord}(\Delta))$ at $t=0$ are of type $(4,6,12)$, indicating a non-minimal singularity in the fiber.


A Weierstrass model with only minimal singularities
is obtained via
a sequence of four blowups,
\bea  \label{blowupchain}
(t,e_i,x,y) &\to& (t e_{i+1}, e_{i} e_{i+1}, x e_{i+1}^2, y e_{i+1}^3 )    \, \qquad  i = 0,1,2,3  \,,
\eea
where we have set $e_0 =u$.
After each step the Weierstrass equation exhibits an overall factor of $e_{i+1}^6$, and to obtain the proper transform of the Weierstrass equation we 
divide by this factor and consider the resulting equation. This amounts to a rescaling $(x,y,z,f,g) \sim (\lambda^2 x, \lambda^3 y, z, \lambda^4 f, \lambda^6 g)$ for $\lambda = e_{i+1}$, which does not affect the Calabi-Yau condition. The process terminates after four steps. 
The resulting fibration is a Weierstrass model $Y_0$ which degenerates into a chain of five components,
\bea
Y_0 = \cup_{i=0}^4 Y^i \,.
\eea
 Each component $Y^i$ is an in general degenerate elliptic fibration over a rational base curve 
 \be
 B^i = \{e_i =0\} \,,
 \ee
where we have defined $e_0 = u$.
Here ``degenerate'' refers to the fact that the Weierstrass model may have Kodaira fibers of Type I$_{n_i}$ over generic points of $B^i$, as explained generally in \cite{Kulikov}, but apart from this it
has only minimal Kodaira fibers over special points of $B^i$. These codimension-one singularities can be resolved, but we keep the description as a Weierstrass model.

After the blowup, the Stanley-Reisner ideal of coordinates that are not allowed to vanish simultaneously contains in particular the elements
\be 
\begin{split} \label{SRIa}
\{(e_4, e_2), (e_4,e_1), (e_4,u), (e_4,s), (e_3,t), (e_3, e_1), (e_3 , u), (e_3, s),  (e_2,u), (e_2,t), (e_2,s), \cr
 (e_1,s), (e_1,t) (e_1,s) (u,t)\}\in {\rm SRI}  \,.
\end{split}
\ee
To analyse the geometry of the Weierstrass models $Y^i$, we restrict $f$, $g$ and $\Delta$ to the individual components by setting $e_i=0$, and using the Stanley-Reisner ideal
(\ref{SRIa}) to set all coordinates that are not allowed to vanish on this locus equal to one.
In particular, for the discriminant factor $\Delta'$ in (\ref{DeltaprimedII}) this gives
\be
\begin{split}   \label{equ_E7E8TypeII}
\Delta'|_{e_4=0} &= e_3 P_2(e_3, t;x,a_0) + y Q_3 (e_3,t;y,a_0)    \cr
\Delta'|_{u=0} & =  e_1  P_1(e_1,s;x,a_0) + y Q_2(e_1,s;y,a_0)    \cr
\Delta'|_{e_i=0} &=    4 a_0^3 y (2 + y)   \,, \qquad   \qquad   i =1,2,3     \,,
\end{split}
\ee
where $P_i$ and $Q_j$ are polynomials of indicated degrees in the first variables and we also indicate their dependence on the parameters of the model.
From this we draw the following conclusions: 
The fiber over generic points of each base component $B^i$ is of Kodaira Type I$_0$. The two end components $Y^0$ and $Y^4$ have 12 singular fibers away from the intersection with the other components
and hence represent rational elliptic, or dP$_9$, surfaces. The component $Y^4$  has a Type III$^\ast$ singularity at $t=0$, corresponding to an algebra $E_7$, and three more I$_1$ singularities at generic points on $B^4$.
Moreover, $Y^0$ exhibits a Type II$^\ast$ singularity at $s=0$, corresponding to an algebra $E_8$, together with 2 more I$_1$ singularities over generic points on $B^0$.
The intermediate components $Y^i$, $i=1,2,3$, are trivial fibrations over the respective $B^i$, and all adjacent components intersect in an elliptic curve. 

The degeneration is therefore an example of a Kulikov Type II model\footnote{It is guaranteed that the model can be brought into the stable form of a Type II.a model \cite{Clingher:2003ui}, but we do not make this further step explicit here.} 
with symmetry group
\bea \label{hatE9hatE9}
G_{\rm \infty} = \hat E_9 \oplus \hat E_9 / \sim  \,.
\eea
As reviewed in Section \ref{Sec_LoopAlgebras},
the string junctions associated with the imaginary roots $\delta_1$
 and $\delta_2$ within $\hat E_9$ can be interpreted as the Kaluza-Klein towers from the decompactification to 10 dimensions in the heterotic duality picture. 
 Indeed, this can be seen explicitly by applying the duality map (\ref{abcdid}) in its expanded version (\ref{expansions-gen}). Up to order one coefficients, one finds 
\be
\begin{split}
 \chi_{10} \sim \chi_{12}   \sim   u^4 \,, \qquad 
\psi_4^3-\psi_6^2 \sim   1 \,, \qquad 
\chi^2_{35}/\chi_{10} \sim    u^{12} \,,
\end{split}
\ee
which indicates a scaling
\be
q_T \sim u   \,, \qquad q_U \sim 1    \,, \qquad \qv \sim 1   \,. 
\ee
This is of course precisely the behaviour $T \to i \infty$ with $U$ (and $\V$) of order one, for the dual heterotic moduli. 
It goes of course without saying that the gauge algebra in the 10d limit of the theory is $G_{\rm 10d}= E_8 \oplus E_8$, the gauge algebra of the 10d heterotic string, which corresponds to the maximal finite Lie algebra
within (\ref{hatE9hatE9}).

\subsection{Type III degenerations}  \label{subsec_TypeIIIhet}

To enhance the non-minimal singularity at $t=0$ further, we must arrange for $4 a^3 + 27 b^2$ to vanish as $u \to 0$, which amounts to setting the parameter $y=0$ in (\ref{hatE7tuning5}).
Now the Weierstrass vanishing orders at $t=0$ become $(4,6,13)$ in the limit $u \to 0$, and a rich structure of physically inequivalent branches opens up.
The non-minimal singularity can still be removed by the same blowup procedure as shown in
(\ref{blowupchain}). The crucial difference is that now the generic fiber of the intermediate surface components $Y^i$, $i=1,2,3$, is of Type I$_{n_i>0}$. This can be inferred from the third line in (\ref{equ_E7E8TypeII}), which shows that the discriminant along the intermediate $B^i$ vanishes if $y=0$.
This  is the hallmark of a Type III.a elliptic Kulikov model \cite{Kulikov}.

Let us now discuss a number of specializations:

\subsubsection*{$b_1$, $x$ generic of ${\cal O}(1)$}

For generic values of the parameters $b_1$, $a_0$, $p$, $x$ after the four blowups (\ref{blowupchain}), the discriminant of the properly transformed Weierstrass model takes the form
\bea \label{discrIII-1}
\Delta = e_1 e_2 e_3 t^9 s^{10} \Delta'\,,
\eea
with 
\be
\begin{split}   \label{equ_E7E8b1genxgen}
\Delta'|_{e_4=0} &= a_0^3 (-12 t^2 x + 3 e_3 t (-3 + 8 i \sqrt{3} x + 4 x^2) + 
   4 e_3^2 (3 i \sqrt{3} + 9 x - 3 i \sqrt{3} x^2 - x^3))   \cr
\Delta'|_{e_3=0} &=  -4 a_0^3 (2 b_1 e_4 + 3 e_2 x)     \cr
\Delta'|_{e_2=0} &= -8 a_0^3 b_1     \cr
\Delta'|_{e_1=0} &=   -4 a_0^{3/2} (2 a_0^{3/2} b_1 u - 3 i \sqrt{3} e_2)  \, \cr
\Delta'|_{e_0=0} &= 3  (9 e_1 + 4 i \sqrt{3} a_0^{3/2} s)  \,.
\end{split}
\ee
The geometry is depicted in Figure \ref{fig:E7E8b1genxgen}.
From (\ref{discrIII-1}) we infer that the intermediate surface components $Y^i$, $i=1,2,3$, have I$_1$ singular fibers in codimension zero, i.e. over generic points of their base $B^i$.
 The end components $Y^0$ and $Y^4$ are dP$_9$ surfaces intersecting the adjacent surfaces $Y^1$ and $Y^3$ in such an I$_1$ fiber. 
 Away from the respective intersection points, there are 11 singular fibers, distributed as a III$^\ast$ Kodaira fiber ($E_7$) at $t=0$ and 2 I$_1$ fibers  over generic points  of $B^4$ and, respectively, as a II$^\ast$ Kodaira fiber ($E_8$) at $s=0$ along with one extra I$_1$ fiber on $B^0$. One of the originally twelve 7-branes on each of the end components has moved to the adjacent component, that is to $Y^1$ or $Y^3$, respectively.
 
\begin{figure}[t!]
\centering
\includegraphics[width=9cm]{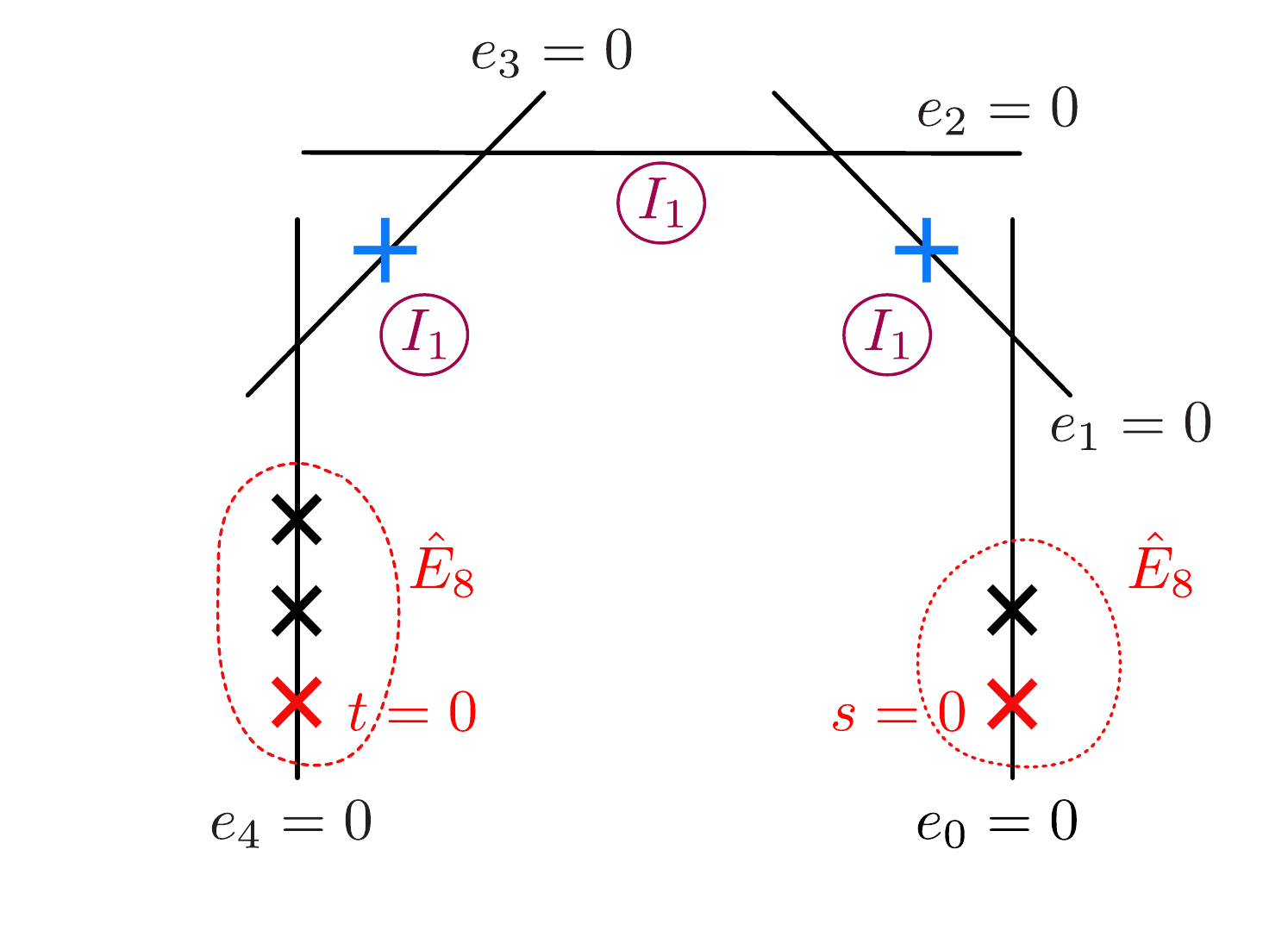}
\caption{
Kulikov Type III degeneration of $E_7 \times E_8$ Weierstrass model (\ref{equ_E7E8b1genxgen}) for generic values of the parameters $b_1$ and $x$.}
\label{fig:E7E8b1genxgen}
\end{figure}
 
 Upon encircling the intersection point $B^0 \cap B^1$  {\it clockwise} on $B^0$, and likewise the intersection $B^3 \cap B^4$ on $B^4$, one picks up a monodromy
 \bea
 M^{-1}_{I_1} = M_{\hat E_8} = \left(\begin{matrix}  1& 1 \\ 0 & 1  \end{matrix}\right) \,.
 \eea
This transformation leaves the string junction $\delta_1 = (1,0)$ invariant, 
which can be identified with the imaginary root of the affine Lie algebra $\hat E_8$. 
This affine algebra is obtained upon colliding the 11 branes located on $B^0$ and $B^4$ away from the intersection points. The blowup has separated the 11 branes in such a way as to give rise only to minimal Kodaira singularities in the fiber. 
The non-abelian part of the symmetry algebra of the theory in the limit is identified as
\bea \label{hatE8E8}
G_{\infty} = \hat E_8 \oplus \hat E_8 / \sim  \,. 
\eea 
As explained in Section \ref{subsec_8dto9d}, $n$ copies of the string junction $\delta_1$, for $n \in \mathbb Z$, give rise to a tower of massless BPS particles in the effective theory. These play the role of
a tower of Kaluza-Klein modes signalling the decompactification of the dual heterotic theory from 8d to 9d.
Furthermore we claim that the non-abelian part of the gauge algebra in the effectively 9d theory attained in the limit is $G_{\rm 9d} = E_8 \oplus E_8$, the maximal non-abelian Lie algebra within (\ref{hatE8E8}).

These two statements can be checked explicitly with the help of the dictionary (\ref{abcdid})  and (\ref{expansions-gen}), which now gives
\be
\begin{split} \label{scalingsTypeIII-2}
\chi_{10} \sim  \chi_{12}   \sim   u^4 \,, \qquad  \psi_4^3-\psi_6^2   \sim    u \,, \qquad 
 \chi^2_{35}/\chi_{10} \sim    u^{14}  \,.
\end{split}
\ee

From the behaviour of $\psi_4^3-\psi_6^2$ in the limit under consideration, we infer that either $\qt \sim u$ or $\qr \sim u$. W.l.o.g we take $\qu\sim u$.
Then 
$\frac{\qt}{\qv}  \sim u^3$ 
by inspection of $\chi_{10}$ and $\chi_{12}$.
With these two scaling relations, the behaviour of $\chi^2_{35}/\chi_{10}$ shows that altogether 
\bea \label{scalingsIII1}
\qu \sim u   \,, \qquad \qt \sim u^3 \,, \qquad \qv  = {\cal O}(1) \,.
\eea
The limit hence corresponds to a partial decompactification to 9d, for which 
\bea
{\rm Im}(T) \sim 3 \, {\rm Im}(U)   \,,   \qquad {\rm Im}(T) \to \infty   \,.
\eea
The vanishing of the heterotic Wilson line in 9 dimensions is reflected by $\V =  {\cal O}(1)$, which is realised in the present model. On this branch in moduli space, it is not possible to produce an additional $SU(2)$ factor because this would require that either $U = T$ or $2\V = T$ (or, after exchange of $T$ and $U$, $2\V =U$) in the dual heterotic model, which is incompatible with (\ref{scalingsIII1}).

Correspondingly in F-theory, generating an additional $SU(2)$ factor would require the collision of two single branes: This is not possible in the Kulikov model realising the branch under consideration with generic values for the parameters $b_1$ and $x$, 
because the two branes are localised on different components.

\subsubsection*{$x=0$, $b_1$ generic of ${\cal O}(1)$}

\begin{figure}[t!]
\centering
\includegraphics[width=18cm]{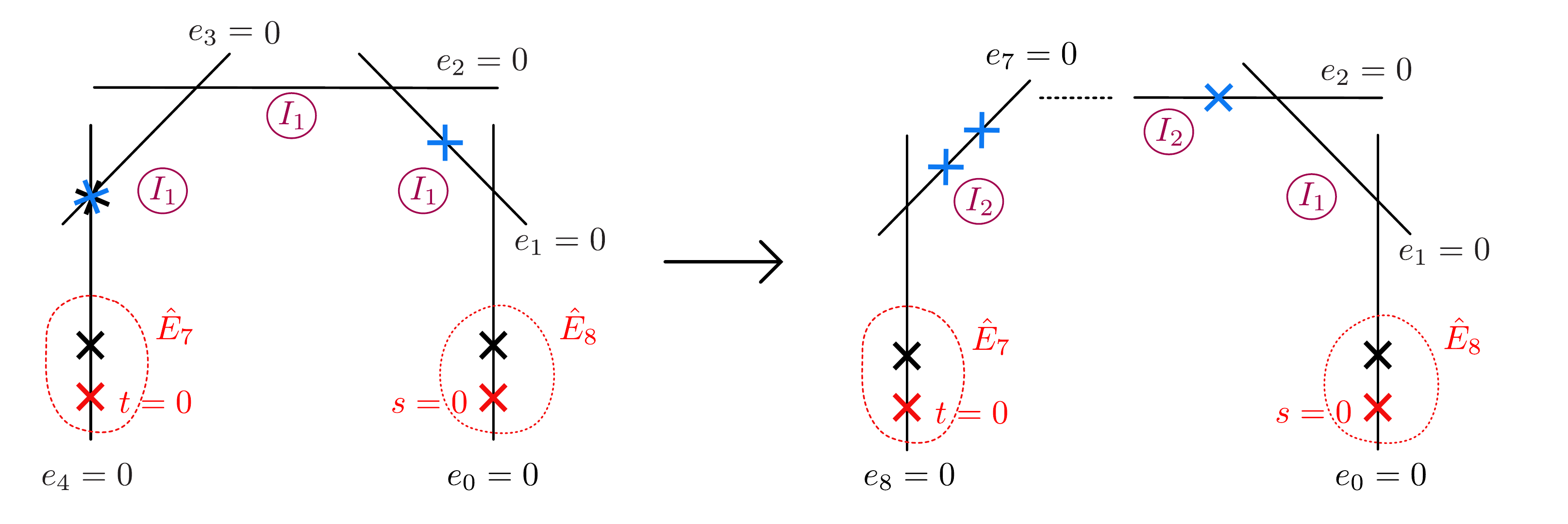}
\caption{
Kulikov Type III degeneration of $E_7 \times E_8$ Weierstrass model (\ref{equ_E7E8b1genxgen}) for generic value of the parameter $b_1$ and $x=0$.  The left shows how for $x=0$ one 7-brane from $B^4$ and one from $B^3$ coalesce at the intersection. The right picture shows the result of a base change and blowups. The intermediate components over $B^3$ up to $B^6$ have generic I$_2$ fibers and are not explicitly drawn.}
\label{fig:E7E8b1genx0}
\end{figure}

A qualitative change occurs if we set the parameter
$x=0$: One 7-brane on the base component $B^4$  and the 7-brane on the component $B^3$ both wander towards the intersection point $B^3 \cap B^4$. See Figure \ref{fig:E7E8b1genx0}.
This has two related effects: First, at the collision point, it seems at first sight that an additional $SU(2)$ gauge symmetry might arise. Second, according to our general logic the singularity at $t=0$ in the limit $u=0$
is of type $\hat E_7$, since the degrees of $(f,g, \Delta)$ decrease to $(4,6,10)$ away from the intersection point.

Concerning the potential $SU(2)$ enhancement, however, we note that the intersection locus $B^3 \cap B^4$ is a nodal singularity of the base 
and hence the physics interpretation of the 7-branes thereon is a priori obscure. To obtain a better understanding we go back to the degenerate Weierstrass model prior to the blowups and perform a base change $u \to u^2$, as explained generally in \cite{Kulikov}. Such a rescaling of the parameter $u$ cannot change the interpretation of the physics. Removing the non-minimal singularities in the fiber of the model after base change now requires a total of 8 blowups with blowup coordinates $e_i$, $i=1, \ldots, 8$.
One obtains a chain of 9 degenerate Weierstrass models, $Y_0 = \cup_{i=0}^8 Y^i$, with I$_{n_i}$ fibers over generic points of the following form:
\bea  \label{basechangeconfig}
{\rm I}_0 - {\rm I}_2- {\rm I}_2- {\rm I}_2- {\rm I}_2- {\rm I}_2- {\rm I}_2- {\rm I}_1- {\rm I}_0  \,.
\eea 
Here the left end component $Y^8$ (with base $B^8 :\{e_8=0\}$) is intersected by $t=0$ and the right end component $Y^0$ by $s=0$.
The base change resolves the relative position of two 7-branes which seem to coalesce for $x=0$: In the new model, both branes are located at two generic points within the base component $B^7$ given by the zeroes of
\bea
9 e_6^2 + 8 b_1 e_8^2 =0  \,.
\eea 
This shows that as long as $b_1 \neq 0$, the two branes cannot coincide and therefore $x=0$ with $b_1 \neq 0$ does not lead to an additional $SU(2)$ factor.

We conclude that for $x=0$ the symmetry group in the Kulikov model under consideration is 
\bea
G_{\infty } = (\hat E_7 \oplus \hat E_8)/ \sim   \,
\eea
and the non-abelian part of the gauge group in the partial decompactification limit to 9d is $G_{\rm 9d} = E_7 \times E_8$.

This behaviour is consistent with the heterotic dual model: The scaling (\ref{scalingsTypeIII-2}) of the parameters of the Weierstrass model remains unchanged by setting $x=0$ except for 
\bea \label{Qexpu15}
Q \sim \chi^2_{35}/\chi_{10}& \sim \qt^3 \qr^3 (\qt -\qr)^2   \qv (\qv + \frac{1}{\qv})^2 \sim    u^{15} \,.
\eea
Naively, this is compatible with the scaling
\bea
\qr \sim u \,, \qquad    \frac{\qt}{\qv} \sim u^3 \,, \qquad \qv \sim u^{1/2}   \,,
\eea
from which one would conclude that $2\V = U$ (corresponding to an $SU(2)$ enhancement), but this is too quick:
For this solution the convergence condition (\ref{conditionforchi35}) is badly violated and the expansion (\ref{Qexpu15}) breaks down. Hence there is no indication of an $SU(2)$ enhancement on the 
heterotic side either and we are left with the non-abelian part of the 9d gauge algebra given by
\bea
G_{\rm 9d} = E_7 \oplus E_8   \,.
\eea

\subsubsection*{$b_1=0$, $x ={\cal O}(1)$ and generic}

If, by contrast, we set $b_1=0$, while keeping $x  ={\cal O}(1)$ and non-zero, we enter a new branch because the 
vanishing order of the discriminant (\ref{discrIII-1}) along the component $e_2=0$ increases to
\bea
\Delta = e_1 e_2^2 e_3 t^9 s^{10} \Delta'
\eea
with 
\be
\begin{split} 
\Delta'|_{e_4=0} &= a_0^3 (-12 t^2 x + 3 e_3 t (-3 + 8 i \sqrt{3} x + 4 x^2) + 
   4 e_3^2 (3 i \sqrt{3} + 9 x - 3 i \sqrt{3} x^2 - x^3)) \,,  \cr
\Delta'|_{e_3=0} &=  -12 a_0^3 x  \,,  \cr
\Delta'|_{e_2=0} &= 12 i a_0^{3/2} (\sqrt{3} b_2 e_1 e_3 + i a_0^{3/2} e_1^2 x + \sqrt{3} e_3^2)  \,,  \cr
\Delta'|_{e_1=0} &= 12 i \sqrt{3} a_0^{3/2} \,.  \cr
\end{split}
\ee

\begin{figure}[t!]
\centering
\includegraphics[width=9cm]{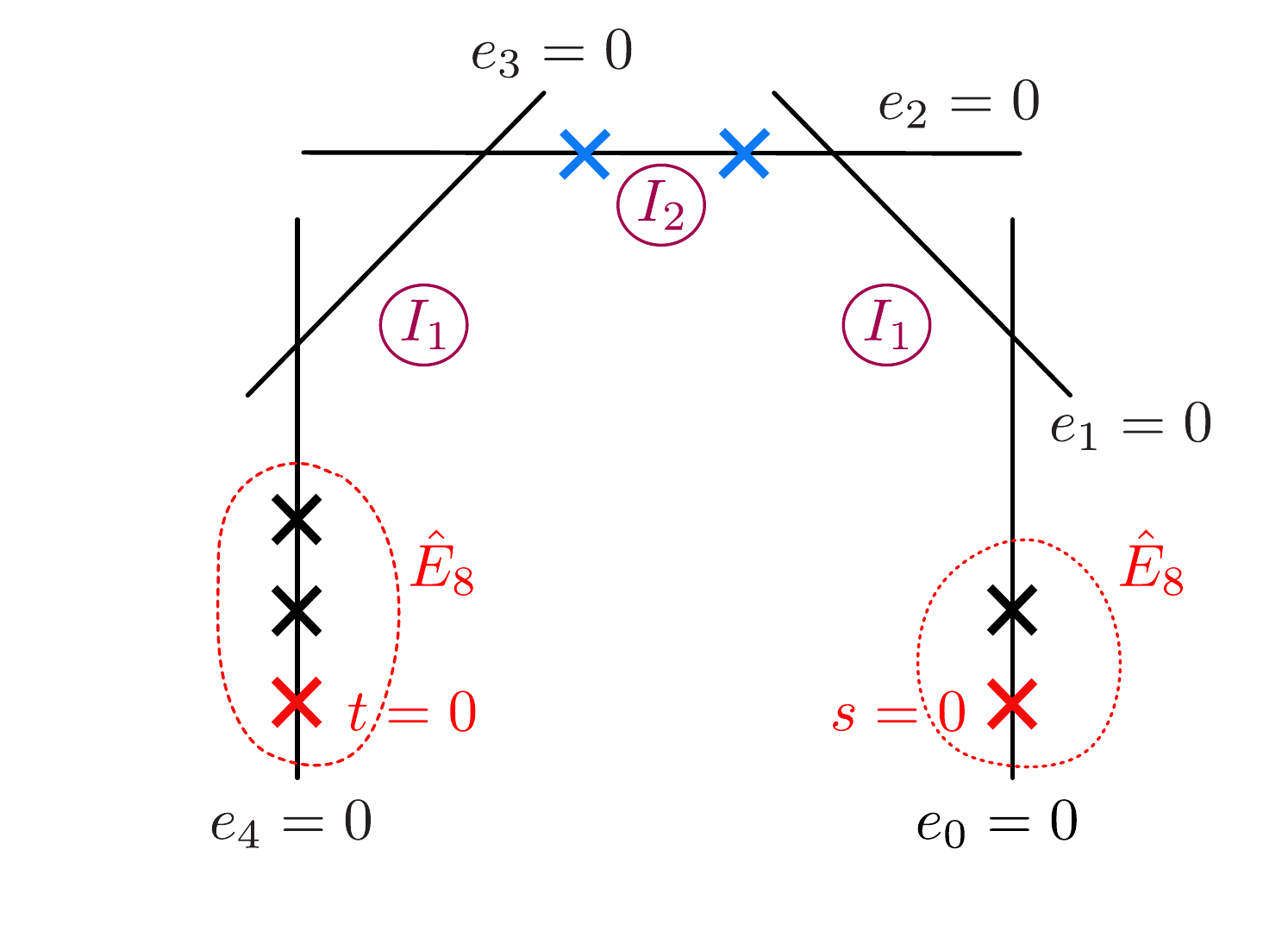}
\caption{
Kulikov Type III degeneration of $E_7 \times E_8$ Weierstrass model (\ref{equ_E7E8b1genxgen}) for generic value of the parameter $x$ and $b_1 =0$.}
\label{fig:E7E8b10xgen}
\end{figure}

The 7-brane from the component $e_3=0$ has moved onto the component $e_2=0$, see Figure~\ref{fig:E7E8b10xgen}.
This is reflected in a change of vanishing orders of the modular forms to
\be
 \chi_{10} \sim \chi_{12} \sim   u^4 \,, \qquad  \psi_4^3-\psi_6^2  \sim   u^2 \,, \qquad \chi^2_{35}/\chi_{10} \sim u^{16} \,.
\ee
Such behaviour requires that the heterotic variables scale as
\bea
\qt \sim u^2   \,, \qquad \qr \sim u^2 \,, \qquad \qv  = {\cal O}(1) \,.
\eea
Again, $\V =  {\cal O}(1)$ implies that the Wilson line in 9 dimensions is ineffective and the gauge algebra is enhanced to $G_{\rm 9d} = E_8 \oplus E_8$ in the 9-dimensional
limit.

The only way to achieve an additional gauge enhancement along this branch is if the two roots of $\Delta'|_{e_2=0}$ coincide, which occurs for
\bea
x = - \frac{i \sqrt{3}b_2^2 }{4 a_0^{3/2}}    \, ,
\eea
leading to the (non-abelian) symmetry group
\bea
G_{\infty} = SU(2) \oplus (\hat E_8 \oplus \hat E_8)/ \sim  \,  \qquad  \Longrightarrow  \,  \, G_{\rm 9d} = SU(2) \oplus E_8 \oplus E_8   \,.
\eea
Since the two coinciding 7-branes belong to the two different groups of 12 branes on K3, each associated with a full $\hat E_9$ loop algebra,
we expect such enhancement to correspond to the locus where, in heterotic coordinates, $T = U$.
This can be achieved on the branch under consideration, where $\qt \sim \qr \sim u^2$.
By contrast, an enhancement by taking $2 \V = T$ cannot be obtained on this branch, where $\V$ and $T$ are separated by two orders in $u$.

The vanishing order of the discriminant polynomial on the branch with $b_1=0$ is
\bea
Q \sim \frac{\chi^2_{35}}{\chi_{10}} \sim x^2 u^{16} \left( 9 a_0^{3/2} b_2^2  - 12 i \sqrt{3} a_0^3 x - 
 2 i \sqrt{3} a_0^{3/2} b_2^2 x - 8 a_0^3 x^2   + {\cal O}(u) \right)  \,.
\eea
The polynomial in brackets vanishes for
\bea
x = - \frac{i \sqrt{3}b_2^2 }{4 a_0^{3/2}}      \qquad {\rm or}   \qquad  x= - \frac{i 3 \sqrt{3}}{2}  \,.
\eea
The first solution indeed corresponds to the $SU(2)$ enhancement locus found above. We interpret the vanishing orders as indicative of the asymptotic scaling 
\bea
\qt \sim u^2 \,,  \quad  \qr = \qt + {\cal O}(u^{5/2})   \,, \quad \qv = {\cal O}(1)   \,.
\eea

\subsubsection*{Branch $b_1 =0$, $x=0$}

Finally, a novel branch opens up for  $b_1=0$ and $x=0$.
The discriminant becomes
\bea
\Delta = e_1 e_2^2 e_3^2 t^9 s^{10} \Delta'
\eea
with 
\be
\begin{split}
\Delta'|_{e_4=0} &= 3 i a_0^3 (4 \sqrt{3} e_3 + 3 i t) \,,  \cr
\Delta'|_{e_3=0} &=  -3 a_0^{3/2} (3 a_0^{3/2} e_2^2 - (4 i \sqrt{3} d_1 + 4 a_0^{3/2} p)     e_2 e_4 -   4 i \sqrt{3} b_2 e_4^2)  \,, \cr
\Delta'|_{e_2=0} &= 12 i \sqrt{3} a_0^{3/2} (b_2 e_1 + e_3 ) \,, \cr
\Delta'|_{e_1=0} &= 12 i \sqrt{3} a_0^{3/2}  \,, \cr
\Delta'|_{u=0} & = 3  (4 i \sqrt{3} a_0^{3/2} + 9 e_1)   \,.
\end{split}
\ee

\begin{figure}[t!]
\centering
\includegraphics[width=9cm]{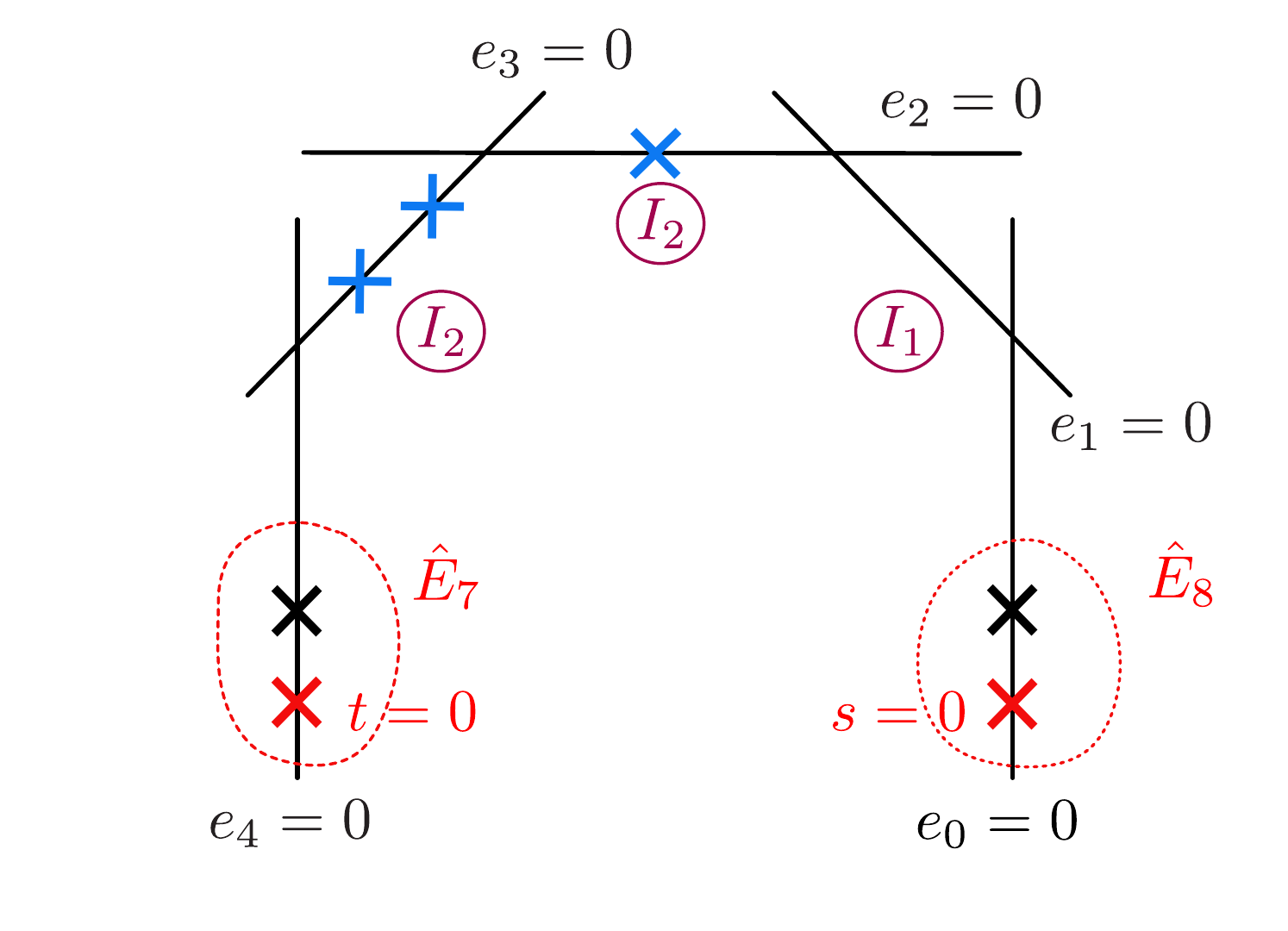}
\caption{
Kulikov Type III degeneration of $E_7 \times E_8$ Weierstrass model (\ref{equ_E7E8b1genxgen}) for $b_1=0$ and $x=0$.}
\label{fig:E7E8b10x0}
\end{figure}

The geometry is depicted  in Figure \ref{fig:E7E8b10x0}.
Consider first the surface component $Y^4$. Apart from the $E_7$ singularity at $t=0$, an additional 7-brane is located at $4 \sqrt{3} e_3 + 3 i t=0$.
The intersection curve $Y^4 \cap Y^3$ is an I$_2$ fiber, as required for consistency of the rational surface $Y^4$, which exhibits a brane content of total monodromy 
$M^{-1}_{{\rm I}_2} = M_{\hat E_7}$ away from the intersection locus.
The two additional 7-branes (which would enhance $\hat E_7$ further to $\hat E_9$) are located at two points of $B^3$, given by 
the vanishing locus of the polynomial $\Delta'|_{e_3=0}$. 

Similarly, on $Y^0$ we recognise an $E_8$ singularity at $s=0$ along with one more I$_1$ fiber at a generic position
given by $4 i \sqrt{3} a_0^{3/2} s + 9 e_1=0$.
Consistently with the total monodromy, the intersection  $Y^0 \cap Y^1$ is an I$_1$ fiber.
The 7-brane which would enhance $\hat E_8$ to $\hat E_9$ is located on $B^2$ at  $b_2 e_1 + e_3=0$.

Let us now analyse the potential $SU(2)$ enhancements from the point of view of the Weierstrass model.
An obvious possibility occurs when the two 7-branes on the base component $B^3$ coalesce, i.e. when the roots of the polynomial $\Delta'|_{e_3=0}$ coincide.
By inspection of $\Delta'|_{e_3=0}$, this happens for
\bea
3 i \sqrt{3} a_0^{3/2} b_2 - 3 d_1^2 + 2 i \sqrt{3} a_0^{3/2} d_1 p + 
 a_0^3 p^2 =0 \,.
\eea
From the previous discussion it is clear that 
the resulting $SU(2)$ factor embeds into one of the two perturbative $E_8$ factors, i.e. the $SU(2)$ enhancement is the result of a Wilson line breaking $E_8 \to E_7 \oplus SU(2)$, leading altogether to
\be
G_{\rm 9d} = E_7 \oplus E_8 \oplus SU(2) \,.
\ee
This locus must correspond to the solution $2\V =T$ (or, after exchange symmetry, $2\V =U$) on the heterotic side.
Note that the scaling of the modular forms
\be
\begin{split}
& \chi_{10} \sim \chi_{12}  \sim   u^4 \,, \qquad 
\psi_4^3-\psi_6^2 \sim   u^2 (b_2 + b_3 u + {\cal O}(u^2)) \,,  \cr
 & \chi^2_{35}/\chi_{10} \sim    u^{18} b_2^2 (3 i \sqrt{3} a_0^{3/2} b_2 - 3 d_1^2 + 2 i \sqrt{3} a_0^{3/2} d_1 p +   a_0^3 p^2 + {\cal O}(u) ) \,.
\end{split}
\ee
takes us out of the domain  of convergence (\ref{conditionforchi35}) of the expansions. This impedes a simple comparison with the heterotic dual theory.

The second possible $SU(2)$ enhancement naively occurs 
if we set $b_2=0$. On this locus, one brane on the base component $B^3$ and the brane on $B^2$ each move towards the intersection point of both  base components.
However, we must again be careful in interpreting this configuration. A base change $u \to u^2$ followed by eight blowups, similar to the one underlying the configuration (\ref{basechangeconfig}), show that for $b_2=0$ no actual gauge enhancement occurs.

The same conclusion holds also for the putative intersection of both special $SU(2)$ loci, i.e. for
\bea
b_2 = 0   \,, \qquad - 3 d_1^2 + 2 i \sqrt{3} a_0^{3/2} d_1 p + 
 a_0^3 p^2 =0 \,.
\eea
While naively one would conclude an enhancement to $E_8 \oplus E_7 \oplus SU(3)$, this is an artefact of the collision of the three branes
at the intersection point of two base components. A base change $u \to u^6$, followed by 24 blowups, reveals that for this special
choice of parameters the gauge algebra remains $G_{\rm 9d} = E_8 \oplus E_7$, with no additional enhancement.

\subsubsection*{Maximal enhancement $G_{\rm 9d} = E_7 \oplus E_8 \oplus SU(3)$}

The picture just described and the pattern of non-abelian enhancements do not change qualitatively 
by further tuning the vanishing orders of the coefficient $b$ in (\ref{hatE7tuning5}) to higher order in $u$.
This raises the question if the maximal enhancement $G_{\rm 9d} = E_7 \oplus E_8 \oplus SU(3)$ can be achieved as an infinite distance limit of F-theory.
This is possible, but in a regime of moduli space outside the realm of (\ref{restriction-mod}).
To find a parameterisation of the limit leading to $G_{\rm 9d} = E_7 \oplus E_8 \oplus SU(3)$, we start from the Weierstrass model (\ref{Weierstrass-E7E8-v1}) and first tune the coefficients such as
to realise an I$_3$ singularity at a generic point ${\cal P}$ away from $s=0$ or $t=0$. To be concrete, let us take ${\cal P} = \{  s-t=0 \}$. Then the I$_3$ singularity is obtained if we take for instance
\be
\begin{split}
&a = \left(\frac{-3}{16}\right)^{1/3}(1 - 4 u) \,, \qquad c =  \left(\frac{-3}{16}\right)^{1/3}   4 u   \,,   \cr
&d = u +   u^2  \,, \qquad b = \frac16 - u - 2  u^2 \,, \qquad e =  u^2   \,.
\end{split}
\ee
In the limit $u \to 0$ a non-minimal singularity with ${\rm ord}(f,g,\Delta)|_{s=u=0} = (4,6,13)$ arises.
As it turns out, to achieve a configuration with no further non-minimalities, one must first perform a base change $u \to u^2$, followed by 4 blowups beginning at $s=u=0$ and 2 more blowups beginning at $t=u=0$. 
This leads to a chain of 7 surface components with generic fibers of the form
\bea
{\rm I}_{0} - {\rm I}_{1} - {\rm I}_{2} - {\rm I}_{3} - {\rm I}_{4} - {\rm I}_{2} - {\rm I}_{0} \,,
\eea
where $t=0$ lies on the right end component and $s=0$ on the left. 
The ${\rm I}_{4}$-surface contains the I$_3$ singularity in the fiber over a special point of its base.
Altogether the geometry realises the enhancement $G_{\infty} = SU(3) \oplus (\hat E_7 \oplus \hat E_8)/\sim$, which is interpreted as a gauge $G_{\rm 9d} = E_7 \oplus E_8 \oplus SU(3)$ in the 9d decompactification limit.

\section{Maximal Non-abelian Gauge Algebras in 9d from Type III.a Kulikov Models}   \label{sec_Maximal9d}

As an  immediate application of the classification of Type III.a Kulikov models in \cite{Kulikov} and their physical
 interpretation given in Section~\ref{subsec_8dto9d}, we can determine the maximal non-abelian gauge algebras in the
 nine-dimensional theories that are obtained as decompactification limits at infinite distance in the complex structure moduli space. As a consistency check, we will find a perfect match with the results of \cite{Font:2020rsk,Bedroya:2021fbu} (and the previous analysis in \cite{Cachazo:2000ey}), which analyse the maximal non-abelian enhancements in heterotic string theory compactified on $S^1$.
Note that throughout this work, we are considering F-theory on K3 surfaces without any non-trivial gauge background turned on, which would reduce the rank of the theory.\footnote{Such rank-reduced setups have been considered recently from the viewpoint of the Swampland program in \cite{Kim:2019vuc,Cvetic:2020kuw,Font:2021uyw,Cvetic:2021sjm}.}
Hence the total rank in the original eight-dimensional setting is $r=18$. It decreases to $r=17$ in the nine-dimensional decompactification limit since one of the abelian gauge groups in eight dimensions corresponds to the KK $U(1)$.

\subsection{A-series and D-series}

The decompactification limits of F-theory to nine dimensions correspond to two kinds of Kulikov Type III.a degenerations:
either both end surface components  $Y^0$ and $Y^P$ are dP$_9$ surfaces with $(12-n_1)$ and $(12-n_{P-1})$ 7-branes, respectively, localised away from the intersection with the neighbouring surface components; or $Y^0$ is a dP$_9$ surface and $Y^P$ is an I$_{n>0}$ surface with 2 $D$-type singularities and possibly additonal $A$-type singularities.

In the first case, the non-abelian part of the symmetry algebra is 
\beq
G_\infty = H \oplus (\hat E_{n} \oplus \hat E_m)/\sim\,, 
\eeq
where $H$ refers to a direct sum of $A$-type Lie algebras and the ranks of the involved E-type algebras are determined as $n=9-n_1$ and $m=9-n_{P-1}$.
Indeed, we know from the discussion of the Type III.a models that the codimension-two singularities on the intermediate surface components $Y^i$ must all be of $A$ type.
To determine the maximal possible non-abelian enhancements in this class of models, we merely need to note that out of the total of 24 7-branes (or singular fibers) of the original K3 surface,
$(n+3)$ constitute the first 
$\hat E_n$ factor and $(m+3)$ the second $\hat E_m$. This identifies
\bea
G_{\infty} = A_{17-n-m} \oplus  (\hat E_{n} \oplus \hat E_m)/\sim
\eea
as the maximal non-abelian symmetry algebra of a Type III.a model, for which the remaining $18-n-m$ branes all coincide to form the $A_{17-n-m}$ configuration. Such configurations realise the maximal enhancement pattern 
\bea
G_{\rm 9d}^{\rm max}  = A_{17 - n- m} \oplus (E_n \oplus E_m)  \,, \qquad 0 \leq n,m \leq 8 \,, \qquad n \neq 2,  m\neq 2 \,   \label{maxAseries}  
\eea
in the asymptotic nine dimensional theory.
The reason why we have excluded the values $n,m =2$ is because these do not yield a {\it maximal} enhancement. Indeed,  recall from (\ref{hatEandnonhat}) that $\hat E_2$ yields a gauge algebra
$A_1 + u(1)$ in 9d, whose non-abelian component agrees with the gauge algebra associated with $\hat E_1$. Therefore the non-abelian gauge algebra of the configuration with $n=2$
can be embedded into one with $n=1$.

It remains to analyse the second kind of Type III.a models, where the end component $Y^0$ is a dP$_9$ surface with symmetry algebra $\hat E_k$ with $k=9-n_1$ and the end component $Y^P$
has generic I$_{n_P>0}$ fibers. 
The maximal non-abelian enhancements occur if both $D$-type algebras from $Y^P$ and the extra $A$-type factors from the components different from $Y^0$ combine into a single $D$-type algebra. Maximality in particular requires that one of the $D$-type configuration consists only of a pair of branes of type $B$ and $C$, in the notation of Section \ref{Sec_LoopAlgebras}, hence realising the trivial algebra $D_0$. Counting branes thus yields for the maximal non-abelian symmetry algebras:
\bea
G_{\infty} =D_{17-k} \oplus \hat E_k    \,.
\eea
This realises the second series for the maximal 9d gauge algebras that can be realized via geometry:
\bea
G_{\rm 9d}^{\rm max} = D_{17-k} \oplus E_k    \,, \qquad  0 \leq k \leq 8 \,,   k \neq 2 \,.\label{maxDseries}     
\eea
Recalling that $E_1 = A_1$, $E_2 = A_1 \oplus u(1)$, $E_3 =  A_2 \oplus A_1$, $E_4 = A_4$, $E_5 = D_5$, one can check explicitly that the 44 inequivalent configurations of the $A$-series (\ref{maxAseries}) and the $D$-series (\ref{maxDseries}) generate the complete list of maximal non-abelian gauge symmetries that can be realized by the heterotic string compactified on $S^1$ \cite{Font:2020rsk}.

We note in passing that the same conclusions can also be reached 
by taking the realisation of the affine algebras $\hat E_k$ of Section \ref{Sec_LoopAlgebras} as the starting point and explicitly constructing the possible maximal brane configurations whose total $SL(2,\mathbb Z)$ monodromy multiplies to unity. 

It is also clear that the series of maximal $A$-type enhancements
can be constructed explicitly by a generalisation of the procedure that was
exemplified for the special case $G_{\infty} = A_{2} \oplus  (\hat E_{7} \oplus \hat E_8)/\sim$
at the end of the previous section.
Instead of spelling this out for the other elements of the $A$-series, let us now sketch the realisation of the maximal $D$-type enhancements,  which engineers the second type of Tye III.a Kulikov models with only one rational elliptic end component.

\subsection{Maximal $D$-series: construction}

The second maximal non-abelian series (\ref{maxDseries}) with a $D_{17 -n} \oplus \hat E_n $ enhancement can be realised along the following lines:
\begin{enumerate}
\item
The starting point is a singularity with Weierstrass vanishing orders ${\rm ord}(f,g,\Delta)|_{t=0} =(2,3,19-n)$, realising a $D_{17-n}$ enhancement at $t=0$. 
\item
In a second step, one engineers a singularity with non-minimal vanishing orders 
\be
{\rm ord}(f,g,\Delta)|_{s=0} = (4,6,*)
\ee
in an infinite distance limit parametrised as $u \to 0$ such that at the same time 
 the fiber over a {\it generic} point degenerates to an I$_k$ singularity. 
 This requires a degeneration of the form (\ref{IIIafurthertuning}).
The infinite distance limit then realises a (local, but not global) weak coupling limit in which the 10d string coupling vanishes asymptotically at least away from a non-minimal singularity.
\item

The non-minimal singularity at $s=0$ in the limit $u \to 0$ is removed by one or several blowups, starting at $s=u=0$.

After the blowups, the point $s=0$ lies on the base of  a rational elliptic component surface. The total degree of $(f,g,\Delta)$ on this component away from the intersection
point with the other surface components is $(4,6,3+n)$. This indicates that the blowups have separated the branes associated with a $\hat E_n$ singularity in the limit $u \to 0$.
The $\hat E_n$ configuration contributes an asymptotically massless tower of BPS states which we interpret as the Kaluza-Klein tower for the partial decompactification to 9d, as before.

\item
The component containing $t=0$, on the other hand, is not a rational elliptic surface because the elliptic fiber degenerates over every point of the base component.
This behaviour is inherited from the I$_k$ degeneration over generic points of the K3 surface prior to the blowup.
For a maximal symmetry algebra, in addition to the overall degeneration, the component should exhibit
 the $(2,3,19-n)$ singularity at $t=0$ as well as one more $D$-type singularity with vanishing orders $(2,3,2)$ at a generic position away from $t=0$. 
As explained at the end of Section \ref{subsec_8dto9d}, the latter singularity, which is a $D_0$ fiber, cannot occur at finite distance for an elliptic K3, but is realisable here due to the local weak coupling nature. 
\end{enumerate}

Let us spell this out in more detail for the example of an asymptotic $G_{\infty} = D_{16} \oplus \hat E_1$ algebra, which realises the maximal non-abelian gauge algebra $G_{\rm 9d} = D_{16} \oplus  A_1 $ in 
the 9d infinite distance limit. 
Conveniently, the $D_{16}$ singularity at $t=0$ can be engineered as a global Tate model \cite{Bershadsky:1996nh},
which is a  parametrisation of the functions $f$ and $g$ of the Weierstrass model as
\bea
f = - \frac{1}{48} (\beta_2^2 - 24 \beta_4)   \,, \qquad  g=  - \frac{1}{864} (- \beta_2^3 + 36  \beta_2 \beta_4 - 216 \beta_6)  
\eea
for 
\be
\beta_2 =  a_1^2 + 4a_2 \,, \qquad   \beta_4 = a_1 a_3 + 2 a_4 \,, \qquad \beta_6 = a_3^2 + 4 a_6 \,.
\ee
Each $a_n$ is a section of the line bundle $K^{-n}$, where by $K$ we denote the canonical bundle of the base.
In the present case, this means that $a_n$ is a polynomial of degree $2n$ in the homogeneous coordinates $[s : t]$ of the base $\mathbb P^1$ of the elliptic K3, which therefore can  be parametrised as 
\bea
a_n = \sum_{i=0}^{2n}  a_{n,i}\,  t^i \, s^{2n-i}  \,.
\eea
A $D$-type singularity at $t=0$ arises if one specialises the $a_n$ such that they vanish to a certain order at $t=0$, concretely \cite{Bershadsky:1996nh}
\bea
 {\rm ord}(a_1,a_2,a_3,a_4,a_6)|_{t=0} = \begin{cases} (1,1,k+1,k+1,2k+1)  & \text{for} \,  D_{2k+2}  \\   (1,1,k,k+1,2k+1) & \text{for} \,   D_{2k+1} \end{cases} \,.
\eea
Applied to $D_{16}$, this means that we must set
\bea
a_{1,0} =0  \,, \quad a_{2,0} =0 \,, \quad  a_3 =0   \,, \quad a_{4} = a_{4,8} t^8   \,, \quad a_6 =0 \,,
\eea
which indeed leads to vanishing orders ${\rm ord}(f,g,\Delta)|_{t=0} = (2,3,18)$ as required for $D_{16}$.

In the next step we engineer a non-minimal singularity with vanishing orders \\${\rm ord}(f,g,\Delta)|_{s=0} = (4,6,*)$ in an infinite distance limit of the form (\ref{IIIafurthertuning}).
To this end we first parametrise the remaining Tate monomials as
\bea
a_{i,j} = \sum_{k\geq 0} a^{(k)}_{i,j}  \, u^k    
\eea
and tune 
\bea
a^{(0)}_{2,3} = - \frac{1}{2} a^{(0)}_{1,1} \, a^{(0)}_{1,2} \,, \qquad a^{(0)}_{2,4} = - \frac{1}{4} (a^{(0)}_{1,2})^2 \,, \qquad a^{(0)}_{4,8} =0  \,.
\eea
With this prescription,
\bea
f|_{ u=0} = - \frac{1}{48} t^2 s^4 p_1(s,t)^2 \,,  \qquad g|_{u=0} =  \frac{1}{864} t^3 s^6 p_1(s,t)^3   
\eea
for a linear polynomial $p_1(s,t) = 4 a_{2,1} s + ((a^{(0)}_{1,1})^2    + 4 a^{(0)}_{2,2}) t$, while the discriminant takes the form
\bea
\Delta = -  \frac{1}{16} (a^{(1)}_{4,8})^2 \,  t^{18} \, u^2  \, \tilde \Delta' \,.
\eea
Note that 
in the limit $u  \to 0$, the fiber over a generic point on the base degenerates into an I$_2$ fiber, as intended.

\begin{figure}[t!]
\centering
\includegraphics[width=7cm]{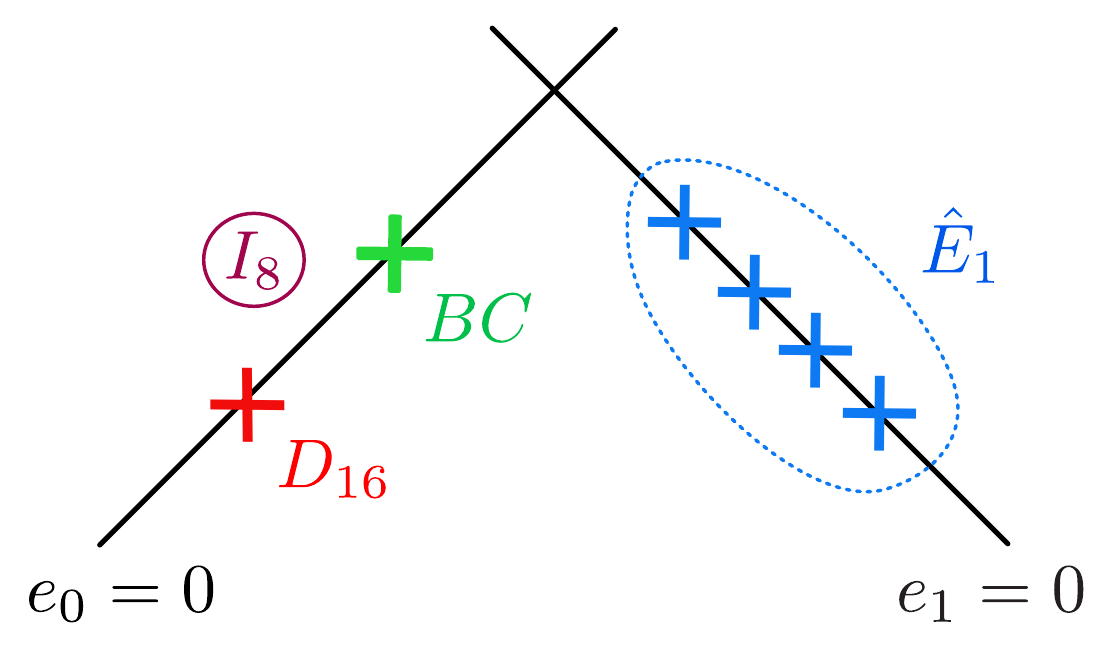} 
\caption{
Kulikov Type III Weierstrass model with $D_{16} \times \hat E_1$ enhancement. \label{Fig:D16hatE1}}
\end{figure}

The  non-minimal singularity at $s=0$ in the limit $u \to 0$ can be removed by a blowup.
The blowup turns out to be simplified if one performs a base change 
\be
u \to  u^4  \,,
\ee
after which 
\bea
\Delta = -  \frac{1}{16} {(a^{(1)}_{4,8})^2} \,  t^{18} \,  u^8 \,  \Delta' \,.
\eea
For this degeneration, the simple blowup
\bea
(s, u ) \to (s \, e_1, u \,  e_1) 
\eea
gives rise to a Type III.a Kulikov model consisting of two intersecting components. See Figure \ref{Fig:D16hatE1} for an illustration; as before, we replace the coordinate symbol $u$ by $e_0$ after the blowup. 
Over the component $e_0=0$, the elliptic fiber degenerates as indicated by the power of $e_0^8$ in the discriminant.
The fibration over the component $e_1=0$, on the other hand, is a rational elliptic surface with 12 singular fibers. Of these, $8$ singular fibers are accounted for by the intersection
with the component $e_0=0$ carrying the I$_8$ singularity. The remaining 4 singular fibers are located away from intersection point $e_1 = e_0=0$.
Only the latter correspond to physical 7-branes located at points of the base.
More precisely, over $e_1=0$ we can set $t=1$ and obtain 
\bea
f|_{e_1 =0} = p_4(s,e_0) \,, \qquad 
g|_{e_1 =0} = s^2 p_2(s,e_0) \,, \qquad 
\Delta|_{e_1 =0} = q_4(s,e_0) \,,
\eea
for some polynomials of the indicated degrees. According to our general prescription,
the rational elliptic surface contributes a $\hat E_1$ factor to the asymptotic symmetry algebra. This corresponds to a non-abelian gauge algebra component $A_1$ in the 9d theory at infinite distance.\footnote{As discussed after~\eqref{hatTildeENenh1}, the $4$ singular fibers may in principle lead to an extensioin to $\hat{\tilde{E}}_1$. We leave it to future work to clarify this ambiguity.}

Over the surface patch with base $e_0=0$ we have, on the other hand, 
\bea
f|_{e_0=0} = t^2 \, p^2_1(e_1,t) \,,  \qquad 
g|_{e_0=0} = t^3 \, p^3_1(e_1,t)  \,, \qquad 
\Delta'|_{e_0=0} =  p^2_1(e_1,t) \,,
\eea
where $\Delta'$ is the residual discriminant after factoring out $ - \frac{1}{16} (a^{(1)}_{4,8})^2 \,  t^{18}$.
Apart from the $(2,3,18)$ singularity at $t=0$ we observe a non-minimal singularity with vanishing orders $(2,3,2)$ at $p_1(e_1,t) =0$.
Altogether, this component carries 20 of the original 24 7-branes on K3, complementing the 4 7-branes which form the $\hat E_1$ singularity on the dP$_9$ surface over $e_1=0$.

This procedure can be generalised to construct also the remaining maximal enhancements in the $D$-series, albeit not all of them as degenerations of global Tate models.

 \section{Emergence of Quartic Gauge Couplings?} \label{emergence}
 
We now change tack and comment on the large distance asymptotics 
 of the quartic gauge couplings in eight dimensions, by mostly making use of known results for the dual heterotic string.
These couplings are the analogs of the well-investigated
quadratic gauge couplings in $N=2$ supersymmetric string theories in four dimensions
(see, for example, the early works \cite{Dixon:1990pc,Harvey:1995fq}). Indeed, 
in eight dimensions it is not the gauge kinetic terms, but rather the quartic couplings that exhibit logarithmic one-loop corrections: the quadratic gauge kinetic terms do not receive any quantum corrections and become irrelevant in the UV.  Distinguishing quadratic and quartic terms in eight dimensions may thus shed
some light on the notion of emergence, from the point of view that not only the quadratic terms but  all couplings might be emergent in string theory. 

At one loop order, the relevant, bosonic structure of the quartic terms takes the schematic form
\be\label{S4}
\cS_{(4)}^{{\rm 8d, 1-loop}} \ \sim\  \int d^8x\sqrt{-G^{(8)}}\left(\frac1{g(t)^2}\, [F^4] + \theta(t) [F\wedge F\wedge 
 F\wedge  F]\right)\,,
 \ee
where $[\ast]$ summarily denotes the possible gauge and Lorentz contractions pertaining to a variety of superinvariants.
The moduli dependent quantum couplings $g(t)$, $\theta(t)$ are known to be BPS-saturated
and one-loop exact in the dual frame of the heterotic string. As such they
can be computed \cite{Bachas:1997mc,Kiritsis:1997hf,Lerche:1998nx,Lerche:1998gz,Foerger:1998kw} by generalizing the methods originally developed in \cite{Dixon:1990pc,Harvey:1995fq}  for
four-dimensional theories, and the notions of Borcherds lifts and Kac-Moody
algebras apply here as well. Computations in F-theory geometry were presented in \cite{Lerche:1998nx,Lerche:1998gz,Lerche:1999hg}, and for Type~I strings e.g.~in \cite{Bachas:1996bp,Bachas:1997mc,Kiritsis:1997hf,Foerger:1998kw,Billo:2009di,Fucito:2009rs}.

We will consider here only abelian gauge fields, namely those which arise from the Kaluza-Klein compactification and which are the superpartners of the moduli $t$. 
They have quite different properties as compared to the gauge fields that
are already present in ten dimensions. As we will see, this is true in particular from the point of view of emergence, which is why we will focus on them.

We will consider the parity-even couplings that arise from KK reduction of the heterotic string action in ten dimensions. The part relevant for us reads schematically  \cite{Gross:1986mw,deRoo:1992zp}
\bea
\cS^{{\rm 10d}}\ \sim&& \!\!\!\!\!
\int d^{10}x\,\sqrt{-G^{(10)}}\left[e^{-2\phi^{(10)}}\left( R+H^2+R^2  +R^4\dots\right)_{{\rm tree}}\right.
\\
&&\qquad\qquad\qquad+\left.\left( R^4+B\wedge R\wedge R\wedge R\wedge R+\dots\right)_{{\rm1-loop}}+\dots\right].\nn
\eea
Here $H=dB+\Omega_L+\Omega_F$ is the Chern-Simons corrected three-form field strength, where
$\Omega_L=\omega\wedge d\omega-\frac23\omega\wedge \omega\wedge \omega$, and $\Omega_F$ is the Chern-Simons form for the 10d gauge fields, which we are not interested in here. The $H^2$ term is related by supersymmetry to the $R^2$ term \cite{Romans:1985xd,Bergshoeff:1988nn}. 
Via field redefinitons the latter can be completed into Gauss-Bonnet form
\be\label{GBterm}
\cS_{GB}^{{\rm 10d}}\ \sim\  \int d^{10}x\,\sqrt{-G^{(10)}}e^{-2\phi^{(10)}}\left( R_{mnrs}R^{mnrs} 
-4  R_{mn} R^{mn}+R^2\right),
\ee
which avoids higher derivative contributions to the quadratic kinetic terms that otherwise would lead to unitarity violating poles in the graviton propagator \cite{Zwiebach:1985uq}.

At tree-level, the quadratic and quartic gauge couplings arise via KK reduction from terms linear and quadratic in the Riemann tensor. 
To illustrate our points and keep the discussion simple, we restrict ourselves to the sub-sector of abelian gauge fields  for the eight-dimensional theory with generic $E_8\times E_8\times U(1)^2$  gauge symmetry; including Wilson lines does not add anything substantial so we leave them out for now.
The corresponding scalar moduli are thus $t\in\{T,U\}$,  in the usual notation.

Upon performing standard KK reduction and choosing a suitable choice of basis for the gauge fields, one finds from the Ricci scalar and the Gauss-Bonnet term
(\ref{GBterm})  the following quadratic and quartic pieces of the tree-level action for the $U(1)_T\times U(1)_U$ gauge fields:
\bea\label{F4action}
\cS_{(2)}^{{\rm 8d,class}}& \sim& \int d^8x \sqrt{-G^{(8)}}\frac1{{g}^2}{M_s}^4
 F_{\mu\nu}^TF^{U,\mu\nu}
\,,\nn
 \\
\cS_{(4)}^{{\rm8d, class}}& \sim& \int d^8x \sqrt{-G^{(8)}}\frac1{{g}^2}\left( 
-\frac{1}{32} t_8^{\mu_1\mu_2\mu_3\mu_4\mu_5\mu_6\mu_7\mu_8} F_{\mu_1\mu_2}^TF_{\mu_3\mu_4}^TF_{\mu_5\mu_6}^UF_{\mu_7\mu_8}^U
 \right)\,.
\eea
Here $t_8$ is the familiar kinematical tensor in eight dimensions \cite{Green:1987mn}, and
\be
\label{8ddil}
\frac1{{g}^2} \ =\  \frac1{g_s^2}\,, \ \ \ \ {\rm with}\ \ \ \
\frac1{g_s^2}\ \equiv\ e^{-2\phi^{(8)}} =\ \left(\frac{M_p}{M_s}\right)^6=\ \imt T\,e^{-2\phi^{(10)}}\,,
\ee
is the eight-dimensional string coupling. It coincides with the quartic gauge coupling at tree level.
Above, we exhibited that the kinetic term has a dimensionful coupling and thus vanishes in the UV. From supersymmetry there are no loop corrections to this term.

There are also corresponding quadratic and quartic derivative terms in the moduli that we do not
consider here.
The quadratic term figures, as usual, as the metric on moduli space, and the familiar statements
about distances in moduli space apply \cite{Ooguri:2006in,Palti:2019pca}.

A major difference as compared to $N=2$ supersymmetric theories in four dimensions lies in the structure of the parity-odd terms. From the $H^2$ term one obtains after reduction $dB\cdot (F_T\wedge A_U)+...$,
which dualizes in eight dimensions to $A_4\wedge F_T\wedge F_U+\dots$, where $A_4$ is the four-form dual of the $B$-field. 

Thus there is no dynamical axion field $\theta$ at tree-level, and this is as expected as the dilaton sits in the gravity multiplet in eight dimensions \cite{Awada:1985ag}. This is in contrast to four dimensions, where the dilaton belongs, together with the $B$-field,  to a linear multiplet, which can be dualized to a vector multiplet that hosts both the dilaton and the axion fields. In $N=1$ language this can be represented by the familiar
complexified coupling
\be
\label{Sdef}
S\ =\ \frac 1{8\pi^2}\theta+ i\frac {1}{g_s^2}\ \,.
\ee
By contrast, in eight dimensions we do not have such a complexified scalar coupling at tree level to begin with. This is in line with the fact that the corresponding coupling constant on the F-theory side is a real parameter, namely the K\"ahler parameter 
$t_b=vol(\IP^1)$ of the base of the elliptic K3 \cite{Morrison:1996pp}.

Nevertheless, at the one-loop level it is known that parity-even and parity-odd $F^4$ terms are
related by supersymmetry \cite{Lerche:1988zy}. Thus suggests packing the effective, field dependent couplings
in (\ref{S4}) together into one complexified parameter
\be
\tau^{{\rm 1-loop}}(T,U)\ =\ \frac 1{8\pi^2}{\theta(T,U)}+ i\frac {1}{{g(T,U)}^2}\,.
\ee
This fits together with the fact that the one-loop quartic gauge couplings are given by covariant derivatives of a holomorphic prepotential, $\cF(t)$ \cite{Lerche:1998nx,Foerger:1998kw}. 
That is, their harmonic part obeys
\be\label{Fijkl}
\tau_{ijkl}^{{\rm 1-loop}}(t){\big|_{harm.}}\ \sim\  \cF_{ijkl}(t)\ \equiv\   \partial_{t_i}\partial_{t_j}\partial_{t_k}\partial_{t_l}\cF(t)\,, \qquad  t_*\in\{T,U\}\,.
\ee
Further, non-harmonic terms arise from the covariant derivatives which reflect massless states running in the loop. 

However, given our lack of knowledge of a possible superspace formulation of the eight-dimensional couplings, it is unclear to us to what extent $\cF(t)$ makes literal
 sense as a superspace Lagrangian that would
reproduce the one-loop couplings in eq.~(\ref{S4}). Therefore we view it simply as a formal generating function of the parity even, one-loop, four point correlation functions for the time being.

The prepotential takes the familiar form of a ``polylogarithmic lift'', and for the geometry under consideration
and positive Weyl chamber given by Im$\T>$Im$U$, it reads \cite{Lerche:1998nx,Foerger:1998kw}
\be\label{FTU}
\cF(\T,\U) = -\frac 1{240\pi}U^5+Q(\T,\U)-\frac{ic(0)\zeta(5)}{128\pi^6}-\frac i{(2\pi)^6}
\sum_{(n,m)>0}c(nm) \,{\mathcal L}i_5({\qt}^n{\qu}^m)\,,
\ee
where $q_T=e^{2\pi i T}$ and $q_U=e^{2\pi i U}$. The coefficients are determined by
the elliptic genus of the 10d heterotic string, which is a
weight $k=-4$, meromorphic modular form:
\be\label{refellgen}
\phi_{-4}(q)\ =\  \frac1{\eta^{24}(q)}{E_4(q)^2}
\ =:\  \sum_{n}  c(n)q^n\,.
\ee
Moreover, $Q(\T,\U)$ denotes certain quartic terms which depend on the particular Weyl chamber and whose precise form does not matter here.

In terms of the prepotential $\cF(T,U)$, the one-loop threshold corrections to the quartic couplings read  \cite{Lerche:1998nx,Foerger:1998kw}:\footnote{In the string basis. We expect from analogous computations in four dimensions \cite{Forger:1997tu} that our conclusions will not change if we transformed to the supergravity basis.}
\be\label{quartictau}
\imt \tau^{{\rm 1-loop}}_{T^{4-s}U^s}(\T,\U)\ =\ 
-\frac i4{D_T}^{4-s}{D_U}^s\cF(T,U)
+ 
\frac i4\left( {\imt U\over \imt T}\right)^{4-2s}\!\!{\bar D_{\bar T}}^{s}{{\bar D_{\bar U}}}^{4-s}\bar\cF(\bar T,\bar U)\,,
\ee
for $s=0,\dots4$. Here 
\be
\label{covder}
 D_{j}\ =\ \left(\partial_{t_j}+\frac {k_j}{2i\imt{t_j}}\right)
 \ee
 is the Serre derivative that maps modular forms of weight $k_j$ to modular forms of weight $k_j+2$.
 
Let us now connect to the  theme of the present work, where
we focus on the infinite distance degenerations corresponding to Kulikov limits of 
\bea
&&\text{Type II.a}:\, \,  \, {\rm Im}T\rightarrow \infty, \, {\rm Im}U\ {\rm finite}\,,
\\
&&\text{Type III.a}:\ {\rm Im}T,{\rm Im}U\rightarrow \infty, \,{\rm Im}T > {\rm Im}U 
 .\nn
\eea
For these limits one might expect the corrections (\ref{quartictau}) to display a logarithmic dependence in $q_T$, corresponding to linear behavior in $T$ (and similarly for $q_ U$ in the Type III limit, which we will not consider in the following). This ought to reflect the  exponential growth of integrated-out KK states associated with decompactification to ten dimensions.

Clearly, since  from (\ref{8ddil}) we have $1/g^2\sim \imt T$, such a behavior occurs already at the classical level, purely from geometry.
In contrast, upon inspection one finds that the one-loop thresholds (\ref{quartictau}) derived from the prepotential (\ref{FTU}) do not show such a divergence.

We emphasize that this applies to the specific abelian gauge symmetries we consider, which arise from KK reduction, and not to non-abelian gauge symmetries that appear already in ten dimensions; for these, the one-loop quartic thresholds generically do blow up linearly in the decompactification limit \cite{Bachas:1997mc,Kiritsis:1997hf}.

This special feature of KK-related gauge symmetries sharpens the general question in what sense  couplings can be thought to emerge from integrating out a tower of massive states at the loop level. It is a well-known fact that in general there is no absolute distinction between which terms may be called classical and which quantum, as these notions get mixed up by dualities. The  same issue arises for the analogous gauge couplings in $N=2$ supersymmetric strings in four dimensions, where one can map from the heterotic to a dual Type IIA string frame. In this duality frame both the classical and one-loop couplings of the heterotic string combine into one classical, genus zero preprotential of the Type IIA string: $\cF_{II}(S,T,U) = \cF_{{\rm het}}^{{\rm class}}(S,T,U)+\cF_{{\rm het}}^{{1-loop}}(T,U)$.
In this way, the conifold singularity, which arises at the classical level on the Type II side at finite distance in the moduli space, reproduces the one-loop correction on the heterotic side  \cite{Strominger:1995cz}.  On the other hand, for the infinite distance limit of the particular correlators we consider, the situation appears different (in both four\footnote{See for example the familiar $STU$-model in four dimensions \cite{Harvey:1995fq,Louis:1996mt,Forger:1997tu}.}and eight dimensions) in that the singularity arises also on the heterotic side only at tree level. The situation being similar also for Type I strings, it thus seems that there is no perturbative duality frame in which the infinite distance singularity arises at the quantum level.

Concentrating on eight dimensions, we
 will now try to refine the picture to reconcile it, at least partially, with the notion
of integrating out massive states at the quantum level.
To this end we consider the one-loop correction to the 
quartic gauge coupling $1/{g}^2$, which is governed by the imaginary part of
\be
\cF_{TTUU}\ =\  {\partial_T}^2{\partial_U}^2\cF(T,U)\,.
\ee
This coupling has been already evaluated in \cite{Lerche:1998nx,Foerger:1998kw},
and knowing the result we can concisely rewrite it to fit our purposes.
For this we push the quartic derivatives
acting on $\cF$ to a double derivative acting on the elliptic genus,
\be
\tilde\phi_0(q)\ =\ (q\partial_q)^2 \ \phi_{-4}(q)
\ =:\ \sum_{n>-1}\tilde c(n)q^n\,,
\ee
where $q\partial_q\equiv\frac1{2\pi i}\partial_\tau$.
Owing to $(p\partial_p)^5{\mathcal L}i_5(p)=-\log(1-p)$,  
this indeed maps back to the four-point function 
\be
\cF_{TTUU}(T,U)\ =\ \frac i{4 \pi^2} \log\,\Psi_{TTUU}\,,
\ee 
via an exponential Borcherds type of lift (\ref{Explift}) given by:
\be
\Psi_{TTUU}\ =\ 
\text{Exp-lift}[\tilde\phi_{0}]=
\prod_{{n,m\in \mathbb Z} \atop (n,m)>0}\left(1-\qt^{n}\qu^{m}\right)^{\tilde c(nm)}\,.
\ee
The lift subsumes the complexities that arise at intermediate steps when performing the one-loop
modular integral.
Note that due to the derivatives,
$\tilde\phi_0(q)$ is not a modular, but rather a quasi-modular form
involving the Eisenstein series $E_2(q)=q \partial_{q}\log \eta^{24}(q)$.
This is signified by the tilde. Accordingly one can split
$\tilde\phi_0(q)$ into a fully modular and a quasi-modular part,
\bea\label{phisplit}
\tilde\phi_0(q)&=&   \phi_0^M(q)+\phi_0^{QM}(q)\,,
\\
 \phi_0^M(q) &=&  \frac1{72\,\eta^{24}(q)}\left(41 E_4^3+31  E_6^2 \right)(q) =
 J(q)- 744 
 \ =:\ \sum_{n\geq-1}\tilde c^M(n)q^n\,,
 \nn
 \\
 \phi_0^{QM} (q) &=& -\frac1{24\,\eta^{24}(q)}
 \left(5 (E_4^3+ E_6^2)-8 E_2 E_4 E_6-2 {E_2}^2 {E_4}^2 \right)(q)
 \ =:\ \sum_{n\geq1}\tilde c^{\,QM}(n)q^n\,,
 \nn
 \eea
and define the respective lifts
\bea\label{PhiTTUU}
\Psi_{TTUU}^{M}& =& {q_T}^{-1}
\prod_{{n,m\in \mathbb Z} \atop (n,m)>0}\left(1-\qt^{n}\qu^{m}\right)^{\tilde c^M(nm)}\,,
\\
\Psi_{TTUU}^{QM}& =& {q_T}
\prod_{{n,m\in \mathbb Z} \atop (n,m)>0}\left(1-\qt^{n}\qu^{m}\right)^{\tilde c^{\,QM}(nm)}\,,
\nn
\eea
such that
\be
\cF_{TTUU}(T,U)\ =\ \frac i{4 \pi^2}\log[J(T)-J(U)] + \frac i{4 \pi^2}  \sigma(T,U)\,,
\ee
with 
\bea
 \log[J(T)-J(U)]   &\equiv& \log\,\Psi_{TTUU}^M\,,
 \\
 \sigma(T,U)  &:=& \log\,\Psi_{TTUU}^{QM}\,.\nn
\eea
The first line reflects a classic identity by Borcherds \cite{BORCHERDS199030} and the second is a definition.
At this point the question arises about the significance of the particular split (\ref{phisplit}), since a priori there is an ambiguity
in adding and subtracting a weight zero modular form between both terms. 
We pushed the $1/q$ pole completely into $\phi_0^M(q)$ so that $\sigma(T,U)$ is a cusp form without singularity in the interior of the moduli space. There is an additional ambiguity in trading a constant between 
$\phi_0^M(q)$ and $\phi_0^{QM}(q)$. This translates to an ambiguity
of adding and subtracting a term proportional to $\log \eta(T)\eta(U)$ between $\log\,\Psi_{TTUU}^M$ and $\log\,\Psi_{TTUU}^{QM}$.  On top of that there is the freedom of trading a factor between
$\Psi_{TTUU}^M$ in $\Psi_{TTUU}^{QM}$, such as shown in eq.~(\ref{PhiTTUU}).

We resolved these ambiguities via imposing modular invariance; that is, by requiring that both $\Psi_{TTUU}^{M}$ and $\Psi_{TTUU}^{QM}$ individually transform 
with bi-modular weight $(0,0)$ under modular transformations in $SL(2,\IZ)_T\times SL(2,\IZ)_U$. While this may seem ad hoc, it is guided by the physical intuition that splitting up $\Psi$ makes best sense if
both components respect the underlying modular gauge symmetries individually. 

Putting everything together,  the complete one-loop contribution to the quartic coupling (\ref{quartictau}) can thus be rewritten as
\be\label{gaugecorr}
\imt \tau^{{\rm 1-loop}}_{TTUU}(T,U)\ =\ -\frac i4{D_T}^2{D_U}^2 \cF +{\rm cc.}\ =\ 
\frac1{16\pi^2}\log[J(T)-J(U)]+\frac1{16\pi^2} \sigma(T,U)-\frac i2G(T,U)+{\rm cc.}\,,
\ee
whose non-harmonic part,
\bea
G(T,U)& =&  \frac{24}{ (\imt T\, \imt U)^2}
\Big(3  -3 i \left(\imt T\, \partial_T+\imt U\,\partial_U\right)
-(\imt T)^2 {\partial_T}^2 -(\imt U)^2 {\partial_U}^2
\\
&& 
-3 \imt T\, \imt U\,\partial_T\partial_U+i \,\imt T\, \imt U \left(\imt T\, {\partial_T}^2{\partial_U}+\imt U\, {\partial_T}{\partial_U}^2\right) 
  \Big)\cF(T,U)\,,
  \nn
\eea
arises from the connection pieces of the
covariant derivatives (\ref{covder}). By construction it restores the modular 
 $SL(2,\IZ)_T\times SL(2,\IZ)_U$ symmetry that is broken by 
$\sigma(T,U)$ as a consequence of the quasi-modular components of $\phi_0^{QM} (q)$. 

It is easy to verify that $\imt \tau^{{\rm 1-loop}}_{TTUU}(T,U)$ does indeed not diverge for $\imt T\rightarrow \infty$, and thus it
seems, at first sight, that the tower of asymptotically massless KK states has not been integrated out.

In order to understand this apparent clash with expectations, recall that
the decomposition in (\ref{gaugecorr})  is familiar from four dimensions \cite{deWit:1995dmj,Kiritsis:1996dn}. There the actual string coupling has an extra one-loop "Green-Schwarz" correction given by (the 4d analog of) $-\imt G(T,U)$, which cancels against the non-harmonic connection term $+\imt G(T,U)$ in the one-loop quadratic gauge coupling. It arises from the duality transformation from the linear to a vector supermultiplet and  compensates the anomalous transformation properties of the complex axio-dilaton field, $S$. 
If we absorb the last two terms of the decomposition (\ref{gaugecorr}) into the axio-dilaton $S$, we thus obtain a holomorphic, modular invariant version of it: $S^{\rm inv}=S+\frac i{8\pi^2}\sigma(T,U)$ \cite{deWit:1995dmj,Kiritsis:1996dn}.

In eight dimensions, the situation is different because the dilaton as a member of the gravity multiplet is a real field, and there is no duality transformation extending it to a complex field. Nevertheless the last two terms of (\ref{gaugecorr}) are universal for one-loop amplitudes and so can be similarly absorbed into a  modular invariant (though this time non-harmonic) string coupling:
\bea
\label{geff}
 \frac 1{g_{{\rm ren}}^2(T,U)}
&:=& \frac 1{g_s^2}\ +\ \Delta(T,U)\,,
\\
\nn
\Delta(T,U) &\equiv&\frac 1{8\pi^2}  \ret \sigma(T,U)+ \imt G(T,U)\,.
\eea
This renormalized coupling has no singularity in the interior of the moduli space and so makes good sense as a physical coupling;  by construction,
$\sigma$ is a cusp form, and the derivatives in $G$ acting on $\cF$ yield higher polylogs
and so do not produce singularities either.

In terms of this renormalized string coupling, the total tree-level plus one-loop quartic gauge coupling then takes the form
\be\label{oneloopthresh}
\imt \tau_{TTUU}(T,U)\ \equiv\ \frac 1{g^2(T,U)}\ = \
 \frac 1{g_{{\rm ren}}^2(T,U)}+\,\ \frac1{16\pi^2}\log|J(T)-J(U)|^2\,.\ee
Its threshold piece captures not only the finite distance singularity at $T=U$ 
(plus its modular images),  which arises from charged BPS states becoming massless there, 
but now also the infinite distance singularity 
\be
\label{LogJdiv}
\frac1{16\pi^2} \log|J(T)-J(U)|^2 \ \longrightarrow\ \frac1{4\pi} \imt\, T\,,\qquad  {\rm as \ }T\rightarrow i\infty\,,
\ee
which does reflect an integrating out of asymptotically massless KK states. This is because it originates from 
the modular $J$-functions and not from geometric zero mode terms.
This resonates well with the notion of emergence.  

We propose that $g_{{\rm ren}}(T,U)$ in (\ref{geff}) is the physically relevant coupling to consider in this context,
even though it is not the standard loop counting parameter in the heterotic string frame. This is in line with the considerations
of ref.~\cite{Kaplunovsky:1995jw,deWit:1995dmj,Kiritsis:1996dn} which apply to four-dimensional theories.
From this point of view, the renormalized coupling defines, via (\ref{8ddil}), a field-dependent, modular invariant
string scale $M^{{\rm ren}}_s(T,U)$ as follows:
\be
\left(\frac{M_p}{M_s^{{\rm ren}}(T,U)}\right)^6\ =\ \left(\frac{M_p}{M_s}\right)^6+\Delta(T,U)\,.
\ee

This is only a mild modification of $M_s$, despite the fact that  hidden in $\Delta(T,U)$
is the infinite distance singularity of $\sigma(T,U)$. The point is that the latter is subdominant to the tree-level singularity.
More precisely, the singularity of $\sigma(T,U)$ is the negative of the singularity of the quartic gauge threshold correction
(\ref{LogJdiv}); both stem from the relative factor $q_T$ in the split (\ref{PhiTTUU}) which was necessary for the individual modular invariance of the two pieces. Requiring positivity of the right hand side of eq.~(\ref{geff}) in the decompactification limit thus imposes the following constraint
(up to order one coefficients):
\be\label{posconstr}
\left( \frac 1{g_{{\rm ren}}^2(T,U)}\right)_{T\rightarrow i\infty}\ \sim\   \imt T\left( e^{-2\phi^{(10)}}-1\right) > 0\,,
\ee
which puts the theory to the weak coupling regime at all energies and so is self-consistent within perturbation theory. This is analogous to the well-known constraint in four dimensions \cite{Kaplunovsky:1995jw} which posits that the tree-level term should always be dominant.
It is consistent with the positivity of the K\"ahler parameter $t_b=vol(\IP^1)$ on the F-theory side upon identifying
$t_b\sim 1/g_s^2-\imt T>0$, which is analogous to the map between type II and heterotic parameters, $T_{{b,\rm II}}=S_{{\rm het}}-T_{\rm het}$ in four dimensions \cite{Louis:1996mt,LopesCardoso:1996nc}. It is also consistent with the choice of positive Weyl chamber, ie., the quintic term, for the prepotential as in (\ref{FTU}), and is fully in line with analogous arguments for the gauge kinetic terms in four dimensions.

The upshot is that (\ref{posconstr}) shows 
that the classical contribution necessarily dominates over the one-loop correction in the decompactification limit $T\rightarrow i\infty$, even after having renormalized the string coupling such as to have a divergent one-loop contribution in the first place.
Thus the putative emergence of the quartic gauge coupling is at best partial.
This obstacle against emergence in a strong sense, namely that
couplings emerge entirely from integrating out states at the quantum level, has been recognized and addressed  before eg. in \cite{Heidenreich:2017sim,Heidenreich:2018kpg,Corvilain:2018lgw,Palti:2019pca}.

 \section{Discussion} \label{disc}

In this article we have interpreted the infinite distance limits in the complex structure moduli space of elliptic K3 surfaces
from the point of view of F-theory in eight dimensions. The geometric classification of such limits in \cite{Kulikov} is in a precise one-to-one correspondence with the possible weak coupling or (partial) decompactification limits, as summarised in Figure \ref{fig:Fmodspace}.
In the decompactification limits, the light towers of KK states are encoded geometrically in M2-branes wrapping vanishing transcendental elliptic curves in M-theory, or equivalently in string junctions 
signalling an enhancement of the symmetry algebra of the theory to a Kac-Moody or loop algebra. Weak coupling limits, on the other hand, exhibit additional towers of states, 
be it excitations of a light emergent string or proper supergravity KK modes.

To corroborate the interpretation of loop algebras as the hallmark of a dual decompactification, we  have translated a representative class of models exhibiting such enhancements into the language of the dual heterotic string, where the mirror map allows for a clear comparison with the heterotic variables. 
The upshot of this analysis is that while at first sight the complex degeneration of the compactification space at infinite distance might obscure the physics encountered on the boundary of moduli space,
one can give, on the contrary, a very clear interpretation in full agreement with the predictions of the Emergent String Conjecture \cite{Lee:2019wij}.

The simplicity of the asymptotic theory in the dual heterotic frame, given by a compactification on a torus, invites a quantitative analysis of the Emergence Proposal \cite{Harlow:2015lma,Heidenreich:2017sim,Heidenreich:2018kpg,Grimm:2018ohb,Corvilain:2018lgw,Palti:2019pca}, applied to the quartic, BPS saturated couplings of the KK $U(1)$s of the eight-dimensional theory.
Building on the explicit computation of the 1-loop corrected quartic couplings in eight dimensions, we have encountered a certain tension, described generally already in\cite{Heidenreich:2017sim,Heidenreich:2018kpg,Corvilain:2018lgw,Palti:2019pca}, with the idea that all couplings are generated by
the integration of massless towers at infinite distance. Contrary to naive expectations, one finds a perfect cancellation between the divergent contributions to the 1-loop KK threshold corrections at infinite distance.
However, by interpreting a subsector of these corrections as contributions to the modular invariant dilaton, one can argue that at least part of the divergent contributions to the KK couplings are obtained at the 1-loop level, albeit only the subleading ones. 

This observation applies also to four dimensions, despite important differences of the analysis including the fact that the relevant couplings in eight dimensions are the quartic, rather than quadratic, gauge couplings.
It would be interesting to see to what extent the findings of \cite{Heidenreich:2021yda} about KK compactifications
and axionic strings in four dimensions carry over to eight dimensions, 
in relation to the parity odd $F^4$ term in (\ref{S4})  (which may couple to ``exotic'' instantons  \cite{Billo:2009gc}).
Its axion coupling $\theta(t)$ formally gauges a $(-1)$-form Chern-Weil symmetry, but as we described, it is not
a fundamental field in the eight-dimensional heterotic string. This seems consistent with \cite{Heidenreich:2021yda} who argue
that in such a case the massless tower is given by KK states rather than by an emergent string.

\subsection*{Acknowledgements}

We thank Markus Dierigl, Hee-Cheol Kim, Stephan Stieberger, Eran Palti and Irene Valenzuela for helpful discussions. The work of S.-J.L. is supported by IBS under the project code, IBS-R018-D1. W.L. thanks the CERN Theory Department for support.
The work of T.W. is supported in part by Deutsche Forschungsgemeinschaft under Germany's Excellence Strategy EXC 2121 Quantum Universe 390833306.

\appendix

 \section{Siegel Modular Forms: Definitions and Properties} \label{app_seigel}

There is much literature on Jacobi and Siegel modular forms, see for example \cite{EichlerZagier,Gritsenko:1996ax,Gritsenko:1996kt,Gritsenko:1996tm} for a selection of mathematical literature and \cite{Mayr:1995rx,LopesCardoso:1996nc,Dabholkar:2012nd,Malmendier:2014uka,Gu:2014ova} for physical applications related to our work.
 Thus we will be very brief and refer to those papers for details;
our main purpose here is to present a collection of formulas and explicit asymptotic expansions that are useful for our purposes.

Siegel modular forms are special examples of more general automorphic forms, 
and have the property that they describe the moduli spaces of genus $g$ Riemann
surfaces. Genus two Siegel forms happen to underlie the 
the breaking of $E_8\times E_8$ to $E_7\times E_8$, which in heterotic language amounts to switching on
a particular Wilson line; a fact that had been observed in physics first in ref.~\cite{Mayr:1995rx}. 
As we put particular emphasis on this geometry in the main text, we concentrate here on $g=2$ Siegel forms,
with the understanding that most of the structure carries over to more general Wilson line configurations.

Genus two Siegel forms of weight $k$, denoted generically by  $\Phi_k(\Omega)$, 
are holomorphic functions on the Siegel upper half-plane, which is isomorphic to the homogeneous domain,
\be
\mathbb H_2 \ \simeq\ \Big\{\Omega=
 \begin{pmatrix} 
     \T & \V \\
    \V & \U \\
   \end{pmatrix}
   \in {\rm Mat}_2(\mathbb C)\,\Big\vert\, {\rm det\,Im}\Omega>0, {\rm Tr\,Im}\Omega>0
   \Big\}\,,
\ee
equipped with the automorphy property $\Phi_k(\Omega)={\rm det}(C\Omega+D)^{-k}\Phi((A\Omega+B)(C\Omega+D)^{-1})$
for $\big(\begin{smallmatrix}A & B\\ C & D\end{smallmatrix}\big)\in \Gamma_2\equiv Sp(4,\mathbb Z)$. 
The analog of the familiar fundamental domain of the modular group, $\Gamma_1\equiv SL(2,\mathbb Z)$, is the
Siegel modular threefold obtained by quotienting out 
\be
\mathcal A_2 = \mathbb H_2/\Gamma_2\,,
\ee
which is the moduli space of principally polarized abelian surfaces.

Just like ordinary Jacobi forms, the Siegel modular forms for $g=2$ form a ring graded by
the weight $k$ that is finitely generated,\footnote{
There is one relation in that the square of $\chi_{35}$ can expressed in terms of the other generators.}
\be\label{Mring}
M_{*}(\Gamma_2)\equiv\bigoplus M_{k}(\Gamma_2)=\mathbb{C}\big[\psi_4, \psi_6, \chi_{10}, \chi_{12}, \chi_{35}\big]\,,
\ee
where  we denote Eisenstein series and cusp forms by $\psi_k=\psi_k(\utau)$ and
$\chi_k=\chi_k(\utau)$, respectively. 
In the following we will provide explicit expressions and a few identities for the generators  of the ring of Siegel modular forms of genus $2$.

An important feature of Siegel modular forms, and more general automorphic forms,  is that they can
be generated as liftings from ordinary,  weak or almost holomorphic Jacobi forms.
Physically this amounts to a map from world-sheet to space-time quantities.
Recall that a Jacobi form of weight $k$ and index $m$ is a function  $\phi_{k,m}(\tau,z ): \mathbb H _1\times  \mathbb C \rightarrow \mathbb C$
which obeys:
 \begin{enumerate}
  \item
    $\phi_{k,m}\left(\frac{a\tau+b}{c\tau+d},\frac{z}{c\tau+d}\right)=(c\tau+d)^ke^{\frac{2\pi
        i m c z^2}{c\tau+d}}\phi_{k,m}(\tau,z)$ for
    $\left(\begin{smallmatrix} a&
        b\\c&d\end{smallmatrix}\right)\in \Gamma_1$,
  \item $\phi_{k,m}(\tau,z+ \lambda \tau+\mu)=e^{-2\pi i m(\lambda^2\tau+2\lambda
      z)}\phi_{k,m}(\tau,z)$,\ {\rm for}\ $\lambda,\mu\in \mathbb Z$,
  \item and has a Fourier expansion\ \
 $   \phi_{k,m}(\tau,z)=\sum_{n\geq 0}\sum_{r^2\leq 4nm} c(n,r)q^n\xi^r$ with integral coefficients $c$,
 where $q\equiv e^{2\pi i \tau}$, $\xi\equiv e^{2\pi i z}$\,.
  \end{enumerate}
For more details concerning Jacobi forms we refer to \cite{EichlerZagier,Dabholkar:2012nd}.

There are two important basic versions of such liftings, both based on Hecke transforms, namely arithmetic (Maa\ss-Gritsenko)
liftings \cite{Gritsenko:1996ax,Gritsenko:1996tm} and exponential (Borcherds) liftings \cite{BORCHERDS199030}.
The arithmetic lifting is provided by the following Hecke operator  acting on index $m=1$ Jacobi forms,
\be
\label{Vmap}
\mathcal V[\,*\,]: J_{k,1} \ \longrightarrow\  M_{k}(\Gamma_2)\,,
\ee
which maps ordinary Eisenstein series and cusp forms into genus $2$ Siegel Eisenstein series and cusp forms, respectively.

Concretely, one starts from an index $m=1$ Jacobi form with even weight $k=2*\geq0$ and consider its integral Fourier expansion
\be
\phi_{k,1}(\tau,z) := \sum_{n\geq0,\, l \in \mathbb Z}c(n,l)q^n\xi^l\,.
\ee
Owing to the defining transformation properties of Jacobi forms mentioned above, 
the coefficients depend only on the discriminant $D=4n-l^2$, i.e., $c(n,l)=c(D)$. 
The action of the Hecke operator $\mathcal V$ in  (\ref{Vmap}) on the coefficients is then given by
\be
\tilde c(n,m,l):=\sum_{a|(n,m,l)}a^{k-1}c\left({nm\over a^2},{l\over a}\right)\,, 
\ee
where  $\tilde c(0,0,0)=-B_{2k}/4k \,c(0)$ for Eisenstein series (and $\tilde c(0,0,0)=0$ for cusp forms).
The image under the lift  $\mathcal V$ is the following Siegel-Maa\ss\,  form of even weight:
\be
\psi_{k}(\Omega)\ =\ \mathcal V[ \phi_{k,1}]\ = \sum_{{n,m,r\in\mathbb Z,\, n,m,\geq0\atop 4nm-l^2\geq 0}}
\tilde c(n,m,l) \,\qt^n\qu^m\qv^l\,,
\ee
where we denote: $\qt\equiv e^{2\pi iT}$, $\qu\equiv e^{2\pi iU}$ and $\qv\equiv e^{2\pi iV}$.

Explictly, for the even generators of $M_{k}$ we first consider
\bea
\phi_{4,1}(\tau,z) &\equiv&  E_{4,1}(\tau,z) = \frac{1}{2} \left(
\theta_2^6\theta_2^2(\tau,z)+\theta_3^6\theta_3^2(\tau,z)+\theta_4^6\theta_4^2(\tau,z) 
\right)
\\
&=&1+ q \left(126+\xi^2+\frac{1}{ \xi^2}+56 \xi+\frac{56}{\xi} \right)+ {\cal O}(q^2)\nn
\eea
and thus
\bea
 \psi_{4}(\utau) &= &\mathcal V[\phi_{4,1}]\ =\ 1+ 240(\qt+\qu) +
2160 ({\qt}^2+ \qu^2)    \\
&& + {\qt} \qu \left(240 \qv^2+\frac{240}{\qv^2}+13440 \qv+\frac{13440}{\qv}+30240\right) 
+ {\cal O}(q^3)\,,
\eea
while
\bea
\phi_{6,1}(\tau,z) &\equiv&  E_{6,1}(\tau,z) =  
\frac{1}{2} \left(
-(\theta_3^4+\theta_4^4)\,\theta_2^6\theta_2^2(\tau,z)+
(\theta_4^4-\theta_2^4)\,\theta_3^6\theta_3^2(\tau,z)+
(\theta_2^4+\theta_3^4)\,\theta_4^6\theta_4^2(\tau,z)\right) \nn
\\
&=&1+ q \left(-330+\xi^2+\frac{1}{ \xi^2}-88 \xi-\frac{88}{\xi} \right)+ {\cal O}(q^2)
\eea
lifts to
\bea
 \psi_{6}(\utau) &=&  \mathcal V[ \phi_{6,1}]\ =\ 1-504(\qt+\qu) 
-16632 ({\qt}^2+{\qu}^2) \\
&& +    {\qt} {\qu} \left(-504 \qv^2-\frac{504}{\qv^2}+44352 \qv+\frac{44352}{\qv}+166320\right)
+ {\cal O}(q^3)\,.
\eea
Moreover, for the cusp forms at weights $10$ and $12$ we consider:
\bea
\phi_{10,1}(\tau,z) &=& \frac1{144}\left( E_{4,1}(\tau,z)E_6(\tau)- E_{6,1}(\tau,z)E_4(\tau)\right) 
\\
&=&q\left(\xi+\frac1\xi-2\right)
+q^2\left(36-\frac2{\xi^2}-2\xi^2-\frac{16}\xi-16\xi\right)
+ {\cal O}(q^3)\nn
\eea
and therefore
\be
\label{Chi10lift}\chi_{10}(\utau) = \mathcal V[ \phi_{10,1}]\ =\ 
\qt\qu\left(\qv+\frac1\qv-2\right)+ 
\qt\qu(\qt+\qu)\left(36-\frac2{\qv^2}-2\qv^2-\frac{16}\qv-16\qv\right)
+ {\cal O}(q^4)\,.
\ee
Finally
\bea
\phi_{12,1}(\tau,z) &=& \frac1{144}\left( E_{4,1}(\tau,z)E_4^2(\tau)- E_{6,1}(\tau,z)E_6(\tau)\right) 
\\
&=&q\left(10+\xi+\frac1\xi\right)
+q^2\left(-132+\frac{10}{\xi^2}+10\xi^2-\frac{88}\xi-88\xi\right)
+ {\cal O}(q^3)\,,   \nn
\eea
which\ yields
\bea
12\chi_{12}(\utau) &=&  \mathcal V[ \phi_{12,1}]\ =\ 
\qt\qu\left(10+\qv+\frac1\qv\right) \\
&& + 
\qt\qu(\qt+\qu)\left(-132+\frac{10}{\qv^2}+10\qv^2-\frac{88}\qv-88\qv\right)
+ {\cal O}(q^4)\,.
\eea
Evidently both cusp forms vanish for large $\imt T$ and $\imt U$.

Apart from the arithmetic liftings instantiated by the Hecke operator $\mathcal V$,
there are also exponential (or Borcherds) liftings that map from Jacobi forms to Siegel
modular forms, or more general automorphic forms. 
The starting point is an (in general weak or nearly holomorphic) Jacobi form of weight zero and index $t$
$$
\phi_{0,t}(\tau,z) := \sum_{n, l \in \mathbb Z}c(n,l)q^n\xi^l\,,
$$ 
with integral Fourier coefficients $c(n,l)=c(4 n-l^2)$.
The exponential lift is then the meromorphic modular form with weight $\frac12c(0,0)$ given by 
\cite{Gritsenko:1996tm}
\be\label{Explift}
\chi_{1/2c(0,0)}(\Omega)\ =\ \text{Exp-lift}[\phi_{0,t}] =\ \qt^\alpha\,\qu^\beta\,\qv^\gamma \prod_{{n,m,l\in \mathbb Z} \atop (n,m,l)>0}\left(1-\qt^{tn}\qu^{m}\qv^l\right)^{c(nm,l)}\,,
\ee
where the positive root condition $(n,m,l)>0$ means that $n>0$, $m,l\in \mathbb Z$ or $n=0$, $m>0$, $l\in \mathbb Z$ or
$m=n=0$, $l<0$. Moreover, $\alpha=\frac14\sum_l l^2c(0,l)$, $\beta=\frac1{24}\sum_l c(0,l)$ and
$\gamma=\frac12\sum_{l>0} lc(0,l)$. For the Wilson line 
geometry we consider, only $t=1$ will be relevant.

Particularly well established in physics \cite{Mayr:1995rx,Dijkgraaf:1996xw}  is the cusp form
$$
\chi_{10}(\utau)={ \chi_{5}}^2(\utau)
$$
which, besides the arithmetic lifting (\ref{Chi10lift}),  also has a representation as exponential lifting
in terms of the following product
\bea\label{chi5def}
\chi_{5}(\utau)&=& \text{Exp-lift}[\phi_{0,1}]
\ =\ 
 (\qt\,\qu\,\qv)^{1/2} \prod_{{n,m,l\in \mathbb Z} \atop (n,m,l)>0}\left(1-\qt^n\qu^m\qv^l\right)^{c_5(4nm-l^2)}
\\
&=&    (\qt\,\qu)^{1/2}(\sqrt\qv-\frac1{\sqrt\qv})\left(1-(\qt+\qu)(10+\qv+\frac1\qv)  +{\cal O}(q^2)\right)\,,
\eea
where
\be\label{phi01}
\phi_{0,1}(\tau,z)\equiv \frac1{\eta^{24}(q)} \phi_{12,1}(\tau,z)=: \sum_{n\geq0,\, l \in \mathbb Z} c_5(4n-l^2)q^n\xi^l
\ee
is one of the standard generators of Jacobi forms. 

Moreover one can apply a certain Hecke transformation (denoted by $T_0(2)$ in \cite{Gritsenko:1996tm}) 
to obtain the following nearly holomorphic weight zero, 
index one Jacobi form
\bea\label{phi012}
\phi_{0,1}^{(2)}(\tau,z) &=& (T_0(2)-2) \phi_{0,1}(\tau,z)
\\
&=&\frac1{18\eta^{24}(q)}\left(11E_4^2E_{4,1}+7 E_6 E_{6,1}\right)(\tau,z)
\\
&=:&  \sum_{n,m}  c_{35}(n,l)q^n\xi^l\,.
\eea
This then  provides the  product representation of the remaining  cusp form: 
\bea\label{chi35}
\chi_{35}(\utau)&=&\text{Exp-lift}[\phi_{0,1}^{(2)}]\ 
=\ 
\qt^2\qu^3\qv\prod_{(n,m,l)>0}\left(1-\qt^n\qu^m\qv^l\right)^{ c_{35}(4nm-l^2)}
\\
&=& \qt^2\qu^2(\qu-\qt)(\qv-\frac1\qv)\left(1-(\qt+\qu)(70+\qv^2+\frac1{\qv^2})+{\cal O}(q^2)\right)\,.\nn
\eea
One can cancel the zero at $\V=0$ by considering
\bea
\chi_{30}(\utau)&:=&\frac{\chi_{35}(\utau)}{ \chi_{5}(\utau)}\ =\ \text{Exp-lift}[\phi_{0,1}^{(2)}-\phi_{0,1}]
\\
&=&\left(\qt^3\qu^3\qv\right)^{1/2}\!(\qu-\qt)(\qv+\frac1{\qv})\left(1-(\qt+\qu)(60+\qv^2+\frac1{\qv^2}-\qv -\frac1\qv)+{\cal O}(q^2)\right)
\nn\eea
which vanishes for $T=U$ (plus its modular images) only.

Finally we note the useful formula for the discriminant:
\bea
\Delta(\utau)&:=&\frac1{1728}(\psi_4^3(\utau)-\psi_6^2(\utau))
\\
&=& \qt+\qu- 24(\qt+\qu)^2-\qt\qu(186-\frac1{\qv^2}-\qv^2+\frac{28}\qv+28\qv)
+{\cal O}(q^2)\nn
\eea
This vanishes if both $\imt T$ and $\imt U$ become large, 
which characterizes a Kulikov Type III limit.

\bibliography{papers}
\bibliographystyle{JHEP}

\end{document}